\documentclass[lualatex,a4paper,11pt,twoside,openany]{article}
\usepackage{amsmath,amssymb}
\usepackage{amsfonts}
\usepackage{bm}
\usepackage{graphicx}
\usepackage{subfigure}
\usepackage{verbatim}
\usepackage{wrapfig}
\usepackage{ascmac}
\usepackage{makeidx}
\usepackage{mathrsfs}
\usepackage{color}

\usepackage{authblk}
\usepackage{caption}

\usepackage{cite}
\usepackage{pdfpages}
\usepackage{flushend}

\usepackage[pdfencoding=auto]{hyperref}
\hypersetup{%
  bookmarksnumbered=true,%
  colorlinks=false,%
  pdftitle={タイトル},%
  pdfauthor={作成者},%
  pdfsubject={サブタイトル},%
  pdfkeywords={キーワード1\000\012キーワード2\000\015キーワード3\000\015\000\012キーワード4}}

\title{
\bf
X-ray multiple-beam ($n$-baem) dynamical diffraction theories,
numerical methods to solve them
and experimental verification by using the synchrotron X-rays
\footnote{
The present manuscript has been translated by the author
for submission 
to arXiv from a review article published in
Journal of the Japanese Society for Synchrotron Radiation Research
(2020) {\bf 33} 61-80
[in Japanese].
}
}

\author{Kouhei OKITSU}
\affil{
Nano-Engineering Research Center,
Institute of Engineering Innovation,
Graduate School of Engineering,
The University of Tokyo,
2-11-16 Yayoi, Bunkyo-ku, Tokyo 113-8656, Japan.
Correspondence e-mail: tkokitsu@g.ecc.u-tokyo.ac.jp
}

\begin{document}

\maketitle

\begin{minipage}{0.9\linewidth}
\begin{abstract}
% \ {\bf 要\ \ 旨}\ \ 
% 完全ないしは完全に近い結晶に入射したX線の振る舞いは，
% 動力学的回折理論(動力学理論)によって記述される。
% 入射し透過するX線と，ひとつのブラッグ反射波のみが強いとする，
% 2波動力学理論には，100年に及ぶ蓄積があるが，
% 3つ以上の波が同時に強い多波($n$波)ケースの研究人口は少ない。
% 筆者は，高木方程式(T-T方程式)を多波ケースに拡張し，数値解法を研究，
% X線の偏光状態を制御した放射光実験により，その検証を行ってきた。
% % エバルト-ラウエの動力学理論(E-L理論)とT-T方程式が，多波ケースにおいても，
% % フーリエ変換で記述される等価な関係にあることを示し，
% % 計算および実験の手法と結果について解説する。
Behavior of X-rays diffracted in a perfect or quasi-perfect crystal
can be described by the dynamical theory of X-ray diffraction.
Study on the two-beam cases
in which only transmitted and one reflected X-ray beams are strong
has a history of one hundred years.
However, the population of researchers who study on
the multiple-beam cases ($n$-beam cases) in which
more than two beams are simultaneously strong is small.
The present author has derived the Takagi-Taupin (T-T) dynamical theory
that can be applied to the $n$-beam cases,
coded the computer programs to solve it and
experimentally verified them by using the synchrotron X-rays.
The equivalence between the Ewald-Laue (E-L)
and the T-T dynamical theories
described by the Fourier transform also for the $n$-beam cases
is explicitly verified in the present paper.
Further, the methods of the computer simulations and the experiments
are also described.

{\bf
Furthermore, a hypothesis concerning the too large values of $R$-factor
in protein crystallography is also described.
This might be extremely important in protein crystallography
in the future.} \\


\end{abstract}
\end{minipage}

% \section{はじめに}
\section{Introduction}

% 完全結晶で回折されるX線の振る舞いを記述する理論は
% 動力学的回折理論(動力学理論)とよばれ，
% 1912年にラウエによって，
% 結晶によるX線の回折現象が発見された直後の
% 1910年代には，
% ダーウィン
% \cite{darwin1914a,darwin1914b}
% やエバルト
% \cite{ewald1917}
% らによって理論の基礎が与えられた。
The theory that describes the behavior of X-rays
diffracted in a perfect or quasi-perfect crystal
is called as the dynamical theory.
Just after the discovery of the X-ray diffraction
by von.\ Laue,
basic theories were given by
Darwin
\cite{darwin1914a,darwin1914b}
and by Ewald
\cite{ewald1917}.
% 今日，最も広く用いられている動力学理論は，
% エバルトによってその基礎が与えられ，
% 1931年，ラウエが完成させた
% \cite{laue1931}
% エバルト-ラウエ動力学理論(E-L理論)である。
The most widely known dynamical theory
is the Ewald-Laue (E-L) theory that has been derived by
applying the two-beam approximation to
the fundamental equation given by von.\ Laue
\cite{laue1931}.
% ラウエによって与えられた動力学理論基本方程式
% \cite{laue1931}
% に，
% 入射X線とひとつの反射X線のみが強いとする2波近似を適用した，
% 2波のE-L理論が，多くの教科書に記述されている
% \cite{miyake1969,azaroff1974,kato1978,pinsker1978,kohra1979,kato1995,authier20% 05,kikuta2011}。
There are several textbooks that describe the dynamical theory
% \cite{miyake1969,azaroff1974,kato1978,pinsker1978,kohra1979,kato1995,authier2005,kikuta2011}.
\cite{zachariasen1945,azaroff1974,pinsker1978,authier2005}.

% 

% 一方，1962年，高木によって発表された
% \cite{takagi1962,takagi1969,takagi1971,taupin1964}
% 高木-トウパンの式(T-T理論,T-T方程式)は，
% 結晶格子歪みを取り扱える理論として，
% 当時黎明期にあった半導体産業から，
% シリコン結晶の評価手段として
% 広く受け入れられ，
% 様々な結晶格子欠陥に対する，X線トポグラフ図形の
% 計算機シミュレーションが行われた
% \cite{epelboin1985,epelboin1987}。
However, the Takagi-Taupin (T-T) equation
that has been derived by Takagi
\cite{takagi1962,takagi1969,taupin1964,kato1973}
was accepted as another form of the dynamical theory.
It can deal with the X-ray wave field
in a distorted crystal.
Various images of the crystal defects
were computer-simulated based on the T-T equation
\cite{epelboin1985,epelboin1987}.

% ひとつの反射面がブラッグの反射条件を満たすとき，
% 逆空間では，逆格子原点H$_0$とひとつの逆格子点H$_1$が
% エバルト球上に存在するが，$\overrightarrow{{\rm H}_0{\rm H}_1}$
% 軸周りに結晶を回転させると，
% H$_0$, H$_1$以外の逆格子点がエバルト球上に載るようにできることは，
% 容易に理解できる。
% $\overrightarrow{{\rm H}_0{\rm H}_1}$軸周りに結晶を回転させ，
% $\overrightarrow{{\rm H}_0{\rm H}_1}$による反射X線の
% 強度を測定するスキャンは，
% 1937年レニンガー
% \cite{renninger1937}
% によって最初に報告されたことから，
% レニンガースキャンとよばれる。
% \index{れにんがーすきゃん@レニンガースキャン}
% \index{れにんがー@レニンガー}
Incidentally, it can easily be understood
that $n$ reciprocal lattice nodes $(n \ge 3)$
can exist on the surface of the Ewald sphere
by rotating the crystal
around the axis of $\overrightarrow{{\rm H}_0{\rm H}_1}$.
Here, ${\rm H}_0$ is the origin of the reciprocal space
and ${\rm H}_1$ is the reciprocal lattice node that causes
the $\overrightarrow{{\rm H}_0{\rm H}_1}$ reflection.
X-ray intensity measurement taken by rotating
the $\overrightarrow{{\rm H}_0{\rm H}_1}$ axis
is called as the Renninger scan
\cite{renninger1937}.

% 
% また，X線スペクトロスコピーにおいては，
% シリコンやダイヤモンドが分光結晶として用いられるが，
% 特定の光子エネルギーでX線反射強度に不連続が見られることがあり，
% グリッチ(glitch)とよばれている。
% 「glitch」は，故障，不具合といった意味の普通名詞である。
% 分光結晶の回転によりエバルト球は伸縮するが，
% 特定の光子エネルギーでH$_0$, H$_1$以外の逆格子点
% がエバルト球の上に載ったときに，グリッチが発生する。
While silicon, diamond and/or germanium crystals
are usually used as the monochromator
in the energy scan of X-ray spectroscopy,
discontinuities of the X-ray intensity
are frequently found and referred to as glitches.
When scanning the photon energy of X-rays by
rotating the monochromator crystal,
the radius of the Ewald sphere changes.
Then, a third reciprocal lattice node
other than the origin of the reciprocal space H$_0$
and reciprocal lattice node H$_1$
giving the primary reflection
causes the glitch when it exists on the surface
of the Ewald sphere.

% H$_0$, H$_1$，$\cdots$, H$_{n-1}$の逆格子点が
% 同時にエバルト球上にあり，$n$個の波が同時に強いケースを
% $n$波ケースとよぶことにする。
% エバルト-ラウエ理論が，$n$波ケースに拡張されたのは，
% 1967-1968年だった
% \cite{joko1967,hildebrandt1967,ewald1968,heno1968}。
% コレラによって数値解が初めて与えられたのは
% 1974年のことである
% \cite{colella1974}。

% \index{らうえ@ラウエ}
% \index{えばると@エバルト}
Let refer to the case that
$n$-reciprocal lattice nodes H$_0$, H$_1$, H$_2$,
$\cdots$, H$_{n-1}$ exist on the surfaces of the Ewald sphere
as the $n$-beam cases.
The E-L two-beam dynamical diffraction theory
was extended such as to deal with the $n$-beam cases
in 1965-1968
\cite{saccocio1965a,saccocio1965b,joko1967,hildebrandt1967,ewald1968,heno1968}.
The numerical method to solve the theory was given by Colella in 1974
\cite{colella1974}.

% 一方，T-T理論の$n$波ケースへの拡張は，随分遅れた。
% 偏光の取扱が厄介であるため，
% 1987年，偏光を無視した3波ケースへの拡張が行われ
% \cite{thorkildsen1987}，
% 1998年に偏光を考慮する3波ケースの方程式
% \cite{larsen1998a}
% が初めて報告された。
% 偏光の効果を考慮しての，
% $n \in \{3, 4, 6, 8, 12\}$の$n$波ケースへの拡張
% \cite{okitsu2003a}
% と，計算機による数値解法の開発，および6波ケースでの
% 放射光によるピンホールトポグラフ実験結果との比較による検証
% \cite{okitsu2003b,okitsu2006,okitsu2011a}
% は2003年，筆者らによる報告が初めてとなった。
% 以降，2012年までに
% $n \in \{3, 4, 5, 6, 8, 12\}$の$n$波ケース
% について，計算機シミュレーションと
% 実験結果のよい一致を報告してきた
% \cite{okitsu2012}。
However,
the extension of the T-T equation
was delayed for many years
due to the complexity
when dealing with the polarization effect of X-rays.
The three-beam T-T equation that neglected the polarization effect
was given by Thorkildsen in 1987
\cite{thorkildsen1987}.
The T-T equation that takes into account the polarization factor
was for the first time given by Larsen and Thorkildsen in 1998
\cite{larsen1998a}.
The present author reported
the T-T equation extended to the $n$-beam cases
for $n \in \{3, 4, 6, 8, 12\}$ in 2003
\cite{okitsu2003a}.
The numerical method to solve it
and experimental verification by using the synchrotron radiation
was given by the present author and coauthors
for a six-beam case
\cite{okitsu2003b,okitsu2006,okitsu2011a}.
The computer-simulated and experimentally obtained results
to be compared with each other
for the $n$-beam cases
were reported
for $n \in \{3, 4, 5, 6, 8, 12\}$ in 2012 by Okitsu, Imai and Yoda
\cite{okitsu2012}.
The excellent agreements were found between the computer-simulated
and the experimentally obtained pinhole topographs.

% 2005年に最初に報告したことであるが
% \cite{okitsu2005a}，
% E-L理論とT-T理論の間には，フーリエ変換で記述される
% 単純な関係があり，等価である。
% いずれもラウエによる動力学理論基本方程式
% \cite{laue1931}
% から導出されているので，当然なのであるが，
% このことが十分に認識されていなかったことが，
% T-T理論の$n$波ケースへの拡張が遅れた主な原因ではないかと思われる。
Between the E-L and the T-T dynamical theories,
there is a simple relation described by the Fourier transform
that has been implicitly recognized but
has been explicitly described for the first time in 2012
\cite{okitsu2012}.
It can be recognized that
this delayed the extension to the $n$-beam cases of the T-T equation
in comarison with the E-L dynamical theory.

% 本稿では，ラウエの動力学理論基本方程式から，
% まず$n$波E-L理論を導出し，
% これをフーリエ変換することで，
% $n$波T-T方程式を導出する。
% 逆に$n$波T-T方程式をフーリエ変換して
% $n$波E-L理論を導出できることから，
% E-L理論とT-T理論が等価であること
% \cite{okitsu2005a,okitsu2012}
% を，明示的に記述する。
In the present paper,
the $n$-beam E-L theory is derived
from Laue's fundamental equation of the dynamical theory
\cite{laue1931}
at first.
Then, the $n$-beam T-T equation is derived by
Fourier-transforming the $n$-beam E-L theory.
The equivalence between the $n$-beam E-L and T-T dynamical theories
is explicitly described
also for an arbitrary number of $n$.

% $n$波の動力学的回折は，
% E-L理論とT-T方程式のどちらでも記述でき，
% 数値解を求めることができる。
% そして，それぞれに，長所，短所がある。
% その事を理解した上で，これらを使い分けるべきだ，
% というのが筆者の見解である。

The $n$-beam dynamical diffraction phenomena of X-rays
can be described by both the T-T and E-L theories
and be numerically solved.
The numerical methods to solve these theories have
advantages and disadvantages when compared with each other.
The present author considers that
they should be used depending to the purpose
with this recognition.

% 世界的に現在最も広く読まれている
% 動力学理論の教科書は，
% オーティエによる著書
% \cite{authier2005}
% だと思われる。
% 500ページ以上の大著であるが，
% 多波ケース($n$波ケース)に関する記述は，わずか24ページである。
% 多波ケースに特化して記述した教科書としては，
% チャンによる2004年の著書
% \cite{chang2004}
% がある。
% ピンスカーの1978年の著書
% \cite{pinsker1978}
% には，多波のE-L理論に関するやや詳しい記述がある。
% レビューとしては，
% ベッカートとヒュマーによるもの
% \cite{weckert1997,weckert1998}，
% コレラによるもの
% \cite{colella1995a,colella1995b}
% がある。
Authier's book describing the dynamical theory
of X-ray diffraction
\cite{authier2005}
is recognized to be the most widely read textbook
that has over 500 pages.
However, it has only 24 pages for description
on the $n$-beam diffraction.
In Pinsker's book
\cite{pinsker1978},
descriptions concerning the $n$-beam E-L theory
are found.
These have been revied by
Weckert and H\"{u}mmer
\cite{weckert1997,weckert1998}
and Colella
\cite{colella1995a,colella1995b}.

% \section{エバルト-ラウエ(E-L)n波理論の導出}

\section{Derivation of the Ewald-Laue (E-L) $n$-beam dynamical theory}

% 次の式は，ラウエによって導出された，
% 動力学理論基本方程式である。

% \index{らうえ@ラウエ}
% \index{えばると@エバルト}

The following equation is the fundamental equation of the dynamical theory
\cite{laue1931}:
\begin{align}
\frac{\ k_{i}^{2} - K^{2} \ }{k_{i}^{2}} \bm{\mathcal D}_{i}
  = \sum_j \chi_{h_{i} - h_{j}} \left[\bm{\mathcal D}_{j}\right]_{\perp {\mathbf k}_{i}}.
\label{eq01:fundamental}
\end{align}
% $k_{i}$は，$i$番目のブロッホ波の波数，波数ベクトルは，
% ${\mathbf k}_{i}$ $=$ ${\mathbf k}_0 + {\mathbf h}_i$である。
% ${\mathbf k}_0$は，前方回折波の波数ベクトル，
% ${\mathbf h}_i$は散乱ベクトルである。
% $K (= 1/ \lambda)$は，入射X線の真空における波数で，
% $\lambda$は波長である。
% $\bm{\mathcal D}_{i}$と$\bm{\mathcal D}_{j}$
% は，$i$番目と$j$番目のブロッホ波の複素振幅ベクトルである。
% $\sum_j$は，$j$に関する無限のサンメーションである。
% $\chi_{h_i - h_j}$は，電気分極率のフーリエ係数，
% $[\bm{\mathcal D}_{j}]_{\perp {\mathbf k}_{i}}$は，
% 複素振幅ベクトル$\bm{\mathcal D}_{j}$の
% ${\mathbf k}_{i}$に垂直な成分である。
Here,
$k_{i}$ is the wavenumber of the $i$th numbered Bloch wave
whose wavevector is
${\mathbf k}_{i}$
($=$ ${\mathbf k}_0 + {\mathbf h}_i$)
of the $i$th numbered Bloch wave.
${\mathbf k}_0$ is the forward-diffracted X-ray beam
in the crystal.
${\mathbf h}_i$ is the scattering vector.
$K (= 1 / \lambda)$ is the the wavenumber of
the incident X-rays in vacuum where
$\lambda$ is the wavelength of them.
$\bm{\mathcal D}_{i}$ and $\bm{\mathcal D}_{j}$
are amplitude vectors of the $i$th and $j$th numbered Bloch waves.
$\sum_j$ is the infinite summation for $j$.
$\chi_{h_i - h_j}$ is the Fourier coefficient of the electric susceptibility.
$[\bm{\mathcal D}_{j}]_{\perp {\mathbf k}_{i}}$
is the vector component
of $\bm{\mathcal D}_{j}$ perpendicular to ${\mathbf k}_{i}$.

% $k_i + K \approx 2 k_i$の近似を式(\ref{eq01:fundamental})に適用して，
% 次の式が得られる。
By applying the approximation of $k_i + K \approx 2 k_i$
to (\ref{eq01:fundamental}),
the following equation can be obtained:
\begin{align}
\xi_i \bm{\mathcal D}_i = \frac{\ K \ }{2} \sum_j
    \chi_{h_i - h_j}
    \left[ \bm{\mathcal D}_j \right]_{\perp {\mathbf k}_{i}},
\label{eq02:fundamental2}\\
{\rm where} \ \ \xi_i = k_i - K. \nonumber
\end{align}
% 電気変位ベクトル$\bm{\mathcal D}_i$，$\bm{\mathcal D}_j$は，
% スカラー振幅の1次結合で次のように表すことができる。
The electric displacement vectors $\bm{\mathcal D}_i$ and
$\bm{\mathcal D}_j$ can be reperesented as linear combinations of
the scalar electric displacements as follows:
\begin{subequations}
\begin{align}
\bm{\mathcal D}_i = {\mathcal D}_i^{(0)} {\mathbf e}_i^{(0)}
                  + {\mathcal D}_i^{(1)} {\mathbf e}_i^{(1)},
\label{eq03:VectorScalara} \\
\bm{\mathcal D}_j = {\mathcal D}_j^{(0)} {\mathbf e}_j^{(0)}
                  + {\mathcal D}_j^{(1)} {\mathbf e}_j^{(1)}.
\label{eq03:VectorScalarb}
\end{align}
\label{eq03:VectorScalar}
\end{subequations}
% ${\mathbf s}_i$が${\mathbf k}_i$方向の単位ベクトルであるとき，
% ${\mathbf s}_i$に垂直な単位ベクトル
% ${\mathbf e}_i^{(0)}$と
% ${\mathbf e}_i^{(1)}$は，
% ${\mathbf s}_i$,
% ${\mathbf e}_i^{(0)}$,
% ${\mathbf e}_i^{(1)}$が，
% 右手直交系をなすように定義する。
% $j$についても同様である。
When ${\mathbf s}_i$ is a unit vector
in the direction of ${\mathbf k}_i$ and then
${\mathbf e}_i^{(0)}$ and ${\mathbf e}_i^{(1)}$
are defined such that
${\mathbf s}_i$, ${\mathbf e}_i^{(0)}$ and ${\mathbf e}_i^{(1)}$
construct a right-handed orthogonal system in this order.
${\mathbf s}_j$, ${\mathbf e}_j^{(0)}$ and ${\mathbf e}_j^{(1)}$
are defined in the same way.

\begin{figure}[htbp]
\begin{center}
% \centering
\includegraphics[width=0.5\textwidth]{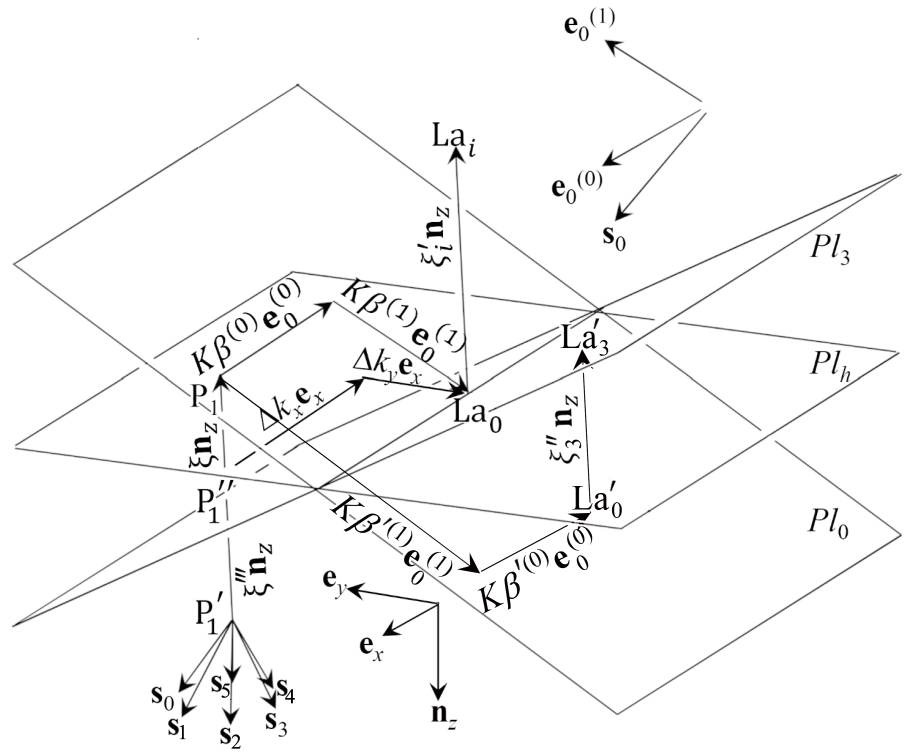}
\end{center}
\captionsetup{width=0.85\textwidth}
\caption[
Geometry around the Laue point ${\rm La}_0$.
$Pl_0$ and $Pl_3$ are planes whose distance from H$_0$ and H$_3$
is $K (= 1/ \lambda)$.
$Pl_h$ is a plane normal to ${\mathbf n}_z$ (downward surface normal).
The Laue point ${\rm La}_0$ and the point
${\rm P}_1^{\prime\prime}$ exist on $Pl_h$.
$Pl_i$ $(i \ne \{ 0, 3 \})$ were not drawn for simplicity.
${\rm La}_i$ and ${\rm La}_i^{\prime}$ are points whose distance from
H$_i$ $(i \in \{ 0,1, \cdots, n-1 \})$ is $K$.
${\rm P}_1^{\prime}$ is the start point of wavevector of the Bloch wave.
${\rm P}_{1,k}^{\prime}$
that appears in equation\ (\ref{eq51_DjPrimeDefinition})
is the $k$th numbered ${\rm P}_{1}^{\prime}$
i.e. the start point of wavevector
of the $k$th numbered Bloch wave
where $k \in \{ 1, 2, 3, \cdots, 2n \}$.
        ]
        {
Geometry around the Laue point ${\rm La}_0$.
$Pl_0$ and $Pl_3$ are planes whose distance from H$_0$ and H$_3$
is $K (= 1/ \lambda)$.
$Pl_h$ is a plane normal to ${\mathbf n}_z$ (downward surface normal).
The Laue point ${\rm La}_0$ and the point
${\rm P}_1^{\prime\prime}$ exist on $Pl_h$.
$Pl_i$ $(i \ne \{ 0, 3 \})$ were not drawn for simplicity.
${\rm La}_i$ and ${\rm La}_i^{\prime}$ are points whose distance from
H$_i$ $(i \in \{ 0,1, \cdots, n-1 \})$ is $K$.
${\rm P}_1^{\prime}$ is the start point of wavevector of the Bloch wave.
${\rm P}_{1,k}^{\prime}$
that appears in equation\ (\ref{eq51_DjPrimeDefinition})
is the $k$th numbered ${\rm P}_{1}^{\prime}$
i.e. the start point of wavevector
of the $k$th numbered Bloch wave
where $k \in \{ 1, 2, 3, \cdots, 2n \}$.
}
\label{Fig01_org_05_RecipPosition}
\end{figure}

% 逆空間に作図した
% Fig.\ \ref{Fig01_org_05_RecipPosition}
% を参照しながら，以下を記述する。
% $n$個の逆格子点が同一円上に存在する場合に限定して，
% 最も対称性が高い立方晶に対して$n$の値を検討すると，
% $n \in \{ 3, 4, 5, 6, 8, 12 \}$となる。
% $n$個以外の逆格子点はエバルト球表面から十分に遠いという近似を適用する。
% ${\rm La}_0$(ラウエ点)は，
% H$_i$ $(i \in \{ 0, 1, \cdots, n - 1 \})$
% からの距離が$K$ $(= 1 / \lambda)$である点，
% $Pl_i$は，H$_i$を中心とする，半径$K$の球面を
% 近似する平面である。
% $Pl_0$と$Pl_3$のみを描いてある。
% ラウエ点は，のちに
% $|\overrightarrow{{\rm La}_0{\rm H}_i}| \ne K$の場合にも
% 適用できるよう理論を拡張するので，あえてLa$_0$とする。
% 
% ${\rm P}_1$は
% $Pl_0$上にあり，
% 入射波の波数ベクトルの始点(終点はH$_0$)である。
% また，偏光因子$S$と$C$を次のように定義する。
With regard to the following description,
Fig.\ \ref{Fig01_org_05_RecipPosition}
should be referred.
When considering the number of $n$ for cubic crystals
with the highest symmetry,
the number of reciprocal lattice nodes $n$ that can simultaneously exist
on the surface of the Ewald sphere,
is restricted to be 3, 4, 5, 6, 8 and 12
by applying the approximation that
the distances of the reciprocal lattice nodes
other than $n$ reciprocal lattice nodes
to be taken into account,
are sufficiently far from the surface of the Ewald sphere.
The Laue point ${\rm La}_0$ is the point whose distance from
the reciprocal lattice nodes H$_i$
$(i \in \{ 0, 1, \cdots, n-1 \})$ is $K$
$(= 1 / \lambda)$.
$\lambda$ is the wavelength of the X-rays in vacuum.
$Pl_i$ is a plane surface that approximates the sphere
whose radius is $K$ and center is H$_i$.
Only $Pl_0$ and $Pl_3$ are drawn in Fig.\ \ref{Fig01_org_05_RecipPosition}.
The Laue point is darely symbolized as La$_0$ such that
the theories can be extended for the cases
that $|\overrightarrow{{\rm La}_0{\rm H}_i}|$
is not exactly equal $K$.

${\rm P}_1$ is the point on $Pl_0$ that is the start point
of the wavevector of the incident X-rays whose end point is H$_0$.
The polarization factors $C$ and $S$ are defined as follows,
\begin{alignat}{1}
& \qquad\qquad
{\mathbf e}_j^{(m)} = S_{i, j}^{(m)} {\mathbf s}_i
+ C_{i, j}^{(0, m)} {\mathbf e}_i^{(0)}
+ C_{i, j}^{(1, m)} {\mathbf e}_i^{(1)}.
\label{eq04:polarizationfactor_a}
\end{alignat}
Therefore,
\begin{subequations}
\begin{alignat}{1}
& \qquad\qquad S_{i, j}^{(m)} = {\mathbf s}_i \cdot {\mathbf e}_j^{(m)},
\label{eq04:polarizationfactor_b} \\
& \qquad\qquad C_{i, j}^{(0, m)} = {\mathbf e}_i^{(0)} \cdot {\mathbf e}_j^{(m)},
\label{eq04:polarizationfactor_c} \\
& \qquad\qquad C_{i, j}^{(1, m)} = {\mathbf e}_i^{(1)} \cdot {\mathbf e}_j^{(m)}.
\label{eq04:polarizationfactor_d}
\end{alignat}
\label{eq04:polarizationfactor}
\end{subequations}
% 式(\ref{eq02:fundamental2})の左辺と右辺に，それぞれ，
% 式(\ref{eq03:VectorScalara})と(\ref{eq03:VectorScalarb})を代入して
% 次の式を得る。
By substituting (\ref{eq03:VectorScalara}) and (\ref{eq03:VectorScalarb})
into the left and right sides of (\ref{eq02:fundamental2}), respectively,
the following equations can be obtained:
\begin{alignat}{1}
% & \xi_i \big(
    \xi_i \big(
      \mathcal{D}_i^{(0)} {\mathbf e}_i^{(0)}
    + \mathcal{D}_i^{(1)} {\mathbf e}_i^{(1)}
        \big)
% \nonumber \\
% & \qquad = \frac{\ K\ }{2}
  &        = \frac{\ K\ }{2}
    \sum_{j = 0}^{n - 1} \chi_{h_i - h_j}
    \big[
      \mathcal{D}_j^{(0)} {\mathbf e}_j^{(0)}
    + \mathcal{D}_j^{(1)} {\mathbf e}_j^{(1)}
    \big]_{\perp {\mathbf k}_{i}}.
\label{eq05_FundEqSclar1}
\end{alignat}
% 式(\ref{eq05_FundEqSclar1})右辺に
% 式(\ref{eq04:polarizationfactor_a})を代入して
% ${\mathbf e}_i^{(l)}$の項を比較することにより，
By substituting (\ref{eq04:polarizationfactor_a}) into
the right side of (\ref{eq05_FundEqSclar1}) and comparing
the terms of  ${\mathbf e}_i^{(l)}$,
\begin{alignat}{1}
\xi_i \mathcal{D}_i^{(l)}
  = \frac{\ K\ }{2}
    \sum_{j = 0}^{n - 1} \chi_{h_i - h_j}
    \sum_{m = 0}^{1}
      C_{i, j}^{(l, m)}
      \mathcal{D}_j^{(m)}.
\label{eq06_FundEqSclar2}
\end{alignat}
% $\overrightarrow{{\rm P}_1^{\prime}{\rm P}_1}$は，
% X線入射側結晶表面の下向き単位法線ベクトル${\mathbf n}_z$
% に平行で，次のように表されるものとする。
$\overrightarrow{{\rm P}_1^{\prime}{\rm P}_1}$ is
parallel to the downward surface normal ${\mathbf n}_z$ of the crystal
and described as follows:
\begin{align}
\overrightarrow{{\rm P}_1^{\prime}{\rm P}_1} = \xi {\mathbf n}_z.
\label{eq07_xidefine}
\end{align}
% $\beta^{(0)}$と$\beta^{(1)}$は，
% X線入射角の
% % 厳密な$n$波条件からの角度のズレで，
% Fig.\ \ref{Fig01_org_05_RecipPosition}を参照して，
$\beta^{(0)}$ and $\beta^{(1)}$ are the angular deviations of the start
point of wavevector of the incident X-rays.
In reference with Fig.\ \ref{Fig01_org_05_RecipPosition},
they are described as follows:
\begin{align}
\overrightarrow{{\rm P}_1{\rm La}_0}
  = K \beta^{(0)} {\mathbf e}_0^{(0)}
  + K \beta^{(1)} {\mathbf e}_0^{(1)}.
\label{eq08_P1La}
\end{align}
% $\xi_i$ $(= k_i - K)$は，
% ${\mathbf s}_i$と${\mathbf k}_i - {\mathbf K}_i$
% の内積をとることにより得られる。
$\xi_i$ $(= k_i - K)$ is obtained from the scalar product
of ${\mathbf s}_i$ and ${\mathbf k}_i - {\mathbf K}_i$.
% ここで，
% ${\mathbf k}_i = \overrightarrow{{\rm P}_1^{\prime}{\rm H}_i}$,
% ${\mathbf K}_i = \overrightarrow{{\rm La}_0{\rm H}_i}$
% である。
Here, ${\mathbf k}_i = \overrightarrow{{\rm P}_1^{\prime}{\rm H}_i}$ and
${\mathbf K}_i = \overrightarrow{{\rm La}_0{\rm H}_i}$.
% 式(\ref{eq07_xidefine})と式(\ref{eq08_P1La})の足し算と，
% ${\mathbf s}_i$の内積をとって，
% 式(\ref{eq04:polarizationfactor_b})を代入すると，
(\ref{eq04:polarizationfactor_b}) can be substituted into
the scalar product of
${\mathbf s}_i$
$\cdot$
$\overrightarrow{{\rm P}_1^{\prime}{\rm La_0}}$
% ${\mathbf s}_i$ and [(\ref{eq07_xidefine}) $+$ (\ref{eq08_P1La})]
to obtain the following equation:
\begin{subequations}
\begin{alignat}{1}
\xi_i
 &= {\mathbf s}_i \cdot \big(
      \overrightarrow{{\rm P}_1^{\prime}{\rm P}_1}
    + \overrightarrow{{\rm P}_1{\rm La}_0}
                        \big)
% \nonumber \\
\label{eq09_XiCal_a} \\
 &= \xi {\mathbf s}_i \cdot {\mathbf n}_z
  + K \beta^{(0)} {\mathbf s}_i \cdot {\mathbf e}_0^{(0)}
  + K \beta^{(1)} {\mathbf s}_i \cdot {\mathbf e}_0^{(1)}
% \nonumber \\
\label{eq09_XiCal_b} \\
 &= \xi \cos \Theta_i + K \beta^{(0)} S_{i, 0}^{(0)}
                      + K \beta^{(1)} S_{i, 0}^{(1)}.
\label{eq09_XiCal_c}
\end{alignat}
\label{eq09_XiCal}
\end{subequations}
% 式(\ref{eq09_XiCal})を式(\ref{eq06_FundEqSclar2})左辺に代入
% すると，次の式が得られる。
(\ref{eq09_XiCal}) can be substituted into
the left side of (\ref{eq06_FundEqSclar2})
to obtain the following equation:
\begin{alignat}{1}
\xi \cos \Theta_i {\mathcal D}_i^{(l)}
%  &+ K \left(
    + K \left(
           S_{i, 0}^{(0)} \beta^{(0)}
         + S_{i, 0}^{(1)} \beta^{(1)}
        \right)
       {\mathcal D}_i^{(l)}
% \nonumber \\
   &=   \frac{\ K \ }{2} \sum_{j = 0}^{n - 1} \chi_{h_i - h_j}
                     \sum_{m = 0}^{1}
                 C_{i, j}^{(l, m)}
                  {\mathcal D}_j^{(m)}.
   \label{eq10:eln-beam}
\end{alignat}
Here, $i, j \in \{ 0,1, \cdots, n-1 \}$,
$n \in \{ 3,4,5,6,8,12 \}$ and
$l, m \in \{ 0, 1 \}$.
% $\Theta_i$は，${\mathbf s}_i$と${\mathbf n}_z$のなす角である。
% さらに，式(\ref{eq10:eln-beam})の両辺を
% $\cos \Theta_i$で割ることにより次の式が得られる。
$\Theta_i$ is the angle spanned by ${\mathbf s}_i$ and ${\mathbf n}_z$.
By deviding the both sides of (\ref{eq10:eln-beam}) by $\cos \Theta_i$,
the following equation can be obtained:
\begin{alignat}{2}
\xi  {\mathcal D}_i^{(l)}
%  &= & &- \frac{K}{\ \cos \Theta_i\ } \left(
   &= \  - \frac{K}{\ \cos \Theta_i\ } \left(
           S_{i, 0}^{(0)} \beta^{(0)}
         + S_{i, 0}^{(1)} \beta^{(1)}
                                       \right)
             {\mathcal D}_i^{(l)}
% \nonumber \\
%  &+ & &\frac{K}{\ 2 \cos \Theta_i\ }
    +    \frac{K}{\ 2 \cos \Theta_i\ }
                     \sum_{j = 0}^{n - 1} \chi_{h_i - h_j}
                     \sum_{m = 0}^{1}
                 C_{i, j}^{(l, m)}
                  {\mathcal D}_j^{(m)}.
                  \label{eq11:eln-beamDash}
\end{alignat}
% 式(\ref{eq11:eln-beamDash})は，
% 行列とベクトルを用いて，次のように表される。
(\ref{eq11:eln-beamDash}) can also be described by using
matrices and a vector as follows:
\begin{align}
\xi {\mathbf E} \pmb{\mathscr{D}}
= {\mathbf A} \pmb{\mathscr{D}}.
\label{eq12:matrix}
\end{align}
% ${\mathbf E}$は，
% $2 n \times 2 n$の単位行列である。
Here, ${\mathbf E}$ is a $2 n \times 2 n$ unit matrix.
% $\pmb{\mathscr D}$は，$2 n$次元の列ベクトルで，
% その$q$番目の要素は
% ${\mathcal D}_j^{(m)}$
% ($q = 2 j + m + 1$)
% である。
$\pmb{\mathscr D}$ is a $2n$-dimensional column vector whose
$q$th element  is ${\mathcal D}_j^{(m)}$
where $q = 2 j + m + 1$.
% 以降，固有ベクトルないしは，固有ベクトルを並べた行列を，
% 花文字で表すこととする。
Hereafter, eigenvector or matrix whose column vectors are eigenvectors,
are symbolized with flower characters.
% ${\mathbf A}$は$2 n \times 2 n$の正方行列で
% その$p$行$q$列目の要素$a_{p, q}$($p = 2 i + l + 1$)は
% 次のように与えられる。
The element of the $p$th raw and $q$th column of
the $2 n \times 2 n$ matrix ${\mathbf A}$,
$a_{p, q}$($p = 2 i + l + 1$) is given as follows:
\begin{alignat}{1}
a_{p, q} &= \frac{\ K \ }{\ 2 \cos \Theta_i \ }
            \chi_{h_i - h_j} C_{i, j}^{(l, m)}
% \nonumber \\
%        &- \frac{\ \delta_{p, q} K}{\ \cos \Theta_i\ }
          - \frac{\ \delta_{p, q} K}{\ \cos \Theta_i\ }
                          \left( S_{i, 0}^{(0)} \beta^{(0)}
                               + S_{i, 0}^{(1)} \beta^{(1)}
                          \right).
\label{eq13:ElementOfMatrix}
\end{alignat}
% ここで，$p = 2 i + l + 1$,
% $\delta_{p, q}$は，クロネッカーデルタである。
% 式(\ref{eq12:matrix})は，
% $2 n$組の固有値$\xi$と固有ベクトル$\pmb{\mathscr D}$を持つ
% 固有値/固有ベクトル問題
% である。
% 固有値$\xi$が，ブロッホ波の波数ベクトルに制限を与え，
% 固有ベクトルが振幅比を与える。
Here, $\delta_{p, q}$ is the Kronecker delta.
(\ref{eq12:matrix}) describes an eigenvalue problem
whose $2 n$ eigenvalues are $\xi$ and
$2n$ eigen vectors are $\pmb{\mathscr D}$.
$\xi$ restricts the wavevector of the the Bloch wave.
$\pmb{\mathscr D}$ are amplitude ratios of the Amplitudes.
% 2波のE-L理論も当然，式(\ref{eq12:matrix})のように
% 記述されるが
% 「固有値/固有ベクトル問題」
% という用語は一般に出てこない。
The two-beam E-L theory can also be described as an eigenvalue problem.
However, any textbook that describes
the two-beam theory as an eigenvalue problem cannot be found.
% 2波E-L理論に関する，ほぼすべての著書において，
% 固有値の集合としての分散面の導出が記述されている。
The conventional description of the two-beam dynamical theory
has two steps, i.e.
description of the dispersion surfaces at first and then
% まず式(\ref{eq12:matrix})を，以下のように変形する。
(\ref{eq12:matrix}) is deformed as follows:
\begin{align}
\big({\mathbf A}-\xi {\mathbf E}\big)
 \pmb{\mathscr D} = {\mathbf O}.
\label{eq14:DispersionSurface}
\end{align}
% ${\mathbf O}$は，すべての成分がゼロの
% $2n$次元の列ベクトルである。
Here, ${\mathbf O}$ are a $2n$-dimensional column vector whose all
elements are zero.
\begin{align}
\det\big({\mathbf A}-\xi {\mathbf E}\big) = 0.
\label{eq15:Determinant}
\end{align}
% 上の式(\ref{eq15:Determinant})は，
% 式(\ref{eq14:DispersionSurface})が
% $\pmb{\mathscr D} = {\mathbf O}$
% 以外の解を持つための条件である。
(\ref{eq15:Determinant}) describes the condition for
(\ref{eq14:DispersionSurface}) has the solution other than zero vector.
% 2波ケースの場合，$\sigma$偏光と$\pi$偏光は，
% 干渉し合うことなく別々に取り扱うことができる。
In the two-beam case,
$\sigma$- and $\pi$-polarized X-rays can be dealt with independently
since they do not interfere with each other.
% さらに，ローレンツ点を定義することにより，
% 式(\ref{eq02:fundamental2})の右辺から
% $j=i$の項を消去するのが一般的である。
Further, the term of $j=i$ of the left side of (\ref{eq02:fundamental2})
is deleted in general by defining the Lorentz point.
% 分散面の方程式(\ref{eq15:Determinant})は，
% $\sigma$偏光と$\pi$偏光に対して，それぞれ双曲線となる。
The dispersion surfaces described with (\ref{eq15:Determinant})
can be approximated by hyperbolic curves
whose cross point of the asymptotes is the Lorentz point.
% このため，2波E-L理論は，近似解とはいえ解析的に解ける。
Then, the two-beam E-L theory can be approximately solved analytically.
% 本稿の記述においては，ローレンツ点の定義を行わない。
In the description of the present paper,
Lorentz point is not defined.
% T-T方程式に$j=i$の項を残すと，
% 任意形状の結晶に対応するという有利な点があるためである
% \cite{okitsu2011a}。
There is an advantage that
the T-T equation explicitly having the term of $j=i$ can deal with
the wave fields in an arbitrary shaped crystal
\cite{okitsu2011a}.
% 多波ケースにおける分散面は，
% $\xi$に関しての$2n$次方程式となり，
% 非常に複雑になる。
The dispersion surfaces for the $n$-beam cases
are described by a complex $2n$th order equation
whose analytical solution cannot be obtained.
% このこともまた，多波動力学理論の発展を遅らせた
% 要因のひとつである。
This is one of the reason for the late development
of the $n$-beam dynamical theory.

% $n$波E-L理論の数値解を最初に与えたのは
% コレラ
% \cite{colella1974}
% である。
% 彼の手法は，ラウエ点を通る，中心H$_i$，半径$K$の球の湾曲をも考慮するものであり% ，
% 式(\ref{eq12:matrix})の
% 固有値/固有ベクトル問題
% を解くよりも，
% ブラッグ条件から大きく離れたところで，
% 高い精度を持つものと考えられる。
The numerical solution of the $n$-beam E-L theory
has been given by Colella in 1974
\cite{colella1974}
for the first time.
Colella's method takes into account the curvature
of the sphere whose radius is $K$ and
center is H$_i$.
This method gives the numerical solution with a higher precision
when the start point of the wavevector of Incident X-rays is far distant
from the Laue Point
in comparison with the method to solve the eigenvalue problem
described by (\ref{eq12:matrix}).

% Fig.\ \ref{Fig01_org_05_RecipPosition}
% には，H$_0$, H$_1$, H$_2$, H$_3$, H$_4$, H$_5$からの距離が
% $K$ $(= 1 / \lambda)$のラウエ点(${\rm La}_0$)に加えて，
% H$_i$ $(i \in \{ 6，7,\cdots, 17 \})$からの距離が
% $K$の${\rm La}_i$を
% 描き加えてある。
% のちの\S \ref{section_05_004_18BeamCase}で示す
% Fig.\ \ref{Fig19_04_18BeamWithCharacters}のような
% $18$波ケースの計算に，
% ${\rm La}_i$の定義が必要になる。
In Fig.\ \ref{Fig01_org_05_RecipPosition},
${\rm La}_i$ whose distance from
H$_i$ $(i \in \{ 6, 7,\cdots, 17 \})$ is $K$,
is also described
in addition to the Laue point ${\rm La}_0$
whose distance from
H$_0$, H$_1$, H$_2$, H$_3$, H$_4$ and H$_5$ is $K$.
For later description in \S \ref{section_05_004_18BeamCase}
regarding to the $18$-beam case
as shown in Fig.\ \ref{Fig19_04_18BeamWithCharacters},
the definition of ${\rm La}_i$ is necessary.

% Fig.\ \ref{Fig01_org_05_RecipPosition}で
% $\overrightarrow{{\rm La}_0{\rm La}_i}$は，
% $\overrightarrow{{\rm P}_1^{\prime}{\rm P}_1}$に平行であり，
In Fig.\ \ref{Fig01_org_05_RecipPosition},
$\overrightarrow{{\rm La}_0{\rm La}_i}$ is parallel to
$\overrightarrow{{\rm P}_1^{\prime}{\rm P}_1}$.
% 式(\ref{eq11:eln-beamDash})の左辺は，
% $\xi {\mathcal D}_i^{(l)}$から
% $(\xi + \xi_i^{\prime}) {\mathcal D}_i^{(l)}$
% (ただし$i < 6$のとき$\xi_i^{\prime} = 0$)
% に置き換える必要がある。
Then, $\xi {\mathcal D}_i^{(l)}$
in the left side of (\ref{eq11:eln-beamDash})
should be replaced with
$(\xi + \xi_i^{\prime}) {\mathcal D}_i^{(l)}$.
Here, $\xi_i^{\prime} = 0$ for $i < 6$.
% 式(\ref{eq12:matrix})の中の$2n \times 2n$行列の要素
% を表す式(\ref{eq13:ElementOfMatrix})は，
% 次のように書き換えなければならない。
(\ref{eq13:ElementOfMatrix}) that gives elements
of $2n \times 2n$ matrix in (\ref{eq12:matrix})
should be rewritten as follows:
\begin{alignat}{1}
a_{p, q} &= \frac{\ K \ }{2 \cos \Theta_i}
            \chi_{h_i - h_j} C_{i, j}^{(l, m)}
% \nonumber \\
%        &- \frac{\delta_{p, q} K}{\ \cos \Theta_i\ }
          - \frac{\delta_{p, q} K}{\ \cos \Theta_i\ }
                          \left( S_{i, 0}^{(0)} \beta^{(0)}
                               + S_{i, 0}^{(1)} \beta^{(1)}
                          \right) - \delta_{p, q} \xi_i^{\prime},
\label{eq16:ElementOfMatrixDash} \\
& {\rm where, }\ \xi_i^{\prime} = 0\ {\rm for}\ i < 6.
\nonumber
\end{alignat}
% 上の式(\ref{eq16:ElementOfMatrixDash})を成分に持つ
% $2n \times 2n$ $(n = 18)$の行列を作り，
% 式(\ref{eq12:matrix})の
% 固有値/固有ベクトル問題
% を解き，
% のちに\S \ref{section_03_002_algorithmEL}に記述する，
% 式(\ref{eq46:BoundCond_a})と
% 式(\ref{Eq47_E-LSolution})から計算した回折プロファイルを
% 高速フーリエ変換して得られたのが，
% \S \ref{section_05_004_18BeamCase}で示す，
% Fig.\ \ref{Fig18_03_18Beam_Exp_Sim_Rev}\ $(b)$である。
Fig.\ \ref{Fig18_03_18Beam_Exp_Sim_Rev}\ $(b)$ has been
obtained in the following sequence i.e.
i) define the $2n \times 2n$ $(n = 18)$ matrix
whose elements are as described in
(\ref{eq16:ElementOfMatrixDash}),
ii) solve the eigenvalue problem of (\ref{eq12:matrix}) and
iii) fast Fourier transform the diffraction curves
obtained based on (\ref{Eq47_E-LSolution}).

% このことには重要な意味がある。
% 筆者らの2012年の論文
% \cite{okitsu2012}まで，
% T-T方程式の解を得るにあたり，
% $n$個の逆格子点がひとつの円周上に
% なければならないという制限を設けてきたが，
% これを取り払ったのである。
This has an important meaning.
In the T-T equation that has been described
in 2012
\cite{okitsu2012},
the $n$ reciprocal lattice nodes should be
on a circle in the reciprocal space.
However, 
both for the E-L and T-T dynamical theories
described in the present paper,
the above restriction has been removed.

% さらに，Fig.\ \ref{Fig01_org_05_RecipPosition}の
% $Pl_0$上の${\rm La}_0$でない場所に${\rm La}_0^{\prime}$を描き，
Further, ${\rm La}_0^{\prime}$ is separately defined
in the vicinity of ${\rm La}_0$.
% $\overrightarrow{{\rm La}_0^{\prime}{\rm La}_i^{\prime}}$
% $=$ $\xi_i^{\prime\prime} {\mathbf n}_z$
% となるように，$Pl_i$上に
% ${\rm La}_i^{\prime}$ $(i \in \{ 0, 1, 2,$
% $\cdots, n - 1 \})$を
% 定義する。
Then, ${\rm La}_i^{\prime}$
$(i \in \{ 0, 1, 2, \cdots, n - 1 \})$
are defined on $Pl_i$ such that
$\overrightarrow{{\rm La}_0^{\prime}{\rm La}_i^{\prime}}$
$=$ $\xi_i^{\prime\prime} {\mathbf n}_z$.
% この場合，$n$波E-L理論を記述する，
% 式(\ref{eq10:eln-beam})に相当する式は，
% 次のようになる。
The $n$-beam E-L theory corresponding to (\ref{eq10:eln-beam})
is described as follows:
\begin{alignat}{1}
& \xi \cos \Theta_i {\mathcal D}_i^{\prime(l)}
%  &+ K \left(
    + K \left(
           S_{i, 0}^{(0)} \beta^{\prime(0)}
         + S_{i, 0}^{(1)} \beta^{\prime(1)}
        \right)
       {\mathcal D}_i^{\prime(l)}
\nonumber \\
   & \qquad \qquad = - \xi_i^{\prime\prime}
        \cos \Theta_i {\mathcal D}_i^{\prime (l)}
% \nonumber \\
%  &+   \frac{\ K \ }{2} \sum_{j = 0}^{n - 1} \chi_{h_i - h_j}
    +   \frac{\ K \ }{2} \sum_{j = 0}^{n - 1} \chi_{h_i - h_j}
                     \sum_{m = 0}^{1}
                 C_{i, j}^{(l, m)}
                  {\mathcal D}_j^{\prime(m)},
                  \label{eq17:eln-beamDash} \\
\qquad \qquad
  & \ \ {\rm where,}\   i, j \in \{ 0,1, \cdots, n-1 \},
\nonumber \\
%                &nは，エバルト球表面近傍の逆格子点の数,
  & \qquad \qquad
                 n\  {\rm\ is\ number\ of\ reciprocal\ lattice\ nodes},
\nonumber \\
  & \qquad \qquad
                 l, m \in \{ 0, 1 \}.
\nonumber
\end{alignat}
% $\overrightarrow{{\rm P}_1{\rm La}_0^{\prime}}$$=$
% $K \beta^{\prime(0)} {\mathbf e}_0^{(0)}$$+$
% $K \beta^{\prime(1)} {\mathbf e}_0^{(1)}$
% である。
Here, $\overrightarrow{{\rm P}_1{\rm La}_0^{\prime}}$$=$
$K \beta^{\prime(0)} {\mathbf e}_0^{(0)}$$+$
$K \beta^{\prime(1)} {\mathbf e}_0^{(1)}$.
% 右辺第1項を左辺に移さなかった理由については，
% 式(\ref{eq33_ELtoTTGeneral01}), (\ref{eq34_ELtoTTGeneral02})を
% 導出した後に記述する。
The reason for that
the first term of the left side of the above equation
will be described after deriving
(\ref{eq33_ELtoTTGeneral01}) and (\ref{eq34_ELtoTTGeneral02}).
% 式(\ref{eq11:eln-beamDash})に相当する式は，次のようになる。
The equation corresponding to (\ref{eq11:eln-beamDash})
is given as follows:
\begin{alignat}{2}
\xi  {\mathcal D}_i^{\prime(l)}
   &= & &- \frac{K}{\ \cos \Theta_i\ } \left(
           S_{i, 0}^{(0)} \beta^{\prime(0)}
         + S_{i, 0}^{(1)} \beta^{\prime(1)}
                                       \right)
             {\mathcal D}_i^{\prime(l)}
         - \xi_i^{\prime\prime}
             {\mathcal D}_i^{\prime(l)}
\nonumber \\
   &+ & &\frac{K}{\ 2 \cos \Theta_i\ }
                     \sum_{j = 0}^{n - 1} \chi_{h_i - h_j}
                     \sum_{m = 0}^{1}
                 C_{i, j}^{(l, m)}
                 {\mathcal D}_j^{\prime(m)}.
                  \label{eq18:eln-beamDashDash}
\end{alignat}
% また，式(\ref{eq16:ElementOfMatrixDash})を
% 次の式で置き換えることになる。
Further, (\ref{eq16:ElementOfMatrixDash})
can be replaced with the following equation:
\begin{alignat}{1}
a_{p, q} &= \frac{\ K \ }{\ 2 \cos \Theta_i \ }
            \chi_{h_i - h_j} C_{i, j}^{(l, m)}
% \nonumber \\
%        &- \frac{\delta_{p, q} K}{\ \cos \Theta_i\ }
          - \frac{\delta_{p, q} K}{\ \cos \Theta_i\ }
                          \left( S_{i, 0}^{(0)} \beta^{\prime(0)}
                               + S_{i, 0}^{(1)} \beta^{\prime(1)}
                          \right) - \delta_{p, q} \xi_i^{\prime\prime}.
\label{eq19:ElementOfMatrixDashDash}
\end{alignat}
% 上の式(\ref{eq19:ElementOfMatrixDashDash})は，
% Fig.\ \ref{Fig19_04_18BeamWithCharacters}のように
% 二重の円周上に逆格子点が存在しなくても，
% 任意の個数の逆格子点が，エバルト球表面近傍に存在するケースについて，
% 式(\ref{eq12:matrix})に代入して，
% 解を求めることができる。
(\ref{eq19:ElementOfMatrixDashDash}) can be solved
by substituting the above equation into (\ref{eq12:matrix})
even when
$n$ ($n$ is an arbitrary number)
reciprocal lattice nodes to be taken into account
that exist in the vicinity of
the surface of the Ewald sphere.

% \subsection{エバルト-ラウエ理論(E-L理論)からの高木方程式(T-T理論)の導出}

\subsection{Derivation of the the Takagi equation (T-T theory) from the Ewald-Laue (E-L) theory}

% この節では，
% 式(\ref{eq10:eln-beam})ないしは
% 式(\ref{eq11:eln-beamDash})で記述される
% $n$波E-L理論から，$n$波T-T方程式を導出する。
% 逆空間における積分と実空間における微分の順序の交換，
% 積分とサンメーションの順序の交換が，
% 議論の骨子となる。
In this section,
the $n$-beam T-T equation
is derived from the $n$-beam E-L theory
described by
(\ref{eq10:eln-beam}) and/or (\ref{eq11:eln-beamDash}).
The exchange of the order of integration in the reciprocal space
and differentiation in the real space
and that of integration and summation
are essential of the discussion.

% ${\mathbf r}$を位置ベクトルだとして，
% 動力学理論の解である全波動場
% $\tilde{\mathbf D} ({\mathbf r})$が，
% 次のように表されるものとする。
Let the whole wave field $\tilde{\mathbf D} ({\mathbf r})$ i.e.
the solution of the dynamical theory
be described as follows:
\begin{align}
\tilde{\mathbf D}({\mathbf r}) = \sum_{i = 0}^{n-1} \sum_{l = 0}^{1}
  {\mathbf e}_i^{(l)} D_i^{(l)}({\mathbf r}) \exp
     \left(
   - {\rm i} 2 \pi \overrightarrow{{\rm La}_0{\rm H}_i} \cdot {\mathbf r}
     \right).
\label{eq20:wavefield}
\end{align}
Here, ${\mathbf r}$ is the location vector.
% 以下の記述のため，
% 位置ベクトル${\mathbf r}$が，
% ${\mathbf s}_i$, ${\mathbf e}_i^{(0)}$, ${\mathbf e}_i^{(1)}$
% の1次結合として
% 次のように表されるものとする。
For the later description,
let ${\mathbf r}$ be described as a linear combination of
${\mathbf s}_i$, ${\mathbf e}_i^{(0)}$ and ${\mathbf e}_i^{(1)}$
as follows:
\begin{alignat}{1}
{\mathbf r} &= s_i {\mathbf s}_i
            + e_i^{(0)} {\mathbf e}_i^{(0)}
            + e_i^{(1)} {\mathbf e}_i^{(1)}.
\label{eq21:rconbination}
\end{alignat}
% 偏光状態$l$の$i$番目の波は，
% ブロッホ波の振幅${\mathcal D}_i^{(l)}$によって
% 次のように記述することができる。
The amplitude of the
$i$th numbered wave with polarization state of $l$ can be described as follows:
\begin{alignat}{2}
{\mathcal D}_i^{(l)} (\Delta {\mathbf k})
  & \exp \left(
          - {\rm i} 2 \pi \overrightarrow{{\rm P}_1^{\prime}{\rm H}_i}
           \cdot {\mathbf r}
         \right)
% &
% &
% \nonumber \\
  &= {\mathcal D}_i^{(l)} (\Delta {\mathbf k}) \exp \big(
          - {\rm i} 2 \pi \Delta {\mathbf k} \cdot {\mathbf r}
                                                    \big)
% &
% &
% \nonumber \\
% &\qquad       \times  \exp \left(
                        \exp \left(
          - {\rm i} 2 \pi \overrightarrow{{\rm La}_0{\rm H}_i} \cdot {\mathbf r}
                             \right),
% &
% &
\label{eq22_BlochWaves} \\
  &{\rm where} \ \ \Delta {\mathbf k}
 = \overrightarrow{{\rm P}_1^{\prime} {\rm La}_0}. \nonumber
% &
% &
\end{alignat}
% 逆空間でフーリエ積分を行う際，
% これまで${\mathcal D}_i^{(l)}$で表してきたブロッホ波の振幅を，
% これが$\Delta {\mathbf k}$の関数であることを明示的に示すため，
% ${\mathcal D}_i^{(l)}(\Delta {\mathbf k})$と記述することにする。
Let us decribe the amplitudes of Bloch waves ${\mathcal D}_i^{(l)}$
as ${\mathcal D}_i^{(l)}(\Delta {\mathbf k})$
to clartify these amplitudes
are functions of $\Delta {\mathbf k}$.
% また，後の記述のため，
% 式(\ref{eq07_xidefine})と式(\ref{eq08_P1La})から，
% $\Delta {\mathbf k}$
% が次のように表されることを確認しておく。
Further, for later description,
let us confirm that $\Delta {\mathbf k}$ are described from
(\ref{eq07_xidefine}) and (\ref{eq08_P1La}) as follows:
\begin{align}
\Delta {\mathbf k} = \xi {\mathbf n}_z
                   + K \beta^{(0)} {\mathbf e}_0^{(0)}
                   + K \beta^{(1)} {\mathbf e}_0^{(1)}.
\label{eq23_Delta_k}
\end{align}
% 式(\ref{eq22_BlochWaves})の
% ${\mathcal D}_i^{(l)}(\Delta {\mathbf k})$は，
% どのような関数であっても構わない。
% 例えば，入射X線の振幅が，
% Fig.\ \ref{Fig01_org_05_RecipPosition}の
% $Pl_0$上の1点でのみゼロでない，
% ディラックのデルタ関数であれば平面波入射，
% 振幅も位相も変わらない一定値の場合は，
% 実空間ではデルタ関数となるため球面波入射となる。
${\mathcal D}_i^{(l)}(\Delta {\mathbf k})$ can be
an arbitrary function of $\Delta {\mathbf k}$.
For example,
${\mathcal D}_i^{(l)}(\Delta {\mathbf k})$ is
the Dirac delta function of $\Delta {\mathbf k}$
for the condition of plane wave incidence.
However,
the constant function whose amplitude and phase do not change depending
on $\Delta {\mathbf k}$ for
the condition of spherical wave incidence.

% 式(\ref{eq20:wavefield})右辺の振幅$D_i^{(l)} ({\mathbf r})$と，
% $i$を$j$に，$l$を$m$に置き換えた振幅$D_j^{(m)} ({\mathbf r})$は，
% ブロッホ波をコヒーレントに重ね合わせたものだと考えられるため，
% 次のように表される。
Since $D_i^{(l)}({\mathbf r})$ and $D_j^{(m)}({\mathbf r})$ are 
the amplitudes of the $i$th and $j$th numbered waves
whose polarization states are $l$ and $m$
are considered to be coherent superpositions of Bloch waves,
they are described as follows:
\begin{subequations}
\begin{alignat}{1}
% & D_i^{(l)} ({\mathbf r})
  & D_i^{(l)} ({\mathbf r})
% \nonumber \\
% & \quad = \int_{\Delta {\mathbf k}}^{D.S.}
          = \int_{\Delta {\mathbf k}}^{D.S.}
  {\mathcal D}_i^{(l)} (\Delta {\mathbf k}) \exp \left(
      - {\rm i} 2 \pi \Delta {\mathbf k} \cdot {\mathbf r}
                            \right)
{\rm d}S,
\label{eq24_integration_a} \\
% & D_j^{(m)} ({\mathbf r})
  & D_j^{(m)} ({\mathbf r})
% \nonumber \\
% & \quad = \int_{\Delta {\mathbf k}}^{D.S.}
          = \int_{\Delta {\mathbf k}}^{D.S.}
  {\mathcal D}_j^{(m)} (\Delta {\mathbf k}) \exp \left(
      - {\rm i} 2 \pi \Delta {\mathbf k} \cdot {\mathbf r}
                            \right)
{\rm d}S.
\label{eq24_integration_b}
\end{alignat}
\label{eq24_integration}
\end{subequations}
% ここで，$\int_{\Delta k}^{D.S.} {\rm d}S$は，
% 分散面全体にわたる積分である。
Here, $\int_{\Delta k}^{D.S.} {\rm d}S$ means
the integration over the dispersion surfaces.
% 分散面と固有ベクトルは，$2n$組あるので，
% それらの番号を$k$$(k \in \{ 1, 2, \cdots 2n \})$として，
% $\sum_{k=1}^{2n}$のサンメーションをとる記述もあり得るが，
% 式(\ref{eq24_integration_a}), (\ref{eq24_integration_b})
% においては，のちの式変形の簡素化のため，
% $\int_{\Delta k}^{D.S.} {\rm d}S$の積分の中に
% $\sum_{k=1}^{2n}$のサンメーションを含むものととして
% 記述している。
Since there are $2n$ couples of dispersion surfaces
and eigenvectors,
they can be described as
$\sum_{k=1}^{2n}$$\int_{\Delta k}^{D.S.} {\rm d}S$.
However, (\ref{eq24_integration}) has been described under the assumption that
$\int_{\Delta k}^{D.S.} {\rm d}S$ means an integration for
$2n$ dispersion surfaces in
(\ref{eq24_integration_a}) and (\ref{eq24_integration_b})
for simplicity in the later deformation of the equations.
% $D_i^{(l)} ({\mathbf r})$と
% $D_j^{(m)} ({\mathbf r})$は，それぞれ，
% $\exp(-{\rm i} 2 \pi \overrightarvearrow{{\rm La}_0{\rm H}_i} \cdot {\mathbf r})$
% ${\mathbf e}_i^{(l)}$と
% % $\exp(-{\rm i} 2 \pi \overrightarrow{{\rm La}_0{\rm H}_j} \cdot {\mathbf r})% $
% % ${\mathbf e}_j^{(m)}$の波を
% 変調する振幅である。
$D_i^{(l)} ({\mathbf r})$ and $D_j^{(m)} ({\mathbf r})$ are the amplitudes
that modulate the waves of
$\exp(-{\rm i} 2 \pi \overrightarrow{{\rm La}_0{\rm H}_i} \cdot {\mathbf r})$
${\mathbf e}_i^{(l)}$
and
$\exp(-{\rm i} 2 \pi \overrightarrow{{\rm La}_0{\rm H}_j} \cdot {\mathbf r})$
${\mathbf e}_j^{(m)}$,
respectively.
% 式(\ref{eq21:rconbination}), (\ref{eq23_Delta_k})
% を式(\ref{eq24_integration_a})に代入して，
% 式(\ref{eq04:polarizationfactor_b})の偏光因子を考慮すると，
By substituting (\ref{eq21:rconbination}) and (\ref{eq23_Delta_k})
and considering the polarization factors
defined as in (\ref{eq04:polarizationfactor_b}),
the following equation can be obtained:
\begin{alignat}{2}
D_i^{(l)} ({\mathbf r})
% &\quad = \int_{\Delta {\mathbf k}}^{D.S.}
  &      = \int_{\Delta {\mathbf k}}^{D.S.}
  {\mathcal D}_i^{(l)} (\Delta {\mathbf k})
                                           \nonumber \\
% &
% &
% \nonumber \\
% &\qquad \times   \exp \Big\{
  &       \times   \exp \Big\{
      - {\rm i} 2 \pi \Big[
                      \Big(
          \xi \cos \Theta_i
% &
% &
% \nonumber \\
% &\qquad\qquad\qquad\quad
        + K \beta^{(0)} S_{i, 0}^{(0)}
% &
% &
% \nonumber \\
% &\qquad\qquad\qquad\quad
%                     \left.
        + K \beta^{(1)} S_{i, 0}^{(1)}
                      \Big) s_i
% &
% &
% \nonumber \\
% &\qquad
%                     \left.
%                     \left.
 + f_i\big(
      e_i^{(0)}, e_i^{(1)}
      \big)
                      \Big]
                            \Big\}
{\rm d}S.
% &
% &
\label{eq25_integration2}
\end{alignat}
% ここで，
% $f_i\big(e_i^{(0)}, e_i^{(1)}\big)$は，
% $e_i^{(0)}$, $e_i^{(1)}$の関数で，
% $s_i$に依存しない項である。
Here, $f_i\big(e_i^{(0)}, e_i^{(1)}\big)$
are functions of $e_i^{(0)}$, $e_i^{(1)}$
that do not depend on $s_i$.
% よって，
% $\partial D_i^{(l)} ({\mathbf r}) / \partial s_i$は，
% 次のように計算される。
Therefore, $\partial D_i^{(l)} ({\mathbf r}) / \partial s_i$
can be calculated as follows:
\begin{subequations}
\begin{alignat}{2}
% &\frac{\partial}{\ \partial s_i \ } D_i^{(l)} ({\mathbf r})
   \frac{\partial}{\ \partial s_i \ } D_i^{(l)} ({\mathbf r})
% &
% &
% \nonumber \\
% &\qquad= \frac{\partial}{\ \partial s_i \ }
  &      = \frac{\partial}{\ \partial s_i \ }
 \int_{\Delta {\mathbf k}}^{D.S.}
 {\mathcal D}_i^{(l)} (\Delta {\mathbf k})
 \exp \left(
   - {\rm i} 2 \pi {\Delta {\mathbf k}} \cdot {\mathbf r}
      \right) {\rm d} S
% &
% &
% \nonumber \\
\label{eq26_diffrentiation_a} \\
% &\qquad=
  &      =
 \int_{\Delta {\mathbf k}}^{D.S.}
 \frac{\partial}{\ \partial s_i \ }
    \left[
 {\mathcal D}_i^{(l)} (\Delta {\mathbf k})
 \exp \left(
   - {\rm i} 2 \pi {\Delta {\mathbf k}} \cdot {\mathbf r}
      \right)
    \right] {\rm d} S
% &
% &
% \nonumber \\
\label{eq26_diffrentiation_b} \\
% &\qquad=
  &      =
 - {\rm i} 2 \pi \int_{\Delta {\mathbf k}}^{D.S.}
                 \left[
          \xi \cos \Theta_i + K
                    \left(
            S_{i, 0}^{(0)} \beta^{(0)}
          + S_{i, 0}^{(1)} \beta^{(1)}
                    \right)
                 \right]
% &
% &
  \nonumber \\
  &\qquad\qquad\qquad\times
 {\mathcal D}_i^{(l)} (\Delta {\mathbf k})\exp \left(
   - {\rm i} 2 \pi {\Delta {\mathbf k}} \cdot {\mathbf r}
      \right)
 {\rm d} S.
% &
% &
\label{eq26_diffrentiation_c}
\end{alignat}
\label{eq26_diffrentiation}
\end{subequations}
% 一方，式(\ref{eq23_Delta_k})の$\Delta {\mathbf k}$の代わりに，
% $\Delta {\mathbf k}^{\prime}$を次のように定義する。
On the other hand,
in place of $\Delta {\mathbf k}$ in (\ref{eq23_Delta_k}),
let define $\Delta {\mathbf k}^{\prime}$ as follows:
\begin{subequations}
\begin{alignat}{1}
\Delta {\mathbf k}^{\prime}
                  &= \overrightarrow{{\rm P}_1^{\prime}{\rm La}_0^{\prime}}
% \nonumber \\
\label{eq27_Delta_kPrime_a} \\
                  &= \xi {\mathbf n}_z
                   + K \beta^{\prime(0)} {\mathbf e}_0^{(0)}
                   + K \beta^{\prime(1)} {\mathbf e}_0^{(1)}.
\label{eq27_Delta_kPrime_b}
\end{alignat}
\label{eq27_Delta_kPrime}
\end{subequations}
% 式(\ref{eq24_integration}),
% (\ref{eq25_integration2})
% で，$\Delta {\mathbf k}$,
% $D_i^{(l)}({\mathbf r})$,
% ${\mathcal D}_i^{(l)}(\Delta {\mathbf k})$,
% $D_j^{(m)}({\mathbf r})$,
% ${\mathcal D}_j^{(m)}(\Delta {\mathbf k})$,
% $\beta^{(0)}$, $\beta^{(1)}$
% をそれぞれ，
% $\Delta {\mathbf k}^{\prime}$,
% $D_i^{\prime(l)}({\mathbf r})$,
% ${\mathcal D}_i^{\prime(l)}(\Delta {\mathbf k}^{\prime})$,
% $D_j^{\prime(m)}({\mathbf r})$,
% ${\mathcal D}_j^{\prime(m)}(\Delta {\mathbf k}^{\prime})$,
% $\beta^{\prime(0)}$, $\beta^{\prime(1)}$
% で置き換えても，同様な変形ができるので，
% 式(\ref{eq26_diffrentiation})と同じ形の
% 次の式が得られる。
In (\ref{eq24_integration}) and (\ref{eq25_integration2}),
$\Delta {\mathbf k}$, ${\mathcal D}_i^{(l)}(\Delta {\mathbf k})$,
$D_j^{(m)}({\mathbf r})$,
${\mathcal D}_j^{(m)}(\Delta {\mathbf k})$,
$\beta^{(0)}$ and $\beta^{(1)}$
can be replaced with $\Delta {\mathbf k}^{\prime}$,
$D_i^{\prime(l)}({\mathbf r})$,
${\mathcal D}_i^{\prime(l)}(\Delta {\mathbf k}^{\prime})$,
$D_j^{\prime(m)}({\mathbf r})$,
${\mathcal D}_j^{\prime(m)}(\Delta {\mathbf k}^{\prime})$,
$\beta^{\prime(0)}$ and $\beta^{\prime(1)}$
to obtain the following equation:
\begin{alignat}{1}
% &\frac{\partial}{\ \partial s_i \ } D_i^{\prime(l)} ({\mathbf r})
   \frac{\partial}{\ \partial s_i \ } D_i^{\prime(l)} ({\mathbf r})
% \nonumber \\
% &\qquad=
  &      =
 - {\rm i} 2 \pi \int_{\Delta {\mathbf k}^{\prime}}^{D.S.}
                 \left[
          \xi \cos \Theta_i + K
                    \left(
            S_{i, 0}^{(0)} \beta^{\prime(0)}
          + S_{i, 0}^{(1)} \beta^{\prime(1)}
                    \right)
                 \right]
\nonumber \\
&\qquad\qquad\qquad\times
 {\mathcal D}_i^{\prime(l)} (\Delta {\mathbf k}^{\prime})
 \exp \left(
   - {\rm i} 2 \pi {\Delta {\mathbf k}^{\prime}} \cdot {\mathbf r}
      \right)
 {\rm d} S.
\label{eq28_diffrentiation2}
\end{alignat}
% 式(\ref{eq26_diffrentiation})の$D_i^{(l)}({\mathbf r})$
% が，$n$個の逆格子点が同一円周上に存在する場合に，
% $\exp\big( -{\rm i} 2 \pi \overrightarrow{{\rm La}_0{\rm H}_i} \cdot {\mathbf % r} \big)$
% ${\mathbf e}_i^{(l)}$の波を変調する振幅なのに対して，
% % 式(\ref{eq28_diffrentiation2})の$D_i^{\prime(l)}({\mathbf r})$は，
% % $\exp\big( -{\rm i} 2 \pi \overrightarrow{{\rm La}_0^{\prime}{\rm H}_i} \cdo% t % {\mathbf r} \big)$
% % ${\mathbf e}_i^{(l)}$を変調する振幅であり，エバルト球表面近傍に存在する
% 逆格子点をすべて考慮する場合へと，
% T-T方程式の適用範囲をのちに一般化するために準備した。
When $n$ reciprocal lattice nodes exist on a circle
in the reciprocal space,
$D_i^{(l)}({\mathbf r})$ in (\ref{eq26_diffrentiation})
is the amplitude
that modulates the oscillation of
$\exp\big( -{\rm i} 2 \pi$
$\overrightarrow{{\rm La}_0{\rm H}_i} \cdot \mathbf{r} \big)$
${\mathbf e}_i^{(l)}$.
However, $D_i^{\prime(l)}({\mathbf r})$
in (\ref{eq28_diffrentiation2})
is the amplitude
that modulates the oscillation of
$\exp\big( -{\rm i} 2 \pi$
$\overrightarrow{{\rm La}_0^{\prime}{\rm H}_i} \cdot {\mathbf r} \big)$
${\mathbf e}_i^{(l)}$.
(\ref{eq28_diffrentiation2})
has been derived to generalize the T-T equation
in the later description
such as to take into account all reciprocal lattice nodes
in the vicinity of the surface of the Ewald sphere.

% 式(\ref{eq10:eln-beam})を式(\ref{eq26_diffrentiation})に代入して
By substituting (\ref{eq10:eln-beam}) into (\ref{eq26_diffrentiation}),
the following equations can be obtained:
\begin{subequations}
\begin{alignat}{2}
% & \frac{\partial}{\partial s_i} D_i^{(l)} ({\mathbf r})
    \frac{\partial}{\partial s_i} D_i^{(l)} ({\mathbf r})
% &
% &
% \nonumber \\
% &\qquad= - {\rm i} \pi K \int_{\Delta {\mathbf k}}^{D.S.}
&        = - {\rm i} \pi K \int_{\Delta {\mathbf k}}^{D.S.}
   \sum_{j = 0}^{n-1} \chi_{h_i - h_j}
   \sum_{m = 0}^{1} C_{i, j}^{(l, m)}
% &
% &
% \nonumber \\
% &\qquad\qquad \times {\mathcal D}_j^{(m)}({\Delta {\mathbf k}})
                       {\mathcal D}_j^{(m)}({\Delta {\mathbf k}})
   \exp \left(
   - {\rm i} 2 \pi {\Delta {\mathbf k}} \cdot {\mathbf r}
        \right)
   {\rm d} S
% &
% &
% \nonumber \\
\label{eq29_ELtoTT_a} \\
% &\qquad= - {\rm i} \pi K
  &      = - {\rm i} \pi K
   \sum_{j = 0}^{n-1} \chi_{h_i - h_j}
   \sum_{m = 0}^{1} C_{i, j}^{(l, m)}
% &
% &
% \nonumber \\
% &\qquad\qquad \times
  \int_{\Delta {\mathbf k}}^{D.S.}{\mathcal D}_j^{(m)} ({\Delta {\mathbf k}})
   \exp \left(
   - {\rm i} 2 \pi {\Delta {\mathbf k}} \cdot {\mathbf r}
        \right)
   {\rm d} S.
% &
% &
\label{eq29_ELtoTT_b}
\end{alignat}
\label{eq29_ELtoTT}
\end{subequations}
% 式(\ref{eq24_integration_b})を式(\ref{eq29_ELtoTT})に代入すると，
% 次の式が得られる
(\ref{eq24_integration_b}) can be substituted into (\ref{eq29_ELtoTT})
to obtain the following equation:
\begin{alignat}{1}
% & \frac{\partial}{\partial s_i} D_i^{(l)}({\mathbf r})
    \frac{\partial}{\partial s_i} D_i^{(l)}({\mathbf r})
% \nonumber \\
% & \qquad = - {\rm i} \pi K \sum_{j = 0}^{n - 1} \chi_{h_i - h_j}
  &        = - {\rm i} \pi K \sum_{j = 0}^{n - 1} \chi_{h_i - h_j}
                           \sum_{m = 0}^{1} C_{i, j}^{(l, m)}
                                            D_j^{(m)} ({\mathbf r}),
\label{eq30_ELtoTTFinal} \\
% & \qquad{\rm where,}   \quad i, j \in \{ 0,1, \cdots, n-1 \},
  &       {\rm where,}   \quad i, j \in \{ 0,1, \cdots, n-1 \},
\nonumber \\
% &        \qquad\qquad\qquad  n \in \{ 3,4,5,6,8,12 \},
  &        \qquad\qquad        n \in \{ 3,4,5,6,8,12 \},
\nonumber \\
% &        \qquad\qquad\qquad  l, m \in \{ 0, 1 \}.
  &        \qquad\qquad        l, m \in \{ 0, 1 \}.
\nonumber
\end{alignat}
% 上の式(\ref{eq30_ELtoTTFinal})は，
% $n$個の逆格子点が同一円周上にある場合に対する
% $n$波T-T方程式である。
The above equation (\ref{eq30_ELtoTTFinal})
is the $n$-beam T-T equation applicable to the cases where
$n$ reciprocal lattice nodes exist on an identical circle
in the reciprocal space.

% ところで，結晶が完全であるとき，
% 電気分極率は，
% $\chi({\mathbf r})$$=$
% $\sum_{i}$$\chi_{h_i}$$\exp\big(-{\rm i}$
% 2 \pi {\mathbf h}_i \cdot {\mathbf r}\big)$
% のように，フーリエ級数で表される。
% しかし，結晶に格子変位${\mathbf u} ({\mathbf r})$があるとき，
% 電気分極率は，近似的に
% $\chi[{\mathbf r} - {\mathbf u}({\mathbf r})]$
% となり，次のように，変形されたフーリエ級数に展開される。
Incidentally, in the case of a perfect crystal,
the electric susceptibility in the crystal $\chi({\mathbf r})$
can be Fourier-expanded to be
$\chi({\mathbf r})$$=$
$\sum_{i}$$\chi_{h_i}$
$\exp\big(-{\rm i} 2 \pi {\mathbf h}_i \cdot {\mathbf r}\big)$.
However,
in the cases that the crystal has the lattice displacement field
${\mathbf u} ({\mathbf r})$,
the electric susceptibility is approximately Fourier-expanded
as follows:
\begin{subequations}
\begin{alignat}{1}
% & \chi \big[
    \chi \big[
        {\mathbf r} - {\mathbf u}({\mathbf r})
       \big]
% \nonumber \\
% &\qquad = \sum_{i} \chi_{h_i}
  &       = \sum_{i} \chi_{h_i}
\exp\big\{
     -{\rm i} 2 \pi {\mathbf h}_i \cdot
          \big[
       {\mathbf r} - {\mathbf u}({\mathbf r})
          \big]
    \big\}
% \nonumber \\
\label{eq31_ChiFourier_a} \\
% &\qquad = \sum_{i} \chi_{h_i}
  &       = \sum_{i} \chi_{h_i}
\exp\big[{\rm i} 2 \pi {\mathbf h}_i \cdot {\mathbf u}({\mathbf r})\big]
\exp\big(-{\rm i} 2 \pi {\mathbf h}_i \cdot {\mathbf r}\big).
\label{eq31_ChiFourier_b}
\end{alignat}
\label{eq31_ChiFourier}
\end{subequations}
% したがって格子変位場${\mathbf u}({\mathbf r})$
% を持つ結晶の場合，
% 電気分極率のフーリエ係数は位置ベクトル${\mathbf r}$の関数となり，
% $\chi_{h_i - h_j}$ $\exp\big[{\rm i}
% 2 \pi ({\mathbf h}_i - {\mathbf h}_j) \cdot {\mathbf u} ({\mathbf r})\big]$
% のように表される。
% よって式(\ref{eq30_ELtoTTFinal})は，格子変位を伴う結晶に対して
% 次のように書き換えられる。
Then, when
the crystal has the lattice displacement field of
${\mathbf u}({\mathbf r})$,
the Fourier coefficient of the electric susceptibility
is a function of the position in the crystal,
can be described as
$\chi_{h_i - h_j}$ $\exp\big[{\rm i}$
$2 \pi ({\mathbf h}_i - {\mathbf h}_j)$
$\cdot {\mathbf u} ({\mathbf r})\big]$.
Therefore, (\ref{eq30_ELtoTTFinal}) can be deformed
for a crystal having the lattice displacement field
as follows:
\begin{alignat}{2}
% &
% & \frac{\partial}{\partial s_i} D_i^{(l)} ({\mathbf r})
    \frac{\partial}{\partial s_i} D_i^{(l)} ({\mathbf r})
% &
% \nonumber \\
% & & = - {\rm i} \pi K
    & = - {\rm i} \pi K
% & \sum_{j = 0}^{n - 1}
    \sum_{j = 0}^{n - 1}
\chi_{h_i - h_j} \exp \big[{\rm i} 2 \pi
({\mathbf h}_i - {\mathbf h}_j) \cdot {\mathbf u} ({\mathbf r})
                      \big]
% \nonumber \\
% & &\times & \sum_{m = 0}^{1}
              \sum_{m = 0}^{1}
   C_{i, j}^{(l, m)} D_j^{(m)} ({\mathbf r}).
\label{eq32_TT}
\end{alignat}
% 上の式(\ref{eq32_TT})は，
% $n$個の逆格子点がひとつの円周上に存在する場合の，
% 結晶格子変位を取り扱える
% $n$波T-T方程式
% \cite{okitsu2006,okitsu2012}
% にほかならない。
The above equation (\ref{eq32_TT})
is nothing but the $n$-beam T-T equation
\cite{okitsu2006,okitsu2012}
that can deal with the X-ray wave fields
in a deformed crystal
in the case that
$n$ reciprocal lattice nodes exist on
an identical circle in the reciprocal space
\cite{okitsu2006,okitsu2012}.

% 次に，エバルト球近傍の逆格子点をすべて考慮する場合の
% $n$波T-T方程式を導出する。
Next,
let us derive the $n$-beam T-T equation
to take into account the all reciprocal lattice nodes
in the vicinity of the surface of the Ewald sphere.

% ${\rm La}_0^{\prime}$は，式(\ref{eq17:eln-beamDash})を導出する
% 直前に記述したように，
% Fig.\ \ref{Fig01_org_05_RecipPosition}
% の$Pl_0$上にある，$0$番目の「一般化されたラウエ点」である。
As decribed just before deriving (\ref{eq17:eln-beamDash}),
${\rm La}_0^{\prime}$ is the 0th-numbered
`generalized Laue point'
that exist on $Pl_0$
in Fig.\ \ref{Fig01_org_05_RecipPosition}.
% 先に記述したように，$0$番目と$i$番目の
% 「一般化されたラウエ点」を結ぶベクトルは
% $\overrightarrow{{\rm La}_0^{\prime}{\rm La}_i^{\prime}}
% = \xi_i^{\prime\prime} {\mathbf n}_z$である。
As described before,
$\overrightarrow{{\rm La}_0^{\prime}{\rm La}_i^{\prime}}$ is
$\xi_i^{\prime\prime} {\mathbf n}_z$
(see Fig.\ \ref{Fig01_org_05_RecipPosition}).
% (Fig.\ \ref{Fig01_org_05_RecipPosition}参照)。
% $n$個の逆格子点が同一円周上にある場合には，
% 式(\ref{eq10:eln-beam})を式(\ref{eq26_diffrentiation})に代入して
% 式(\ref{eq29_ELtoTT})を得たが，
For the case that
$n$ reciprocal lattice nodes exist on an identical circle
in the reciprocal space,
(\ref{eq10:eln-beam}) has been substituted into
(\ref{eq26_diffrentiation}) to obtain (\ref{eq29_ELtoTT}).
% エバルト球近傍の逆格子点をすべて考慮できるようにするには，
% 式(\ref{eq17:eln-beamDash})を式(\ref{eq28_diffrentiation2})に代入
% する必要がある。
To generalize the $n$-beam T-T equation such as to take into account
the all reciprocal lattice nodes in the vicinity of
the surface of the Ewald sphere,
(\ref{eq17:eln-beamDash})
can be substituted into (\ref{eq28_diffrentiation2})
to obtain the following equation:
% 得られる式は，次の通りである。
\begin{subequations}
\begin{alignat}{1}
% & \frac{\partial}{\partial s_i} D_i^{\prime(l)} ({\mathbf r})
    \frac{\partial}{\partial s_i} D_i^{\prime(l)} ({\mathbf r})
% \nonumber \\
% &\qquad = {\rm i} 2 \pi \xi_i^{\prime\prime} \cos \Theta_i
  &      =  {\rm i} 2 \pi \xi_i^{\prime\prime} \cos \Theta_i
   \int_{\Delta {\mathbf k}^{\prime}}^{D.S.}
        {\mathcal D}_i^{\prime(l)} ({\Delta {\mathbf k}^{\prime}})
   \exp \big[
          - {\rm i} 2 \pi \Delta {\mathbf k}^{\prime} \cdot {\mathbf r}
        \big] {\rm d} S
\nonumber \\
  &     - {\rm i} \pi K \int_{\Delta {\mathbf k}^{\prime}}^{D.S.}
           \sum_{j = 0}^{n-1} \chi_{h_i - h_j}
           \sum_{m = 0}^{1} C_{i, j}^{(l, m)}
% \nonumber \\
% &\qquad\qquad\quad \times {\mathcal D}_j^{\prime(m)}
% ({\Delta {\mathbf k}^{\prime}})
                            {\mathcal D}_j^{\prime(m)}
  ({\Delta {\mathbf k}^{\prime}})
   \exp \left(
   - {\rm i} 2 \pi {\Delta {\mathbf k}^{\prime}} \cdot {\mathbf r}
        \right)
   {\rm d} S
\label{eq33_ELtoTTGeneral01} \\
% &\qquad = {\rm i} 2 \pi \xi_i^{\prime\prime} \cos \Theta_i
  &       = {\rm i} 2 \pi \xi_i^{\prime\prime} \cos \Theta_i
                     D_i^{\prime(l)} ({\mathbf r})
% \nonumber \\
% &\qquad - {\rm i} \pi K \sum_{j = 0}^{n - 1}
          - {\rm i} \pi K \sum_{j = 0}^{n - 1}
                             \chi_{h_i - h_j}
                             \sum_{m = 0}^{1} C_{i, j}^{(l, m)}
                             D_j^{\prime(m)} ({\mathbf r}),
\label{eq34_ELtoTTGeneral02} \\
&\qquad  {\rm where,}  \  i, j \in \{ 0,1, \cdots, n-1 \},
\nonumber \\
% &\qquad\qquad\quad        nは，エバルト球表面近傍の逆格子点の数,
  &\qquad\qquad\quad n\ {\rm is\ the\ number\ of\ reciproca\ lattice\ nodes,}
\nonumber \\
&\qquad\qquad\quad        l, m \in \{ 0, 1 \}.
\nonumber
\end{alignat}
\label{eq33_ELtoTTGeneral}
\end{subequations}
% 上の式(\ref{eq33_ELtoTTGeneral01})，(\ref{eq34_ELtoTTGeneral02})
% を導出するにあたり，
% $\Delta {\mathbf k}^{\prime}_i = \overrightarrow{{\rm P}_1^{\prime}{\rm La}_i^% {\prime}}$
% と置き，式(\ref{eq24_integration_a})右辺の積分の中身を
% ${\mathcal D}_i^{\prime(l)}(\Delta{\mathbf k}_i^{\prime})
% \exp(-{\rm i}2\pi\Delta{\mathbf k}_i^{\prime}\cdot{\mathbf r})$
% とすると，一見計算が簡単なように思える。
At first glance, it might seem like
the calculation can be simplified by
putting that
$\Delta {\mathbf k}^{\prime}_i =$
$\overrightarrow{{\rm P}_1^{\prime}{\rm La}_i^{\prime}}$
to change the contents of the integration of the right side
of (\ref{eq24_integration_a})
to be
${\mathcal D}_i^{\prime(l)}(\Delta{\mathbf k}_i^{\prime})$
$\exp(-{\rm i}2\pi\Delta{\mathbf k}_i^{\prime}\cdot{\mathbf r})$.
% しかしこの場合，方程式の対称性保存のため，
% 式(\ref{eq24_integration_b})右辺の積分の中身を
% ${\mathcal D}_j^{(m)}(\Delta{\mathbf k}_j^{\prime})
% \exp(-{\rm i}2\pi\Delta{\mathbf k}_j^{\prime}\cdot{\mathbf r})$
% とする必要があり，
% 式(\ref{eq33_ELtoTTGeneral01})第2項から
% 式(\ref{eq34_ELtoTTGeneral02})第2項への変形ができず，うまくいかない。
However, in this case,
the second term of (\ref{eq33_ELtoTTGeneral01})
cannot be deformed to be
the second term of (\ref{eq34_ELtoTTGeneral02})
since the contents of the integration in the right side
of (\ref{eq24_integration_b})
should be
${\mathcal D}_j^{(m)}(\Delta{\mathbf k}_j^{\prime})$
$\exp(-{\rm i}2\pi\Delta{\mathbf k}_j^{\prime}\cdot{\mathbf r})$
to keep the symmetry of the equation.
% 式(\ref{eq17:eln-beamDash})の右辺第1項を
% 左辺に移すと，左辺は${\mathbf s}_i \cdot \Delta{\mathbf k}_i^{\prime}$となるが% ，
% これを行わず，左辺を${\mathbf s}_i \cdot \Delta{\mathbf k}^{\prime}$のままにし% たのは，
% このためである。
For this reason,
the first term of the right side of (\ref{eq17:eln-beamDash})
has not been transferred to the left side.
% さらに，式(\ref{eq31_ChiFourier})を考慮して，
Then, based on (\ref{eq31_ChiFourier}),
the following equation is obtained:
\begin{alignat}{1}
% & \frac{\partial}{\partial s_i} D_i^{\prime(l)} ({\mathbf r})
    \frac{\partial}{\partial s_i} D_i^{\prime(l)} ({\mathbf r})
% \nonumber \\
% &\qquad = {\rm i} 2 \pi \xi_i^{\prime\prime} \cos \Theta_i
  &       = {\rm i} 2 \pi \xi_i^{\prime\prime} \cos \Theta_i
                     D_i^{\prime(l)} ({\mathbf r})
\nonumber \\
% &\qquad - {\rm i} \pi K \sum_{j = 0}^{n - 1}
  &       - {\rm i} \pi K \sum_{j = 0}^{n - 1}
\chi_{h_i - h_j}  \exp \big[{\rm i} 2 \pi
({\mathbf h}_i - {\mathbf h}_j) \cdot {\mathbf u} ({\mathbf r})
                      \big]
% \nonumber \\
% &\qquad\qquad \times \sum_{m = 0}^{1} C_{i, j}^{(l, m)}
                       \sum_{m = 0}^{1} C_{i, j}^{(l, m)}
                     D_j^{\prime(m)} ({\mathbf r}).
\label{eq35_ELtoTTGeneralWithDisplacement}
\end{alignat}
% 上の式(\ref{eq35_ELtoTTGeneralWithDisplacement})は，
% エバルト球近傍の逆格子点をすべて考慮し，
% 結晶格子歪みを取り扱える$n$波T-T方程式である。
The above equation (\ref{eq35_ELtoTTGeneralWithDisplacement})
is the $n$-beam T-T equation
that describes the X-ray wave fields
in a crystal that has lattice displacement field
by taking into account all reciprocal lattice nodes
in the vicinity of the surface of the Ewald sphere.

% また，平面波入射条件の際に使いやすい方程式も導出しておく。
% 式(\ref{eq34_ELtoTTGeneral02})における
% $D_i^{\prime(l)}({\mathbf r})$，$D_j^{\prime(m)}({\mathbf r})$は，
% それぞれ，
% $\exp(-{\rm i}$$2 \pi$$\overrightarrow{{\rm La}_0^{\prime}{\rm H}_i}$$\cdot$ ${\mathbf r})$
% ${\mathbf e}_i^{(l)}$，
% $\exp(-{\rm i}$$2 \pi$$\overrightarrow{{\rm La}_0^{\prime}{\rm H}_j}$$\cdot$ $% {\mathbf r})$
% ${\mathbf e}_j^{(m)}$
% の波を変調する振幅である。
Now let us derive the equation applicable for the case that
plane wave X-rays are incident on the crystal.
$D_i^{\prime(l)}({\mathbf r})$ and $D_j^{\prime(m)}({\mathbf r})$
in (\ref{eq34_ELtoTTGeneral02})
are X-ray amplitudes that modulate the oscillations of
$\exp(-{\rm i}$$2 \pi$
$\overrightarrow{{\rm La}_0^{\prime}{\rm H}_i}$
$\cdot$ ${\mathbf r})$
${\mathbf e}_i^{(l)}$
and
$\exp(-{\rm i}$$2 \pi$
$\overrightarrow{{\rm La}_0^{\prime}{\rm H}_j}$$\cdot$
${\mathbf r})$
${\mathbf e}_j^{(m)}$, respectively.
% $D_i^{\prime\prime(l)}({\mathbf r})$，$D_j^{\prime\prime(m)}({\mathbf r})$は，% それぞれ，
% $\exp(-{\rm i}$$2 \pi$$\overrightarrow{{\rm P}_1{\rm H}_i}$$\cdot$ ${\mathbf r% })$
% ${\mathbf e}_i^{(l)}$と，
% $\exp(-{\rm i}$$2 \pi$$\overrightarrow{{\rm P}_1{\rm H}_j}$$\cdot$ ${\mathbf r% })$
% ${\mathbf e}_j^{(m)}$の波を変調する振幅であるとして，
% Fig.\ \ref{Fig01_org_05_RecipPosition}を参照すると，
% 次の式が成り立つことが理解できる。
$D_i^{\prime\prime(l)}({\mathbf r})$
and
$D_j^{\prime\prime(m)}({\mathbf r})$
are X-ray amplitudes that modulate the oscillations of
$\exp(-{\rm i}$$2 \pi$
$\overrightarrow{{\rm P}_1{\rm H}_i}$
$\cdot$ ${\mathbf r})$
${\mathbf e}_i^{(l)}$
and
$\exp(-{\rm i}$$2 \pi$
$\overrightarrow{{\rm P}_1{\rm H}_j}$
$\cdot$ ${\mathbf r})$
${\mathbf e}_j^{(m)}$.
In reference to
Fig.\ \ref{Fig01_org_05_RecipPosition},
we can understand the following relations
\begin{subequations}
\begin{alignat}{1}
D_i^{\prime(l)}({\mathbf r})
& = D_i^{\prime\prime(l)}({\mathbf r})
  \exp\big(
        - {\rm i}2\pi \overrightarrow{{\rm P}_1{\rm La}_0^{\prime}}
           \cdot {\mathbf r}
      \big)
\label{eq36_UnitaryTransform_a} \\
& = D_i^{\prime\prime(l)}({\mathbf r})
  \exp\big[
        - {\rm i}2\pi K \big(
                  \beta^{\prime(0)} {\mathbf e}_0^{(0)}
                + \beta^{\prime(1)} {\mathbf e}_0^{(1)}
                        \big)
           \cdot {\mathbf r}
      \big],
\label{eq36_UnitaryTransform_b} \\
D_j^{\prime(m)}({\mathbf r})
& = D_j^{\prime\prime(m)}({\mathbf r})
  \exp\big(
        - {\rm i}2\pi \overrightarrow{{\rm P}_1{\rm La}_0^{\prime}}
           \cdot {\mathbf r}
      \big).
\label{eq36_UnitaryTransform_c}
\end{alignat}
\label{eq36_UnitaryTransform}
\end{subequations}
% 式(\ref{eq36_UnitaryTransform_a})の偏微分$\partial / \partial s_i$を
% 計算すると，
The both sides of (\ref{eq36_UnitaryTransform_b})
can be partially differentiated as follows:
\begin{alignat}{1}
\frac{\partial}{\ \partial s_i\ } D_i^{\prime(l)} ({\mathbf r})
  & = \Big[
        \frac{\partial}{\ \partial s_i\ } D_i^{\prime\prime(l)} ({\mathbf r})
      \Big]
      \exp\big(
          -{\rm i} 2 \pi \overrightarrow{{\rm P}_1{\rm La}_0^{\prime}}
                         \cdot {\mathbf r}
          \big)
\nonumber \\
  & - {\rm i} 2 \pi K \big(
                        \beta^{\prime(0)} S_{i, 0}^{(0)}
                      + \beta^{\prime(1)} S_{i, 0}^{(1)}
                      \big)
% \nonumber \\
% & \qquad \times D_i^{\prime\prime(l)} ({\mathbf r}) \exp\big(
                  D_i^{\prime\prime(l)} ({\mathbf r}) \exp\big(
      - {\rm i} 2 \pi \overrightarrow{{\rm P}_1{\rm La}_0^{\prime}}
                         \cdot {\mathbf r}
                                                          \big).
\label{eq37_Differentiation}
\end{alignat}
% 式(\ref{eq36_UnitaryTransform})と式(\ref{eq37_Differentiation})を
% 式(\ref{eq34_ELtoTTGeneral02})に代入して次の式を得る。
(\ref{eq36_UnitaryTransform_a}),
(\ref{eq36_UnitaryTransform_b})
and (\ref{eq37_Differentiation})
can be substituted into (\ref{eq34_ELtoTTGeneral02})
to obtain the following equation:
\begin{alignat}{1}
% & \frac{\partial}{\ \partial s_i\ } D_i^{\prime\prime(l)} ({\mathbf r})
    \frac{\partial}{\ \partial s_i\ } D_i^{\prime\prime(l)} ({\mathbf r})
% \nonumber \\
% & \qquad = {\rm i} 2 \pi \xi_i^{\prime\prime} \cos \Theta_i
  &        = {\rm i} 2 \pi \xi_i^{\prime\prime} \cos \Theta_i
                                 D_i^{\prime\prime(l)} ({\mathbf r})
% \nonumber \\
% & \qquad + {\rm i} 2 \pi K \big(
           + {\rm i} 2 \pi K \big(
                        \beta^{\prime(0)} S_{i, 0}^{(0)}
                      + \beta^{\prime(1)} S_{i, 0}^{(1)}
                           \big) D_i^{\prime\prime(l)} ({\mathbf r})
\nonumber \\
% & \qquad - {\rm i} \pi K \sum_{j = 0}^{n - 1} \chi_{h_i - h_j}
  &        - {\rm i} \pi K \sum_{j = 0}^{n - 1} \chi_{h_i - h_j}
                         \sum_{m = 0}^{1} C_{i, j}^{(l, m)}
                                          D_j^{\prime\prime(m)} ({\mathbf r}).
\label{eq38_TTPlaneWave}
\end{alignat}
% 上の式(\ref{eq38_TTPlaneWave})は，
% $\exp\big($$-{\rm i}2\pi$$\overrightarrow{{\rm P}_1{\rm H}_i}$$\cdot$${\mathbf%  r}$$\big)$
% ${\mathbf e}_i^{(l)}$,
% $\exp\big($$-{\rm i}2\pi$$\overrightarrow{{\rm P}_1{\rm H}_j}$$\cdot$${\mathbf%  r}$$\big)$
% ${\mathbf e}_j^{(m)}$
% の波を変調する振幅
% $D_i^{\prime\prime(l)}$$({\mathbf r})$および
% $D_j^{\prime\prime(m)}$$({\mathbf r})$
% に対して成り立つ
% $n$波T-T方程式である。
$D_i^{\prime\prime(l)}$$({\mathbf r})$ and
$D_j^{\prime\prime(m)}$$({\mathbf r})$
in the above equation (\ref{eq38_TTPlaneWave})
are X-ray amplitudes that modulate the oscillations of
$\exp\big($$-{\rm i}2\pi$
$\overrightarrow{{\rm P}_1{\rm H}_i}$
$\cdot$${\mathbf r}$$\big)$ ${\mathbf e}_i^{(l)}$
and
$\exp\big($$-{\rm i}2\pi$
$\overrightarrow{{\rm P}_1{\rm H}_j}$
$\cdot$${\mathbf r}$$\big)$ ${\mathbf e}_j^{(m)}$.
% 上の式(\ref{eq38_TTPlaneWave})を，
% 波数ベクトル$\overrightarrow{{\rm P}_1{\rm H}_0}$の平面波入射の条件で
% 解く場合% ，
% 結晶入射側表面で，$D_0^{\prime\prime(0)}({\mathbf r})$または
% $D_0^{\prime\prime(1)}({\mathbf r})$に，一定の値の境界条件を与えることになる。When solving the above equation (\ref{eq38_TTPlaneWave})
Under the condition that plane wave X-rays
whose wavevector is $\overrightarrow{{\rm P}_1{\rm H}_0}$
are incident on the crystal,
a constant value of $D_0^{\prime\prime(0)}({\mathbf r})$
and/or $D_0^{\prime\prime(1)}({\mathbf r})$
should be given as the boundary condition
depending on the polarization state of the incident X-rays.

% Fig.\ \ref{Fig01_org_05_RecipPosition}のように，
% 結晶面下向き単位法線ベクトル
% ${\mathbf n}_z$
% に対して垂直な単位ベクトル
% ${\mathbf e}_x$, ${\mathbf e}_y$を定義し，位置ベクトルを，
% ${\mathbf r}$$=$$x {\mathbf e}_x$$+$$y {\mathbf e}_y$$+$$z{\mathbf n}_z$
% % と記述すると，式(\ref{eq38_TTPlaneWave})を解いて得られる
% % 波動場は，$x$, $y$に依存せず，$z$のみの関数となり，
% % $D_i^{\prime\prime(l)} (z)$, $D_j^{\prime\prime(m)} (z)$と表される。
% また，$\partial D_i^{\prime\prime(l)} ({\mathbf r}) / \partial s_i$
% は，次のようになる。
When the downward surface normal of the crystal is ${\mathbf n}_z$
and unit vectors
${\mathbf e}_x$ and ${\mathbf e}_y$ are
defined such that
${\mathbf e}_x$, ${\mathbf e}_y$ and ${\mathbf n}_z$
construct a right-handed orthogonal system in this order
and the location vector
${\mathbf r}$$=$
$x {\mathbf e}_x$
$+$$y {\mathbf e}_y$$+$$z{\mathbf n}_z$,
the X-ray amplitudes obtained by solving (\ref{eq38_TTPlaneWave})
are the function of only $z$
not depending on the values of $x$ and $y$.
Therefore, they can be described as
$D_i^{\prime\prime(l)} (z)$ and $D_j^{\prime\prime(m)} (z)$.
Then,
$\partial D_i^{\prime\prime(l)} ({\mathbf r}) / \partial s_i$
in the left side of (\ref{eq38_TTPlaneWave})
can be deformed as follows:
\begin{subequations}
\begin{alignat}{1}
 \frac{\ \partial D_i^{\prime\prime(l)} ({\mathbf r})\ }{\ \partial s_i\ }
 &= {\mathbf s}_i \cdot \Big(
                  \frac{\ {\rm d} D_i^{\prime\prime(l)} (z)}{{\rm d} z}
                        \Big) {\mathbf n}_z
% \nonumber \\
\label{eq39_Gradient_a} \\
 &= \cos \Theta_i \frac{\ {\rm d} D_i^{\prime\prime(l)} (z)}{{\rm d} z}.
\label{eq39_Gradient_b}
\end{alignat}
\label{eq39_Gradient}
\end{subequations}
% 式(\ref{eq39_Gradient})を式(\ref{eq38_TTPlaneWave})に代入すると，
% 次のような連立常微分方程式を得る。
(\ref{eq39_Gradient})
can be substituted into (\ref{eq38_TTPlaneWave})
to obtain the ordinary differential equation as follows:
\begin{alignat}{1}
% & \frac{{\rm d}}{\ {\rm d} z\ } D_i^{\prime\prime(l)} (z)
% \nonumber \\
\frac{{\rm d}}{\ {\rm d} z\ } D_i^{\prime\prime(l)} (z)
% & \qquad = {\rm i} 2 \pi \xi_i^{\prime\prime}
&          = {\rm i} 2 \pi \xi_i^{\prime\prime}
                                 D_i^{\prime\prime(l)} (z)
% \nonumber \\
% & \qquad + \frac{{\rm i} 2 \pi K}{\ \cos \Theta_i\ }
           + \frac{{\rm i} 2 \pi K}{\ \cos \Theta_i\ }
                           \big(
                        \beta^{\prime(0)} S_{i, 0}^{(0)}
                      + \beta^{\prime(1)} S_{i, 0}^{(1)}
                           \big) D_i^{\prime\prime(l)} (z)
\nonumber \\
% & \qquad - \frac{{\rm i} \pi K}{\ \cos \Theta_i\ }
&          - \frac{{\rm i} \pi K}{\ \cos \Theta_i\ }
                         \sum_{j = 0}^{n - 1} \chi_{h_i - h_j}
                         \sum_{m = 0}^{1} C_{i, j}^{(l, m)}
                                          D_j^{\prime\prime(m)} (z).
\label{eq40_TTPlaneWave2}
\end{alignat}
% 式(\ref{eq40_TTPlaneWave2})は，
% 固有値/固有ベクトル問題
% に書き換えることができ，
% これを解くことにより解が得られる。
% このことからも，E-L理論がT-T理論と等価であることがわかるのだが，
% ここでは，詳しくは記述しない。
(\ref{eq40_TTPlaneWave2}) can be described also as
an eigenvalue problem
whose numerical solution can be obtained
by using the numerical subroutine libralies e.g. LAPACK.
Also from this,
the equivalence between the E-L and T-T theories
can be verified.
However, the details of this are not described here.

% \subsection{高木方程式(T-T理論)からのエバルト-ラウエ理論(E-L理論)の導出}

\subsection{Derivation of the Ewald-Laue (E-L) theory
from the Takagi-Taupin (T-T) equation}

% この節では，
% 式(\ref{eq10:eln-beam})ないしは
% 式(\ref{eq11:eln-beamDash})で表される$n$波E-L理論が
% $n$波T-T方程式(\ref{eq30_ELtoTTFinal})から導出できることを示す。
In this section,
the $n$-beam E-L theory
described in (\ref{eq10:eln-beam}) and/or (\ref{eq11:eln-beamDash})
is derived from the $n$-beam T-T equation
described in (\ref{eq30_ELtoTTFinal}).

% \centering
\begin{figure}
\begin{center}
\includegraphics[width=0.47\textwidth]{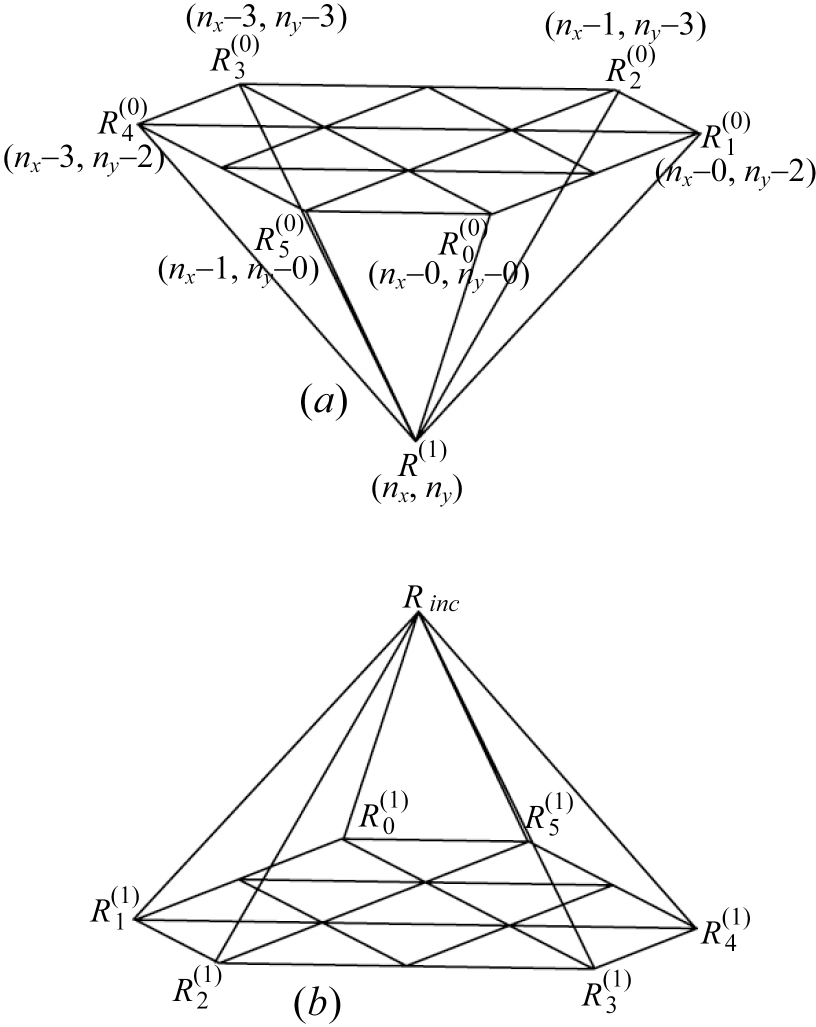}
\end{center}
\captionsetup{width=0.85\textwidth}
\caption[
These figures show small hexagonal pyramids
used when solving the $n$-beam T-T equation (\ref{eq30_ELtoTTFinal})
in a six-beam case whose results are shown in
Fig.\ \ref{Fig12_Omusubi06_Beam}
(reproduction of Fig.\ 1 in Okitsu, Imai and Yoda (2012)).
        ]
        {
These figures show small hexagonal pyramids
used when solving the $n$-beam T-T equation (\ref{eq30_ELtoTTFinal})
in a six-beam case whose results are shown in
Fig.\ \ref{Fig12_Omusubi06_Beam}
[reproduction of Fig.\ 1 in Okitsu, Imai and Yoda (2012)].
}
\label{Fig02_Omusubi_3D_Caption}
\end{figure}

% 平面波X線が結晶に入射し，
% 分散面上の$2n$個のタイポイントを励起するとき，
% 総波動場$\tilde{\bm{\mathcal D}}$は，ブロッホ波のサンメーションとして
% 次のように表される。
When plane wave X-rays are incident on the crystal
to excite $2n$ tie points on the dispersion surfaces,
total wave field $\tilde{\bm{\mathcal D}}$
is described as the summation of Bloch waves as follows:
\begin{alignat}{2}
&
\tilde{\bm{\mathcal D}}
  = \sum_{i = 0}^{n - 1} \sum_{l = 0}^{1}
  {\mathbf e}_i^{(l)} {\mathcal D}_i^{(l)}(\Delta {\mathbf k})
&
&
  \exp \left(
       - {\rm i} 2 \pi \Delta {\mathbf k} \cdot {\mathbf r}
       \right)
% \nonumber \\
% &
% &
% & \times \exp \left(
           \exp \left(
             - {\rm i} 2 \pi \overrightarrow{{\rm La}_0{\rm H}_i}
             \cdot {\mathbf r}
               \right).
\label{eq41_WholeWaveField}
\end{alignat}
Here,
$D_i^{(l)}({\mathbf r})$$=$
${\mathcal D}_i^{(l)}(\Delta {\mathbf k})
\exp ( - {\rm i} 2 \pi \overrightarrow{{\rm P}_1{\rm H}_i}
\cdot {\mathbf r})$
and
$D_j^{(m)}({\mathbf r})$ $=$
${\mathcal D}_j^{(m)}(\Delta {\mathbf k})$
$\exp ( - {\rm i}$$2$$\pi$$\overrightarrow{{\rm P}_1{\rm H}_j}$
$\cdot$${\mathbf r})$.
Therefore,
\begin{alignat}{1}
& \frac{\ \partial \ }{\partial s_i} \left[
   {\mathcal D}_i^{(l)}(\Delta {\mathbf k})
                        \exp \left(
             - {\rm i} 2 \pi \Delta {\mathbf k} \cdot {\mathbf r}
                             \right)
                                   \right]
\nonumber \\
& \qquad = - {\rm i} \pi K
  \sum_{j = 0}^{n - 1} \chi_{h_i - h_j}
% \nonumber \\
% & \qquad\qquad \times \sum_{m = 0}^{1}
                      \sum_{m = 0}^{1}
      C_{i, j}^{(l, m)}
         \left[
     {\mathcal D}_j^{(m)}(\Delta {\mathbf k}) \exp \left(
                       - {\rm i} 2 \pi \Delta {\mathbf k} \cdot {\mathbf r}
                               \right)
         \right].
\label{eq42:PlaneWaveTT}
\end{alignat}
% 一方，式(\ref{eq42:PlaneWaveTT})の左辺は，
% 式(\ref{eq25_integration2})を導出したのと同じ手続きで，
% 次のようにも書ける。
However, the left side of (\ref{eq42:PlaneWaveTT})
can be deformed in the same procedure used
when deriving (\ref{eq25_integration2})
as follows:
\begin{subequations}
\begin{alignat}{1}
& \frac{\ \partial \ }{\partial s_i}\Big[
   {\mathcal D}_i^{(l)}(\Delta {\mathbf k})
                        \exp \big(
             - {\rm i} 2 \pi \Delta {\mathbf k} \cdot {\mathbf r}
                             \big) \Big]
\nonumber \\
& \quad = {\mathcal D}_i^{(l)}(\Delta {\mathbf k})
  \frac{\partial}{\partial s_i} \exp \Big\{
      - {\rm i} 2 \pi           \Big[
                           \big(
        \xi \cos \Theta_i
% \nonumber \\
% & \qquad\qquad\qquad\qquad\qquad\qquad +
      + K \beta^{(0)} S_{i, 0}^{(0)}
% \nonumber \\
% & \qquad\qquad\qquad\qquad\qquad\qquad
      + K \beta^{(1)} S_{i, 0}^{(1)}
                      \big) s_i
% \nonumber \\
% & \qquad\qquad\qquad
  + f_i\big(
      e_i^{(0)}, e_i^{(1)}
      \big)
                                \Big]
                                     \Big\}
\label{eq43:PlaneWaveTT2_a} \\
% \nonumber \\
% & \quad = - {\rm i} 2 \pi \Big(
& \quad = - {\rm i} 2 \pi \Big(
          \xi \cos \Theta_i
        + K \beta^{(0)} S_{i, 0}^{(0)}
        + K \beta^{(1)} S_{i, 0}^{(1)}
                          \Big)
% \nonumber \\
% & \qquad \times
   {\mathcal D}_i^{(l)}(\Delta {\mathbf k})
                        \exp \big(
             - {\rm i} 2 \pi \Delta {\mathbf k} \cdot {\mathbf r}
                             \big).
\label{eq43:PlaneWaveTT2_b}
\end{alignat}
\label{eq43:PlaneWaveTT2}
\end{subequations}
% 式(\ref{eq42:PlaneWaveTT})と式(\ref{eq43:PlaneWaveTT2})の右辺どうし
% を比較することにより，
% 式(\ref{eq10:eln-beam})と同じ式が得られる。
% $n$波E-L理論とT-T方程式
% ($n \in \{ 3,4,5,6,8,12 \}$)
% の等価性は，
% 式(\ref{eq24_integration})，式(\ref{eq25_integration2})で定義される
% フーリエ変換で記述されることが証明された。
The right hands of (\ref{eq42:PlaneWaveTT}) and (\ref{eq43:PlaneWaveTT2})
can be compared to obtain the same equation as (\ref{eq10:eln-beam}).
The equivalence between the E-L and T-T theories
that can be described by the Fourier transform,
has been verified explicitly
from the descriptions of the previous and present subsections.

% 筆者が知る限り，
% E-L理論とT-T理論の等価性に関する記述は，
% オーティエの著書
% \cite{authier2005}
% \S 11.3で，2波理論についてわずかに言及されているのが，
% 唯一のものである。
As far as the present author knows,
just a short statement concerning this relation
between the E-L and T-T theories for the two-beam case
can be found only in \S 11.3 of Authier's book
\cite{authier2005}.

% \section{$n$波動力学理論の数値解法}

% \label{algorithm}
% 
% \subsection{$n$波高木方程式(T-T理論)の数値解法}
% 
% \label{algorithmTT}

\section{Numerical method to solve the $n$-beam dynamical theories}

\label{algorithm}

\subsection{Numerical method to solve the $n$-beam T-T equation}

\label{algorithmTT}

% Figs.\ \ref{Fig02_Omusubi_3D_Caption}\ $(a)$,
% \ref{Fig02_Omusubi_3D_Caption}\ $(b)$を参照して，
% $n=6$のときの
% $n$波T-T方程式(\ref{eq30_ELtoTTFinal})を数値的に解く際のアルゴリズムを説明する% 。
In reference to Fig.\ \ref{Fig02_Omusubi_3D_Caption}\ $(a)$ and
\ref{Fig02_Omusubi_3D_Caption}\ $(b)$,
the algorithm to solve the $n$-beam T-T equation
as described in (\ref{eq30_ELtoTTFinal}) for
$n = 6$
is explained.
% 後に示す
% Fig.\ \ref{Fig12_Omusubi06_Beam}の計算機シミュレーション画像を得る場合の
% 手法に例をにして記述する。
The following method was used to obtain the computer-simulated images
shown in Fig.\ \ref{Fig12_Omusubi06_Beam}.
% Fig.\ \ref{Fig02_Omusubi_3D_Caption}\ $(a)$の
% ベクトル$\overrightarrow{R_i^{(0)}R^{(1)}}$は，
% ${\mathbf s}_i$に平行である。
$\overrightarrow{R_i^{(0)}R^{(1)}}$
in Fig.\ \ref{Fig02_Omusubi_3D_Caption}\ $(a)$
is parallel to ${\mathbf s}_i$.
% ベクトル$\overrightarrow{R_i^{(0)}R^{(1)}}$
% の長さが$\left|-1/(\chi_0 K)\right|$と比較して十分小さいとき，
% 完全結晶に対する
% $n$波T-T方程式(\ref{eq30_ELtoTTFinal})は次の式で近似できる。
When the length of $\overrightarrow{R_i^{(0)}R^{(1)}}$
is sufficiently small compared with the value
of $\left|-1/(\chi_0 K)\right|$,
the $n$-beam T-T equation can be approximated by
the following equation:
\begin{alignat}{1}
\frac{D_i^{(l)}(R^{(1)})-D_i^{(l)}(R_i^{(0)})}
{\left| \overrightarrow{R_i^{(0)}R^{(1)}} \right|}
% \nonumber \\
% & \quad = - {\rm i} \pi K \sum_{j=0}^{n-1}\chi_{h_i-h_j}
  &       = - {\rm i} \pi K \sum_{j=0}^{n-1}\chi_{h_i-h_j}
% \nonumber \\
% & \quad\quad \times
 \sum_{m=0}^{1} C_{i, j}^{(l, m)}
     \frac{\ D_j^{(m)}(R_i^{(0)}) + D_j^{(m)}(R^{(1)})\ }{2}.
\label{eq44:sabun}
\end{alignat}
% 上の式(\ref{eq44:sabun})は，
% $2 n$個の未知数$D_i^{(l)}(R^{(1)})$
% $(i \in \{ 0, 1, \cdots, n - 1 \},\ l \in \{ 0, 1 \})$
% を持つ$2 n$連立1次方程式であり，
% LAPACKのZGeTRFとZGeTRSなどを用いて，
% 計算機で数値解を求めることができる。
(\ref{eq44:sabun}) describes
$2 n$-dimensional simultaneous linear equations
for $(i \in \{ 0, 1, \cdots, n - 1 \},\ l \in \{ 0, 1 \})$
and can be numerically solved by using
subroutine
{e.g.}
ZGeTRF and ZGeTRS
in the lapack
(Linear Algebra subroutine library Package).

\begin{figure}[!t]
% \centering
\begin{center}
\includegraphics[width=0.47\textwidth]{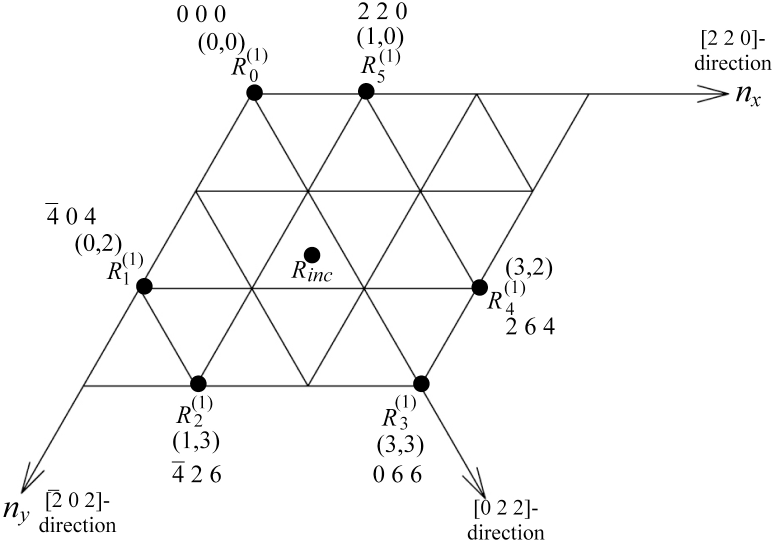}
\end{center}
\captionsetup{width=0.85\textwidth}
\caption[
This figure shows a top view of Fig.\ \ref{Fig02_Omusubi_3D_Caption}\ $(b)$
(reproduction of Fig.\ 2 in Okitsu {\it et al.} (2012)).
        ]
        {
This figure shows a top view of Fig.\ \ref{Fig02_Omusubi_3D_Caption}\ $(b)$
(reproduction of Fig.\ 2 in Okitsu {\it et al.} (2012)\cite{okitsu2012}).
}
\label{Fig03_Omusubi_Caption_Rev01}
\end{figure}

% Fig.\ \ref{Fig03_Omusubi_Caption_Rev01}は，
% Fig.\ \ref{Fig02_Omusubi_3D_Caption}\ $(b)$を
% 上から見たところである。
Fig.\ \ref{Fig03_Omusubi_Caption_Rev01} is the top view
of Fig.\ \ref{Fig02_Omusubi_3D_Caption}\ $(b)$.
% このケースでは，
% $0\ 0\ 0$前方回折波と
% $\overline{4} \ 0 \ 4$,
% $\overline{4} \ 2 \ 6$,
% $0 \ 6 \ 6$,
% $2 \ 6 \ 4$,
% $2 \ 2 \ 0$反射X線が同時に強い
% (Fig.\ \ref{Fig12_Omusubi06_Beam}参照)。
In this case,
$0\ 0\ 0$-forward diffracted,
$\overline{4} \ 0 \ 4$,
$\overline{4} \ 2 \ 6$,
$0 \ 6 \ 6$,
$2 \ 6 \ 4$
and
$2 \ 2 \ 0$-reflected X-ray beams
are simultaneously strong
(see Fig.\ \ref{Fig12_Omusubi06_Beam}).
% Fig.\ \ref{Fig03_Omusubi_Caption_Rev01}の
% $n_x$方向と$n_y$方向のなす角は$120^{\circ}$である。
The angle spanned by the directions of $n_x$ and $n_y$
is $120^{\circ}$.
% Fig.\ \ref{Fig02_Omusubi_3D_Caption}\ $(b)$の
% ベクトル$\overrightarrow{R_{inc}R_i^{(1)}}$
% $(i \in \{ 0,1,2,3,4,5 \})$は，
% $0\ 0\ 0$前方回折波と
% $\overline{4} \ 0 \ 4$,
% $\overline{4} \ 2 \ 6$,
% $0 \ 6 \ 6$,
% $2 \ 6 \ 4$,
% $2 \ 2 \ 0$反射X線の波数ベクトルに平行である。
Vectors
$\overrightarrow{R_{inc}R_i^{(1)}}$
$(i \in \{ 0,1,2,3,4,5 \})$
in Fig.\ \ref{Fig02_Omusubi_3D_Caption}\ $(b)$
are parallel to the directions of
propagations of
$0\ 0\ 0$-forward diffracted,
$\overline{4} \ 0 \ 4$,
$\overline{4} \ 2 \ 6$,
$0 \ 6 \ 6$,
$2 \ 6 \ 4$ and
$2 \ 2 \ 0$-reflected
X-ray beams.
% 4次元の配列$D_{even}(i,l,n_x,n_y)$
% $[i \in \{ 0, 1, \cdots, n-1 \}$,
% $l \in \{ 0, 1 \}$,
% $n_x \in \{ \cdots, -2, -1, 0, 1, 2, \cdots \}$,
% $n_y \in \{ \cdots, -2, -1, 0, 1, 2, \cdots \}]$
% を用意し，
% 入射X線の偏光状態に応じて，
% $(i,l,n_x,n_y)=(0,0,0,0)$ [入射X線の偏光状態が0]または
% $(i,l,n_x,n_y)=(0,1,0,0)$ [入射X線の偏光状態が1]のとき，
% $D_{even}(i,l,n_x,n_y) = 1$を与え，
% それ以外の場合は
% $D_{even}(i,l,n_x,n_y) = 0$を，
% 結晶表面のX線入射点における境界条件として与える。
A four-dimensional array $D_{even}(i,l,n_x,n_y)$
$[i \in \{ 0, 1, \cdots, n-1 \}$,
$l \in \{ 0, 1 \}$,
$n_x \in \{ \cdots, -2, -1, 0, 1, 2, \cdots \}$,
$n_y \in \{ \cdots, -2, -1, 0, 1, 2, \cdots \}]$
should be prepared such that
the calculated X-ray amplitudes are saved.
Here, $i$ is the ordinal number of the X-ray wave,
$l$ ($l \in \{ 0, 1 \}$) is the polarization state,
$n_x$ and $n_y$ are two-dimensional positions
on the layer in the crystal.
% 結晶を十分に多くの層に分割して，
% 上の層から下の層へと計算を実行してゆくが，
% 表面から1層下のX線振幅$D_{odd}(j, m, n_x, n_y)$は，
% Fig.\ \ref{Fig02_Omusubi_3D_Caption}\ $(a)$に示すように
% $D_{odd}(i, l, n_x-0, n_y-0)$,
% $D_{odd}(i, l, n_x-0, n_y-2)$,
% $D_{odd}(i, l, n_x-1, n_y-3)$,
% $D_{odd}(i, l, n_x-3, n_y-3)$,
% $D_{odd}(i, l, n_x-3, n_y-2)$,
% $D_{odd}(i, l, n_x-1, n_y-0)$
% から式(\ref{eq44:sabun})
% を解くことによって求められる。
The crystal is divided to sufficiently large number of layers.
The calculations are repeated layer by layer
toward the depth direction.
As shown in Fig.\ \ref {Fig02_Omusubi_3D_Caption}\ $(a)$,
X-ray amplitudes $D_{odd}(j, m, n_x, n_y)$ just below
the crystal surface can be obtained from
$D_{odd}(i, l, n_x-0, n_y-0)$,
$D_{odd}(i, l, n_x-0, n_y-2)$,
$D_{odd}(i, l, n_x-1, n_y-3)$,
$D_{odd}(i, l, n_x-3, n_y-3)$,
$D_{odd}(i, l, n_x-3, n_y-2)$, and
$D_{odd}(i, l, n_x-1, n_y-0)$
by solving (\ref{eq44:sabun}).
% Fig.\ \ref{Fig02_Omusubi_3D_Caption}\ $(b)$
% のような「ボルマン$n$角錐(ピラミッド)」の外には波動場が存在しないので，
% 計算はピラミッド内を末広がりにスキャンするように行う。
Outside the Borrmann pyramid as shown in
Fig.\ \ref{Fig02_Omusubi_3D_Caption}\ $(b)$,
the wave fields do not exist.
The calculation is performed by scanning insid the Bormann pyramid.
% $\chi_{h_i - h_j}$の値は，
% XOP version2.3
% \cite{sanchez1998}
% を用いて求めた。
The values of  $\chi_{h_i - h_j}$ were calculated
by using XOP version 2.3
\cite{sanchez1998}

% 一方，平面波入射条件に対応する常微分方程式
% (\ref{eq40_TTPlaneWave2})を解く場合の，
% 式(\ref{eq44:sabun})に相当する差分方程式は，
% 次のようになる。
The difference equation that approximates
the standard differential equation (\ref{eq40_TTPlaneWave2})
is given as follows:
\begin{alignat}{1}
\frac{D_i^{\prime\prime(l)}(z+\Delta z)-D_i^{\prime\prime(l)}(z)}
{\Delta z}
% \nonumber \\
% & \quad = {{\rm i 2 \pi}} \Big[
  &       = {{\rm i 2 \pi}} \Big[
          \xi_i^{\prime\prime}
 + \frac{K}{\ \cos \Theta_i\ }  \big(
                          \beta^{\prime(0)} S_{i, 0}^{(0)}
                        + \beta^{\prime(1)} S_{i, 0}^{(1)}
                                \big)
                          \Big]
% \nonumber \\
% & \quad\quad \times
     \frac{\ D_i^{\prime\prime(l)}(z) + D_i^{\prime\prime(l)}
       (z + \Delta z)\ }{2}
\nonumber \\
% & \quad - \frac{{\rm i} \pi K}{\ \cos \Theta_i\ }
% \sum_{j=0}^{n-1}\chi_{h_i-h_j}
  &       - \frac{{\rm i} \pi K}{\ \cos \Theta_i\ }
    \sum_{j=0}^{n-1}\chi_{h_i-h_j}
% & \quad\quad \times
 \sum_{m=0}^{1} C_{i, j}^{(l, m)}
     \frac{\ D_j^{\prime\prime(m)}(z)
           + D_j^{\prime\prime(m)}(z + \Delta z)\ }{2}.
\label{eq45:sabun}
\end{alignat}
% 式(\ref{eq45:sabun})は，平面波X線入射，
% すべての反射がラウエケース限定ではあるが，
% 式(\ref{eq44:sabun})の解を求めるときのように
% 体積積分を行わないので，短時間で解を求めることができる。
(\ref{eq45:sabun}) can be solved in a short time
only when incident plane-wave X-rays excite $n$
diffracted X-ray beams all in Laue geometries.

% \subsection{$n$波エバルト-ラウエ(E-L)理論の数値解法}
\subsection{Numerical method to solve the $n$-beam E-L theory}

% \label{section_03_002_algorithmEL}
\label{section_03_002_algorithmEL}

% 式(\ref{eq12:matrix})の
% 行列${\mathbf A}$に
% 式(\ref{eq13:ElementOfMatrix})を代入し，
% 例えばLAPACKのZGeEVを用いて
% $k$番目($k \in \{ 1, 2, \cdots, 2n \}$)の固有値
% $\xi_k$と固有ベクトル$\pmb{\mathscr{D}}_k$を
% 求めることができる。
After substituting (\ref{eq13:ElementOfMatrix})
into matrix ${\mathbf A}$ of (\ref{eq12:matrix}),
$k$th ($k \in \{ 1, 2, \cdots, 2n \}$)
eigenvalue $\xi_k$ and eigenvector $\pmb{\mathscr{D}}_k$
can be solved by using subroutine libraries
{\rm e.g.} LAPACK.

% これにより，ブロッホ波を構成する波の波数ベクトルと
% $q$番目の波$\big(q = 2 j + m + 1\big)$の振幅比
% が求められるわけであるが，
% 次に，$2n$個のブロッホ波をどのような配合比
% で合成すれば，境界条件を満足するかを考慮する必要がある。
In this way,
wave vectors and amplitude ratios
of the $q$th $\big(q = 2 j + m + 1\big)$
Bloch wave.
On the other hand, mixing ratio of $2n$ Bloch waves
should be calculated such as to
satisfy the boundary condition.

% 列ベクトル$\pmb{\mathscr D}_k$の$q$番目の要素${\mathscr D}_{q, k}$
% ($={\mathscr D}_{j, k}^{(m)}$)
% を
% $q$行$k$列目の要素とする$2n \times 2n$行列
% $\pmb{\mathscr D}$を作り，
% 次のような方程式を立てる。
After making $2n \times 2n$ matrix $\pmb{\mathscr D}$
whose element of $q$th raw
and $k$th column is ${\mathscr D}_{q, k}$
($={\mathscr D}_{j, k}^{(m)}$),
the following equation is obtained:
\begin{subequations}
\begin{alignat}{1}
\pmb{\mathscr{D}} {\boldsymbol \alpha}^{(0)}
  = \big( 1, 0, 0, \cdots, 0, 0 \big)^{T},
\label{eq46:BoundCond_a}
\\
\pmb{\mathscr{D}} {\boldsymbol \alpha}^{(1)}
  = \big( 0, 1, 0, \cdots, 0, 0 \big)^{T}.
\label{eq46:BoundCond_b}
\end{alignat}
\label{eq46:BoundCond}
\end{subequations}
% 上の式(\ref{eq46:BoundCond_a})と(\ref{eq46:BoundCond_b})は，
% $l$ $(l \in \{ 0, 1 \})$の偏光状態のX線を入射したときの
% 入射側結晶表面での境界条件である。
The above equations
(\ref{eq46:BoundCond_a}) and (\ref{eq46:BoundCond_b})
are the boundary conditions that should be given
for the entrance surface
when an X-ray beam whose polarization state
is $l$ $(l \in \{ 0, 1 \})$ is incident
on the crystal.
% これらの式を解くことにより，
% 列ベクトル${\boldsymbol \alpha}^{(l)}$の
% $k$番目の要素$\alpha_k^{(l)}$
% が求められる。
These equations can be solved
to obtain the $k$th element $\alpha_k^{(l)}$
of the column vector ${\boldsymbol \alpha}^{(l)}$.
% これは，$k$番目のブロッホ波の配合比なので，
These are mixing ratios of $k$th Bloch waves.
% $l$の偏光状態の入射X線に励起され，
% 結晶裏面から出射する$j$番目$m$偏光の
% X線の振幅$\mathcal{D}_{j}^{(l, m)}(exit)$
% $\big[ =\mathcal{D}_q^{(l)}(exit) \big]$は，
% 次の式で求められる。
Then,
the amplitudes $\mathcal{D}_{j}^{(l, m)}(exit)$
$\big[ =\mathcal{D}_q^{(l)}(exit) \big]$
of the $j$th numbered X-ray beam whose polarization state is $m$,
can be obtained as follows:
\begin{subequations}
\begin{alignat}{1}
{\mathcal D}_{j}^{(l, m)}(exit)
  &= {\mathcal D}_{q}^{(l)}(exit)
% \nonumber \\
\label{Eq47_E-LSolution_b} \\
  &= \sum_{k=1}^{2n} \alpha_k^{(l)} \mathscr{D}_{j, k}^{(m)}
  \exp[ -{\rm i} 2 \pi \xi_k T_z].
\label{Eq47_E-LSolution_b}
\end{alignat}
\label{Eq47_E-LSolution}
\end{subequations}
% ここで，$T_z$は結晶の厚さである。
Here, $T_z$ is the thickness of the crystal.

% 式(\ref{eq13:ElementOfMatrix})の右辺第2項には，
% 入射X線の，厳密な$n$波条件からの角度のズレ
% を表すパラメーター$\beta^{(0)}$と$\beta^{(1)}$
% が含まれるので，
The second terms $\beta^{(0)}$ and $\beta^{(1)}$ in the left side
of (\ref{eq13:ElementOfMatrix})
are angular deviations
from the exact $n$-beam condition.
% $\left| {\mathcal D}_{j}^{(l, m)}(exit) \right|^2
% \big({\mathbf n}_z \cdot {\mathbf s}_j \big) / \big({\mathbf n}_z \cdot {\math% bf s}_0\big)$
% をとることにより，$j$番目の波に対して，
% Fig.\ \ref{Fig10_4Beam_RockingCurve_deg_000_Rotate000}のような
% 2次元の回折強度曲線が得られる。
Two-dimensional rocking curves are obtained as
shown in Fig.\ Fig.\ \ref{Fig10_4Beam_RockingCurve_deg_000_Rotate000}
by calculating
$\left| {\mathcal D}_{j}^{(l, m)}(exit) \right|^2$
$\big({\mathbf n}_z \cdot {\mathbf s}_j \big)$
$/$
$\big({\mathbf n}_z \cdot {\mathbf s}_0\big)$
as the intensities of the $j$th numbered X-ray beam.
% $({\mathbf n}_z \cdot {\mathbf s}_j) / ({\mathbf n}_z \cdot {\mathbf s}_0)$
% は，0番目と$j$番目のビーム断面積の違いを考慮するための補正項，
% Fig.\ \ref{Fig10_4Beam_RockingCurve_deg_000_Rotate000}では，
% $j = 0$である。
$({\mathbf n}_z \cdot {\mathbf s}_j)$
$/$
$({\mathbf n}_z \cdot {\mathbf s}_0)$
is the correction term for taking into account the difference in
the area of the cross section for the X-ray beams.
$j = 0$ in the case of
Fig.\ \ref{Fig10_4Beam_RockingCurve_deg_000_Rotate000}.

% 記述を省略するが，
% $j$番目の波がブラッグケースになる場合，
% 結晶の裏面での振幅の総和がゼロであるという
% $\sum_{k = 1}^{2n}\alpha_k^{(l)}{\mathscr D}_{j,k}^{(m)}
% \exp\big( -{\rm i} 2 \pi \xi_k T_z \big) = 0$の
% 境界条件を与えることになる。
When the $j$th numbered X-rays are reflected
in the Bragg geometry,
the boundary condition should be given to be
$\sum_{k = 1}^{2n}\alpha_k^{(l)}{\mathscr D}_{j,k}^{(m)}$
$\exp\big( -{\rm i} 2 \pi \xi_k T_z \big) = 0$
such that
the summation of the amplitudes is zero.

% 筆者らの2019年の論文
% \cite{okitsu2019b,okitsu2019c}
% では，
% $n$波E-L理論の解を高速フーリエ変換することにより
% ピンホールトポグラフ図形
% を得ている。
In the present author's and his coauthors papers
published in 2019
\cite{okitsu2019b,okitsu2019c}
report pinhole topograph images
obtained by fast Fourier-transforming the solution
of the E-L theory.
% この手法は，
% Kohn \& Khikhlukha
% \cite{kohn2016}
% およびKohn
% \cite{kohn2017}
% によって考案されたものであり，彼らは対称6波ラウエケース
% についての計算機シミュレーションを報告している。
This method was developed by
Kohn \& Khikhlukha
\cite{kohn2016}
and by Kohn
\cite{kohn2017}.
They reported computer-simulated pinhole topographs
for a symmetric six-beam case.
% 筆者らはこの手法を，
% 結晶が平行平板でない場合の
% 非対称8波ラウエケース
% \cite{okitsu2019b}，
% $n$個の逆格子点が同一円周上にない場合の
% 18波ケース
% \cite{okitsu2019c}
% に拡張し，ピンホールトポグラフのX線強度分布
% $\big|D_j^{(m)}(x_{exit}, y_{exit})\big|^2$
% を求めた。
The present author and his coauthors extended this method
such as to deal with a 18-beam case in which
18 reciprocal lattice nodes do not exist on
an identical circle in the reciprocal space.
% ここで，X線の出射側表面の位置ベクトル
% ${\mathbf r}_{exit}$を，次のように表すものとする。
Here, let the location vector on the exit surface
${\mathbf r}_{exit}$ be described as follows:
\begin{alignat}{1}
{\mathbf r}_{exit} & = x_{exit} {\mathbf e}_x
                      + y_{exit} {\mathbf e}_y
                      + T_z {\mathbf n}_z.
\label{eq48_exitSurface}
\end{alignat}
% また，Fig.\ \ref{Fig01_org_05_RecipPosition}より，
Further, in reference to Fig.\ \ref{Fig01_org_05_RecipPosition},
the following equation is obtained:
\begin{figure*}[!t]
% \centering
\begin{center}
\includegraphics[width=0.71\textwidth]{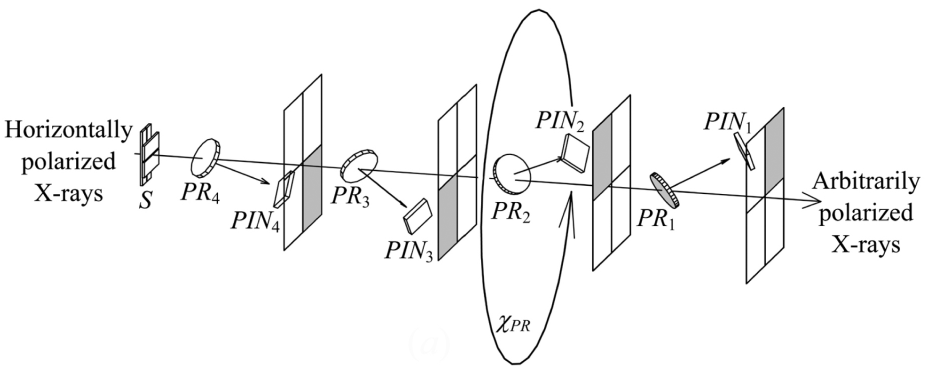}
\end{center}
\caption[
Schematic drawing of
the `rotating four-quadrant phase retarder system'
(reproduction of Fig.\ 3 in Okitsu {\it et al.} (2006)).
        ]
        {
Schematic drawing of
the `rotating four-quadrant phase retarder system'
(reproduction of Fig.\ 3 in Okitsu {\it et al.} (2006) \cite{okitsu2006}).
}
\label{Fig04_PhaseRetarderSystemSchematic}
\end{figure*}
\begin{figure}[!t]
% \centering
\begin{center}
\includegraphics[width=0.41\textwidth]{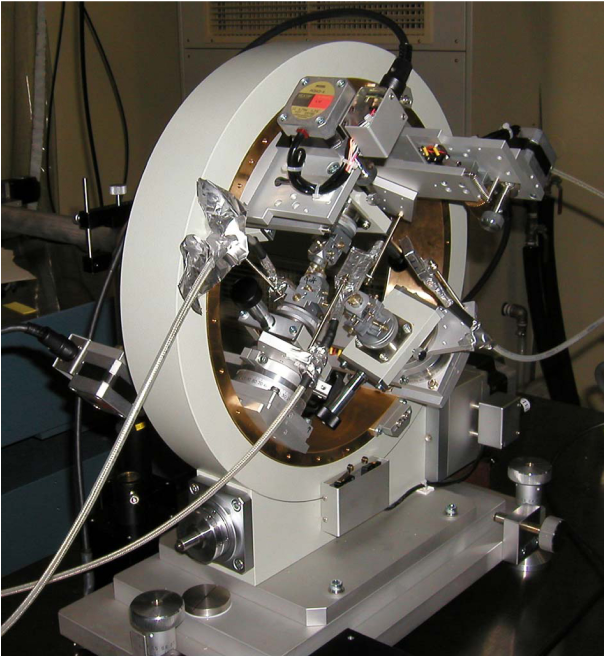}
\end{center}
\captionsetup{width=0.85\textwidth}
\caption[
Photograph of
the `rotating four-quadrant phase retarder system'
(reproduction of Fig.\ 3\ $(b)$ in Okitsu {\it et al.} (2012)).
        ]
        {
Photograph of
the `rotating four-quadrant phase retarder system'
(reproduction of Fig.\ 3\ $(b)$ in Okitsu {\it et al.} (2012) \cite{okitsu2012}).
}
\label{Fig05_PhaseRetarderSystemPhotograph}
\end{figure}
\begin{alignat}{1}
\overrightarrow{{\rm P}_1^{\prime\prime}{\rm La}_0}
 &= \Delta k_x {\mathbf e}_x + \Delta k_y {\mathbf e}_y.
\label{eq49_DeltaKxKy}
\end{alignat}
% 高速フーリエ変換により求めるべき振幅
% $D_j^{(m)}(x_{exit}, y_{exit})$が
% $\exp\big( -{\rm i} 2 \pi
% \overrightarrow{{\rm La}_0{\rm H}_j}\cdot{\mathbf r}\big)
% {\mathbf e}_j^{(m)}$
% の波を変調し，
The X-ray amplitude $D_j^{(m)}(x_{exit}, y_{exit})$
to be calculated by using the fast Fourier transform (FFT)
modulates the wave of
$\exp\big( -{\rm i} 2 \pi$
$\overrightarrow{{\rm La}_0{\rm H}_j}\cdot{\mathbf r}\big)$
${\mathbf e}_j^{(m)}$.
% かつ振幅${\mathcal D}_j^{(m)} (\Delta {\mathbf k})$の
% コヒーレントな重ね合わせであることを考慮すると，
% 次のように計算できる。
Considering that
this amplitude is the coherent superposition of
${\mathcal D}_j^{(m)} (\Delta {\mathbf k})$,
the next equation can be obtained:
\begin{alignat}{1}
& D_j^{(m)}(x_{exit}, y_{exit}) \exp\big(
        -{\rm i} 2 \pi
         \overrightarrow{{\rm La}_0{\rm H}_j} \cdot {\mathbf r}_{exit}
                                    \big)
\nonumber \\
& \quad = \int_{\Delta{\mathbf k}}^{D.S.}
          {\mathcal D}_j^{(m)} (\Delta{\mathbf k})
          \exp\big[
      -{\rm i} 2 \pi \big(
        \overrightarrow{{\rm P}_1^{\prime}{\rm P}_1^{\prime\prime}}
      + \overrightarrow{{\rm P}_1^{\prime\prime}{\rm La}_0}
                     \big) \cdot {\mathbf r}_{exit}
              \big]
% \nonumber \\
% & \qquad \times \exp\big(
                  \exp\big(
        -{\rm i} 2 \pi
         \overrightarrow{{\rm La}_0{\rm H}_j} \cdot {\mathbf r}_{exit}
                    \big) {\rm d}S.
\label{eq50_DjmCalculation}
\end{alignat}
% ここで，${\mathcal D}_j^{\prime(m)}(\Delta k_x, \Delta k_y)$を
% 次のようにおく。
Here, let ${\mathcal D}_j^{\prime(m)}(\Delta k_x, \Delta k_y)$
be as follows:
\begin{alignat}{1}
{\mathcal D}_j^{\prime(m)}(\Delta k_x, \Delta k_y)
&  = \sum_{k=1}^{2n} \alpha_k^{(l)}
              {\mathscr D}_{j, k}^{(m)}(\Delta {\mathbf k})
     \exp\big(
     -{\rm i} 2 \pi \xi_k^{\prime\prime\prime} T_z
         \big),
\label{eq51_DjPrimeDefinition} \\
& \qquad {\rm where}, \ \overrightarrow{{\rm P}_{1, k}^{\prime}{\rm P}_1^{\prime\prime}}
         = \xi_k^{\prime\prime\prime} {\mathbf n}_z.
\nonumber
\end{alignat}
% 式(\ref{eq51_DjPrimeDefinition})のサンメーション
% $\sum_{k=1}^{2n}$は，式(\ref{eq50_DjmCalculation})右辺の積分の中に
% 含ませていたものを，展開したものである。
$\sum_{k=1}^{2n}$ in (\ref{eq51_DjPrimeDefinition})
has been obtained by extracting from
the integration in the right side of (\ref{eq50_DjmCalculation}).
% 式(\ref{eq51_DjPrimeDefinition})を
% 式(\ref{eq50_DjmCalculation})に代入して，
% 式(\ref{eq48_exitSurface})と
% 式(\ref{eq49_DeltaKxKy})を考慮すると，
By substituting (\ref{eq51_DjPrimeDefinition})
into (\ref{eq50_DjmCalculation})
and considering (\ref{eq48_exitSurface}) and (\ref{eq49_DeltaKxKy}),
the following equations can be obtained:
\begin{subequations}
\begin{alignat}{1}
  & D_j^{(m)}(x_{exit}, y_{exit})
\nonumber \\
  & \quad = \int_{\Delta {\mathbf k}}^{D.S.}
          \sum_{k = 1}^{2n} \alpha_k^{(l)}
               {\mathscr D}_{j,k}^{(m)}(\Delta {\mathbf k})
                    \exp\big(
      -{\rm i} 2 \pi \xi_k^{\prime\prime\prime} T_z
                        \big)
% \nonumber \\
% & \quad\quad \times \exp\big(
                      \exp\big(
      -{\rm i} 2 \pi
               \overrightarrow{P_1^{\prime\prime}La_0}
                 \cdot {\mathbf r}_{exit}
                        \big) {\rm d}S
\label{eq52_DjmExit_a} \\
% \nonumber \\
  & \quad = \int_{\Delta k_x} \int_{\Delta k_y}
% &       = \int_{\Delta k_x} \int_{\Delta k_y}
            {\mathcal D}_j^{\prime(m)}(\Delta k_x, \Delta k_y)
% \nonumber \\
% & \quad\quad \times \exp\big[
                      \exp\big[
      -{\rm i} 2 \pi \big(
               \Delta k_x x_{exit} + \Delta k_y y_{exit}
                     \big)
                        \big] {\rm d} \Delta k_y
                              {\rm d} \Delta k_x.
\label{eq52_DjmExit_b}
\end{alignat}
\label{eq52_DjmExit}
\end{subequations}
% 上の式(\ref{eq52_DjmExit})は，標準的な2次元フーリエ変換である。
% すなわち，
% ピンホールトポグラフのX線振幅
% $D_j^{(m)}(x_{exit}, y_{exit})$を得るべく
% 高速フーリエ変換にかけるのは，
% 式(\ref{eq51_DjPrimeDefinition})で定義される
% ${\mathcal D}_j^{\prime(m)}(\Delta k_x, \Delta k_y)$
% である。
The above equation (\ref{eq52_DjmExit})
is a standard two-dimensional Fourier transform.
The X-ray amplitude $D_j^{(m)}(x_{exit}, y_{exit})$
can be obtained by fast Fourier-transforming
${\mathcal D}_j^{\prime(m)}(\Delta k_x, \Delta k_y)$
defined by (\ref{eq51_DjPrimeDefinition}).

% \section{実験}

\section{Experimental}

\label{experimental}

% \subsection{X線移相子システム}

\subsection{Phase retarder system}

% $4$波，
% $5$波，
% $6$波，
% $8$波ケースのピンホールトポグラフの実験は，
% SPring-8，BL09XUのビームラインにおいて，
% 水冷式のダイヤモンドモノクロメーターで，
% 18.245 keVに単色化されたX線を用いて行われた。
The experiments to obtain
$4$-,
$5$-,
$6$- and
$8$-beam X-ray pinhole topographs
were performed at BL09XU of SPring-8
by using the synchrotron X-rays.
They were monochromatized to 18.245 keV
by using the water-cooled diamond monochromator.
% 結晶に入射するX線は
% 「回転型四象限移相子システム」
% \cite{okitsu2006,okitsu2012}
% で偏光状態をコントロールした。
The polarization state of the incident X-rays
was controlled by using
`the rotating four-quadrant phase-retarder system'
\cite{okitsu2006,okitsu2012}.
% この偏光制御システム開発には，前段階があった。
There were previous states before
developing this polarization control system
% 軸収差を補償する「二象限X線移相子システム」
% \cite{okitsu2002}，
% 軸収差と色収差の両方を
% 補償する
% 「四象限X線移相子システム」である
% \cite{okitsu2002,okitsu2003c}。
i.e.
`the two-quadrant X-ray phase retarder system'
to compensate for the off-axis aberration
\cite{okitsu2001}
and `the four-quadrant X-ray phase retarder system'
to compensate for both
the off-axis and chromatic aberrations
\cite{okitsu2002}.
% これらは，筆者が考案，設計し，手作りで製作した。
These were invented, designed and manufactured
by the present author.
% 評価実験は，上ヱ地とともに行い，良好な結果を得た。
The experimental estimations were performed
by the present author and Ueji to obtain
the excellent results at the Photon Factory of KEK.
% さらに発展させ，
% 入射X線の光軸周りに回転させることにより
% 任意の偏光状態を生成可能にしたのが，
% Figs. \ref{Fig04_PhaseRetarderSystemSchematic},
% \ref{Fig05_PhaseRetarderSystemPhotograph}に示すシステムである。
These systems were further improved
as shown in Figs.\ \ref{Fig04_PhaseRetarderSystemSchematic}
and \ref{Fig05_PhaseRetarderSystemPhotograph}
such as to rotate around the optical axis to
generate arbitrary polarized X-rays.

% 透過型X線移相子
% \cite{hirano1991,ishikawa1991,hirano1992,ishikawa1992,hirano1993,hirano1995,gi% les1994a,giles1994b}
% は，それ以前に検討された反射型X線移相子
% \cite{hart1978,annaka1980,annaka1982,golovchenko1986,mills1987}
% と比較して
% 一様な偏光状態のX線を得る手段として画期的なものであった。
% それでもなお，入射X線の角度発散と波長広がりにより
% 偏光状態に不均一(収差)が残る，という問題があった。
% 透過型X線移相子を複数の象限への反射を与えるように
% 重ねて用いることにより
% 収差が補償され，より均一な偏光状態が得られる。
% また移相子結晶の実効厚を稼げるため，
% 高エネルギー領域での偏光コントロールに特に有利となる。
The transmission-type phase retarder
\cite{hirano1991,ishikawa1991,hirano1992,ishikawa1992,hirano1993,hirano1995,giles1994a,giles1994b}
was innovative polarization-controll system.
This can generate uniform phase shift between
$\sigma$- and $\pi$-polarized X-rays
compared with the reflection-type phase retarders
\cite{hart1978,annaka1980,annaka1982,golovchenko1986,mills1987}.
Nevertheless, there was a problem
of inhomogeneity (aberrations) of the phase shift
owing to the angular divergence and energy spread of
the incident X-rays.
However, further homogeneous value of phase shift
can be obtained by overlapping the phase retarder crystals
such that the planes of incidence of them
are inclined by 45$^{\circ}$ and 225$^{\circ}$
from the horizontal plane
(two-quadrant system)
\cite{okitsu2001}
and by 45$^{\circ}$, 135$^{\circ}$, $225^{\circ}$
and 315$^{\circ}$
(four-quadrant system)
\cite{okitsu2002}.
The two- and four-quadrant phase retarder system
are particularly effective in the high energy region
since the large value of total thickness of the diamond crystals
can decrease the residual ununiformity of phase shift for
$\sigma$- and $\pi$-polarized X-rays.

% Fig.\ \ref{Fig04_PhaseRetarderSystemSchematic}は，
% この移相子システムの模式図，
% Fig.\ \ref{Fig05_PhaseRetarderSystemPhotograph}は，写真である。
% [1\ 0\ 0]方位の4枚のダイヤモンド結晶
% $PR_n$ $\big(n \in \{ 1,2,3,4 \}\big)$
% で構成されており，
% 厚さはそれぞれ，
% 1.545, 2.198, 1.565, 2.633 mmである。
% 非対称ラウエケース，$1\ 1\ 1$反射条件の近傍で用いた。
% 偏光コントロールの詳細については，
% 筆者らの2006年の論文
% \cite{okitsu2006}
% に記述してある。
Fig.\ \ref{Fig04_PhaseRetarderSystemSchematic} is
a schematic drawing of the phase retarder system.
Fig.\ \ref{Fig05_PhaseRetarderSystemPhotograph}
is its photograph.
This system consists of $PR_n$ $\big(n \in \{ 1,2,3,4 \}\big)$.
These are [1\ 0\ 0]-oriented
diamond crystals whose thickness are
1.545, 2.198, 1.565 and 2.633 mm
and used in the vicinity of the angles
to give $1\ 1\ 1$ reflection
in an asymmetric Laue geometry.
How to control this system
has been described in detail
in the paper published in 2006
by the present author and his coauthors
\cite{okitsu2006}.

\begin{figure}[!t]
% \centering
\begin{center}
\includegraphics[width=0.45\textwidth]{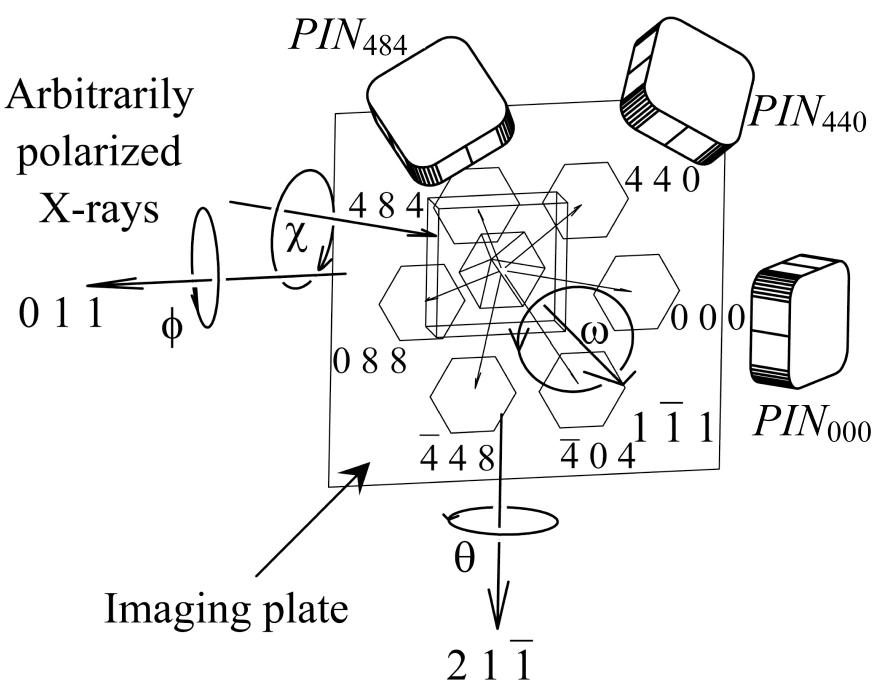}
\end{center}
\captionsetup{width=0.85\textwidth}
\caption[
A schematic drawing of the goniometer on which
the sample crystal was mount
(reproduction of Fig.\ 7 in Okitsu {\it et al.} (2006))
        ]
        {
A schematic drawing of the goniometer on which
the sample crystal was mount
(reproduction of Fig.\ 7 in Okitsu {\it et al.} (2006))
\cite{okitsu2006}].
}
\label{Fig06_au5037_Fig07}
\end{figure}

% $n \in \{ 3, 12, 18 \}$の
% $n$波ケースの実験においては，
% 移相子システムを使わず，
% 18.245 keV ($n = 3$)，
% 22.0 keV ($n \in \{12, 18\}$)に単色化された水平偏光のX線を，
% そのまま結晶に入射した。
In the three-, 12- and 18-beam pinhole topograph experiments,
the horizontally polarized monochromatized synchrotron X-rays
were incident on the crystal
without using the phase retarder system.
The photon energy for the three-beam case
was 18.245 keV.
That for 12- and 18-beam cases was 22.0 keV.

% \subsection{サンプルとして用いた結晶と方位調整}

\subsection{Silicon crystals uses as sample crystals
and their position angle adjustment}

% Fig.\ \ref{Fig06_au5037_Fig07}は，
% 筆者らの2006年の論文
% \cite{okitsu2006}
% のFig.\ 7から転載したものである。
Fig.\ \ref{Fig06_au5037_Fig07} was reproduction
of Fig.\ 7
in the paper published by the present author
and his coauthors in 2006
\cite{okitsu2006}.
% $n \in \{ 3,4,5,6,8,12,18 \}$の
% $n$波ケースに対して，
% $[1\ \overline{1}\ 1]$方位，高純度高抵抗のFZシリコン結晶を用いた。
The sample crystals used in the $n$-beam pinhole topograph experiment
for $n \in \{ 3,4,5,6,8,12,18 \}$
were $[1\ \overline{1}\ 1]$-oriented floating zone silicon crystals
with high purity and high resistance.
% 結晶の厚さは，$12$波，$18$波ケースでは10.0 mm，
% その他のケースでは9.6 mmであった。
The thicknesses of the sample crystals
were 10.0 mm in 12- and 18-beam cases and
9.6 mm in the other cases.
% $\chi$, $\phi$, $\omega$, $\theta$の4軸ゴニオメーターにマウントされ，
% これらの軸の方位は
% Fig.\ \ref{Fig06_au5037_Fig07}に示したとおりである。
The sample crystals were mounted on a goniometer
that has four axes of
$\chi$, $\phi$, $\omega$ and $\theta$.
Their angles were controlled as shown
in Fig.\ \ref{Fig06_au5037_Fig07}.
% $0\ 0\ 0$前方回折波と
% ふたつの反射X線を，$PIN$フォトダイオードでモニターし，
% それらが同時に最強になるように，ゴニオメーターの
% 回転軸を調整した。
$0\ 0\ 0$-forward diffracted and two reflected beam intensities
were monitored with PIN photodiodes.
The angles of the goniometer were controlled
such that their intensities have maximum value.
% 入射X線の下流側から，透過X線の光路と一致するように
% レーザービームをセットし，
% 結晶のX線入射位置に鏡を置き，
% X線の反射方向と完全に同じ方向にレーザーを反射するよう，
% 鏡の角度をゴニオメーターで調整した。
The positions of the diodes were adjusted such that
the laser beam is incident on the detector.
Before that, the rotation angles of the axes of
the goniometer were calculated such as to
reflect the laser beam in the beam direction of
the reflected X-rays.
% 鏡を載せたゴニオメーターの回転角は予め計算しておくが，
% この計算無しで$15 \times 15$ mm程度のダイオード受光面に
% 反射X線を捉えることは不可能である。
It is impossible to adjust the positions of the diodes
whose detection area was $15 \times 15$ mm
such that the reflected X-rays were incident on them.

% % ビームサイズは，$25 \times 25 \mu$mのサイズに絞り，
% 結晶後方に，結晶の出射側表面と平行になるように
% イメージングプレートを置き，
% $n$個の前方回折および反射X線図形を同時に撮影した。
The dimension of the incident X-ray beam was set to be
$25 \times 25 \mu$m with a four-quadrant slit system
placed upstream of the phase retarder system.
$N$ pinhole topograph images of forward diffracted
and reflected X-rays
were simultaneously taken on the imaging plate
set behind the sample crystal.

% \section{実験と計算機シミュレーションの結果}

\section{Results of the experiment and computer-simulation}

% \subsection{3波ケース}

\subsection{Three-beam case}

\begin{figure}[!t]
% \centering
\begin{center}
\includegraphics[width=0.51\textwidth]{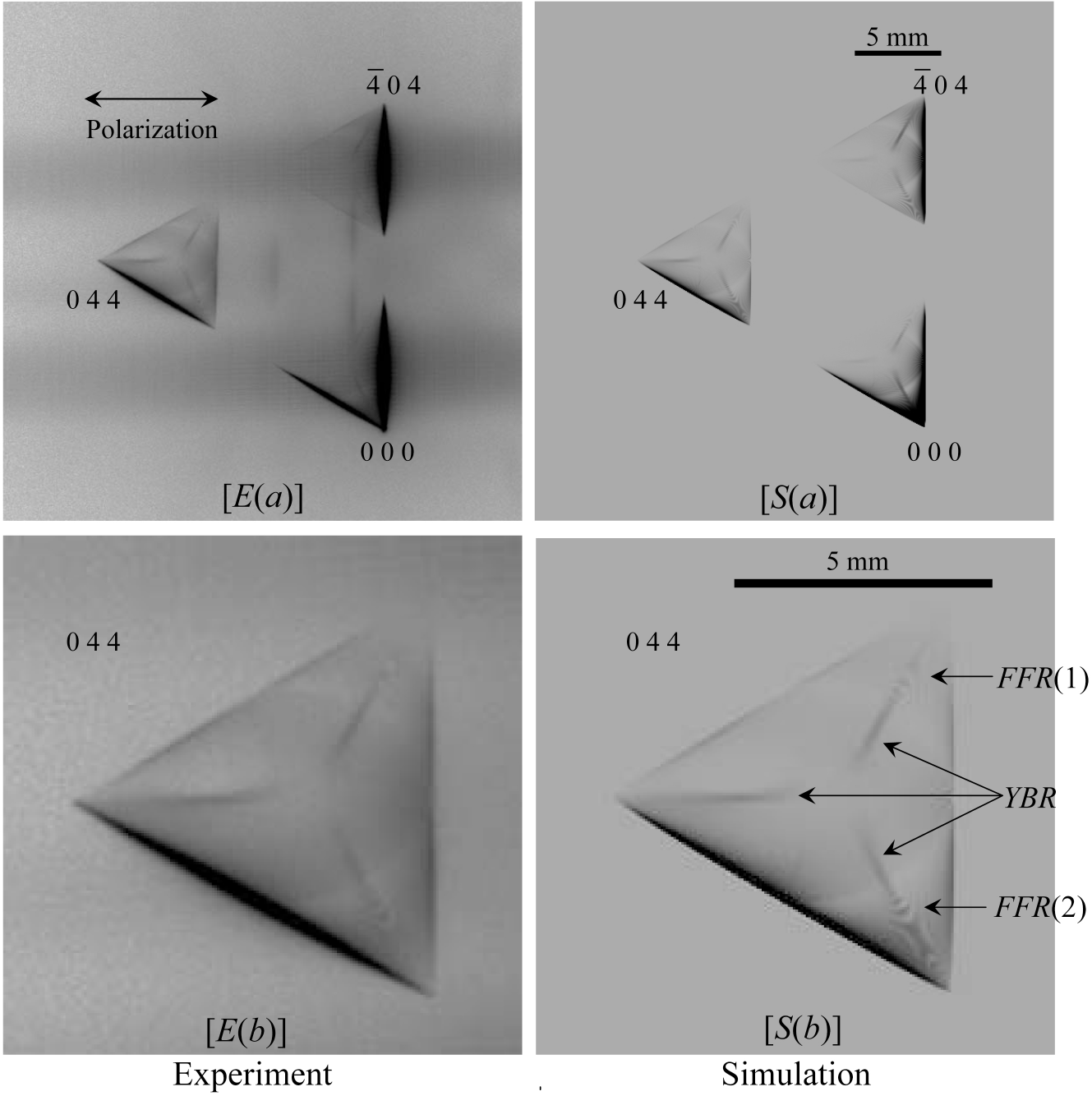}
\end{center}
\captionsetup{width=0.85\textwidth}
\caption[
$E(a)$ and $S(a)$
are experimentally obtained
and computer-simulated three-beam X-ray pinhole topographs
with an incidence of horizontal-linearly polarized X-rays whose photon energy
was 18.245 keV.
$E(b)$ and $S(b)$ are $0 \ 4 \ 4$ reflected X-ray images
enlarged from
$E(a)$ and $S(a)$, respectively(reproduction of Fig.\ 5 in Okitsu {\it et al.} (2012)).
        ]
        {
[$E(a)$] and [$S(a)$]
are experimentally obtained
and computer-simulated three-beam X-ray pinhole topographs
with an incidence of horizontal-linearly polarized X-rays whose photon energy
was 18.245 keV.
[$E(b)$] and [$S(b)$] are $0 \ 4 \ 4$ reflected X-ray images
enlarged from
[$E(a)$] and [$S(a)$], respectively
(reproduction of Fig.\ 5 in Okitsu {\it et al.} (2012)\cite{okitsu2012}).
}
\label{Fig07_Exp_Simu_3beam}
\end{figure}

% Figs.\ \ref{Fig07_Exp_Simu_3beam}\ [$E(a)$],
% \ref{Fig07_Exp_Simu_3beam}\ [$S(a)$]は，
% $0\ 0\ 0$前方回折波，
% $0\ 4\ 4$と$\overline{4}\ 0\ 4$反射波の
% それぞれ，実験と計算によるピンホールトポグラフ図形である
% \cite{okitsu2012}。
Figs.\ \ref{Fig07_Exp_Simu_3beam}\ [$E(a)$] and
\ref{Fig07_Exp_Simu_3beam}\ [$S(a)$] are the
experimentally obtained and computer-simulated images
of $0\ 0\ 0$-forward-diffracted,
$0\ 4\ 4$-reflected and
$\overline{4}\ 0\ 4$-reflected X-ray topographs
\cite{okitsu2012}.
% Figs.\ \ref{Fig07_Exp_Simu_3beam}\ [$E(b)$],
% \ref{Fig07_Exp_Simu_3beam}\ [$S(b)$]は，
% $0\ 4\ 4$反射波を，それぞれ，
% Figs.\ \ref{Fig07_Exp_Simu_3beam}\ [$E(a)$],
% \ref{Fig07_Exp_Simu_3beam}\ [$S(a)$]から
% 拡大して表示したものである。
Figs.\ \ref{Fig07_Exp_Simu_3beam}\ [$E(b)$] and
\ref{Fig07_Exp_Simu_3beam}\ [$S(b)$]
are enlargements of $0\ 4\ 4$-reflected X-ray images of
Figs.\ \ref{Fig07_Exp_Simu_3beam}\ [$E(a)$] and
\ref{Fig07_Exp_Simu_3beam}\ [$S(a)$]
% Fig.\ \ref{Fig07_Exp_Simu_3beam}\ [$S(b)$]に矢印で示した
% 細かいフリンジ[Fine Fringe Region ($FFR(1)$)]と($FFR(2)$)，
% Y字型のパターン[`Y-shaped' Bright Region $(YBR)$]が，
% Fig.\ \ref{Fig07_Exp_Simu_3beam}\ [$E(b)$]にも見られ，
% シミュレーションと実験はよく一致している。
Fine fringe regions ($FFR(1)$), ($FFR(2)$) and
Y-shaped bright region $(YBR)$ indicated by arrows
in Fig.\ \ref{Fig07_Exp_Simu_3beam}\ [$S(b)$]
are found also in
Fig.\ \ref{Fig07_Exp_Simu_3beam}\ [$E(b)$],
which shows the excellent agreement between
the computer-simulated and experimentally obtained results.

% \subsection{4波ケース}

\subsection{Four-beam case}

\begin{figure}[!t]
% \centering
\begin{center}
\includegraphics[width=0.51\textwidth]{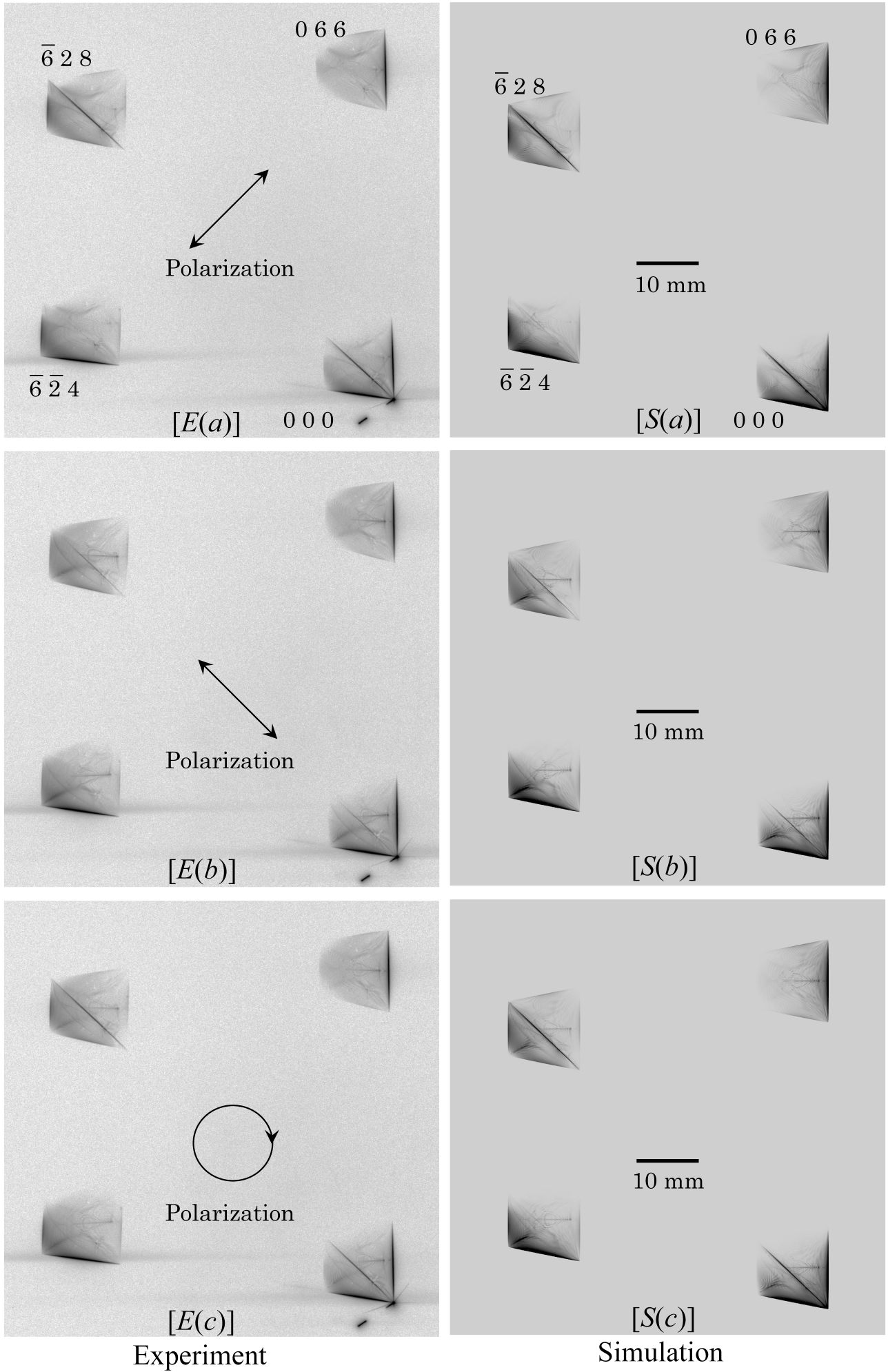}
\end{center}
\captionsetup{width=0.85\textwidth}
\caption[
$E(x)$ and $S(x)$ ($x \in \{ a, b, c \}$) are experimentally obtained
and computer-simulated four-beam X-ray pinhole topographs
with an incidence of $+45^{\circ}$-inclined-linearly,
$-45^{\circ}$-inclined-linearly and
right-screwed-circularly
polarized X-rays whose photon energy
was 18.245 keV
(reproduction of Fig.\ 6 in Okitsu {\it et al.} (2012)).
        ]
        {
[$E(x)$] and [$S(x)$] ($x \in \{ a, b, c \}$) are experimentally obtained
and computer-simulated four-beam X-ray pinhole topographs
with an incidence of $+45^{\circ}$-inclined-linearly,
$-45^{\circ}$-inclined-linearly and
right-screwed-circularly
polarized X-rays whose photon energy
was 18.245 keV
(reproduction of Fig.\ 6 in Okitsu {\it et al.} (2012)\cite{okitsu2012}).
}
\label{Fig08_FourBeam}
\end{figure}

% Figs.\ \ref{Fig08_FourBeam}\ [$E(x)$], \ref{Fig08_FourBeam}\ [$S(x)$]
% ($x \in \{ a,b,c \}$)は，実験および計算で得られた
% $0\ 0\ 0$前方回折波，
% $\overline{6}\ \overline{2}\ 4$,
% $\overline{6}\ 2\ 8$,
% $0\ 6\ 6$反射波の
% トポグラフ図形である
% \cite{okitsu2012}。
Figs.\ \ref{Fig08_FourBeam}\ [$E(x)$], \ref{Fig08_FourBeam}\ [$S(x)$]
($x \in \{ a,b,c \}$)
are experimentally obtained and computer-simulated images
of $0\ 0\ 0$-forward diffracted,
$\overline{6}\ \overline{2}\ 4$-,
$\overline{6}\ 2\ 8$- and
$\overline{6}\ 2\ 8$-reflected X-rays.
% $(a)$, $(b)$, $(c)$は，移相子システムにより
% コントロールされた入射X線の偏光状態が
% 異なっており，
% 下流から見て，水平から$+45^{\circ}$傾いた直線偏光，
% $-45^{\circ}$傾いた直線偏光，
% 右ネジ円偏光である。
$(a)$, $(b)$ and $(c)$
are different from each other
in polarization state of the incident X-rays.
These were obtained
for $(a)$: $+45^{\circ}$-inclined linear polarization,
$(b)$: $-45^{\circ}$-inclined linear polarization and
$(c)$: right-screwed circular polarization
when viewed from the downstream direction.

\begin{figure}[!t]
% \centering
\begin{center}
\includegraphics[width=0.51\textwidth]{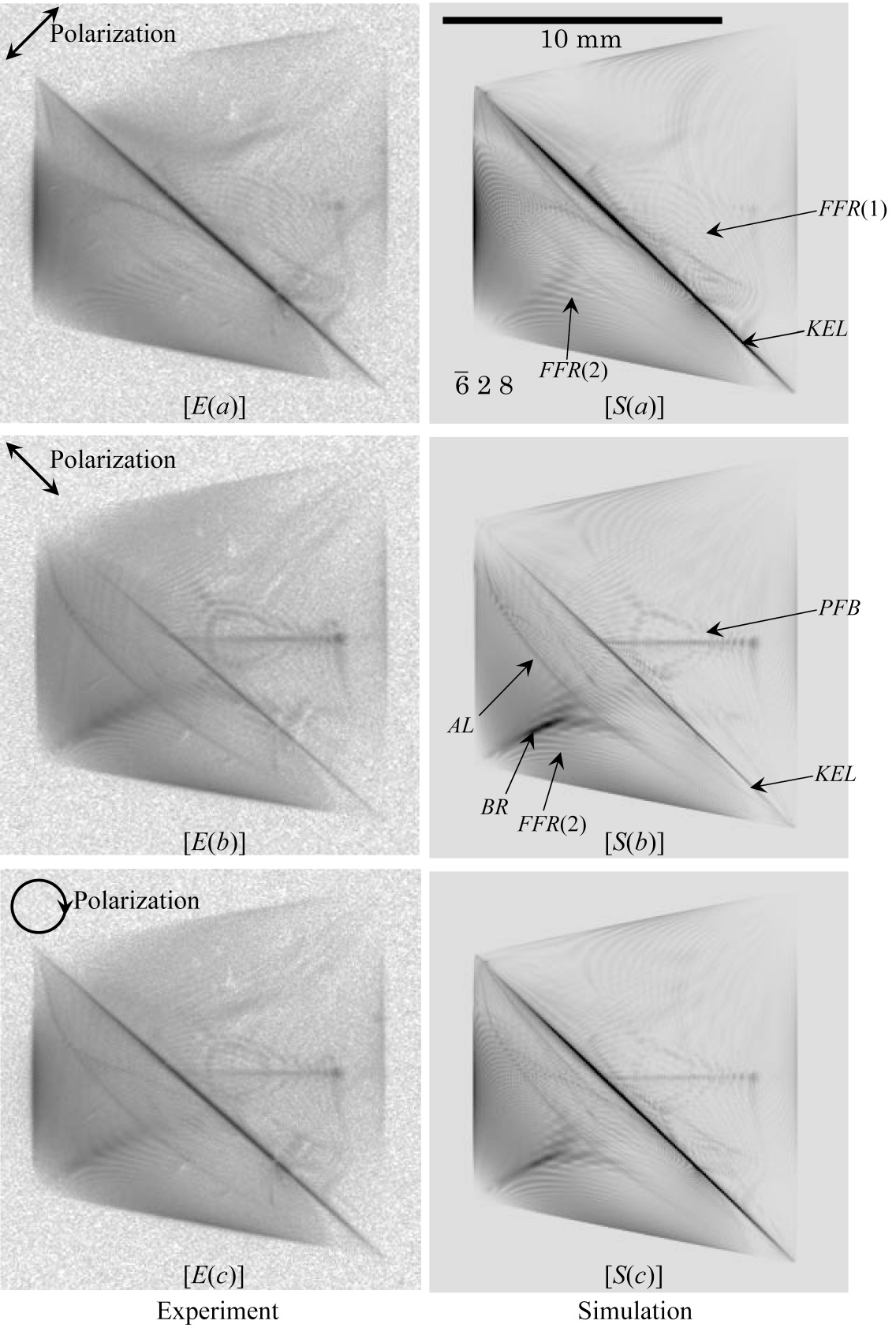}
\end{center}
\captionsetup{width=0.85\textwidth}
\caption[
$E(x)$ and $S(x)$ ($x \in \{ a, b, c \}$) are enlargements of
$\overline{6} \ 2 \ 8$ reflected X-ray images in
Fig.\ \ref{Fig08_FourBeam} $E(x)$ and $S(x)$
(reproduction of Fig.\ 7 in Okitsu {\it et al.} (2012))
        ]
        {
[$E(x)$] and [$S(x)$] ($x \in \{ a, b, c \}$) are enlargements of
$\overline{6} \ 2 \ 8$ reflected X-ray images in
Fig.\ \ref{Fig08_FourBeam} [$E(x)$] and [$S(x)$]
(reproduction of Fig.\ 7 in Okitsu {\it et al.} (2012))
\cite{okitsu2012}].
}
\label{Fig09_FourBeamEnlarged}
\end{figure}

% Figs.\ \ref{Fig09_FourBeamEnlarged}\ [$E(x)$], \ref{Fig09_FourBeamEnlarged}\ [% $S(x)$]
% ($x \in \{ a,b,c \}$)は，それぞれ
% Figs.\ \ref{Fig08_FourBeam}\ [$E(x)$], \ref{Fig08_FourBeam}\ [$S(x)$]
% の$\overline{6}\ 2\ 8$反射波を拡大したものである
% \cite{okitsu2012}。
Figs.\ \ref{Fig09_FourBeamEnlarged}\ [$E(x)$] and
\ref{Fig09_FourBeamEnlarged}\ [$S(x)$]
($x \in \{ a,b,c \}$) are
enlargement of $\overline{6}\ 2\ 8$-reflected X-ray images
in Figs.\ \ref{Fig08_FourBeam}\ [$E(x)$]
and \ref{Fig08_FourBeam}\ [$S(x)$]
\cite{okitsu2012}.
% 細かいフリンジ[Fine Fringe Region ($FFR(1)$)]
% は，Figs.\ \ref{Fig09_FourBeamEnlarged}\ [$E(a)$],
% \ref{Fig09_FourBeamEnlarged}\ [$S(a)$]
% のいずれにも観察される。
Fine Fringe Region ($FFR(1)$) can be found both in
Figs.\ \ref{Fig09_FourBeamEnlarged}\ [$E(a)$] and
\ref{Fig09_FourBeamEnlarged}\ [$S(a)$].
% 細かいフリンジ[Fine Fringe Region ($FFR(2)$)]
% は，Figs.\ \ref{Fig09_FourBeamEnlarged}\ [$E(x)$],
% \ref{Fig09_FourBeamEnlarged}\ [$S(x)$]
% ($x \in \{ a, b, c \}$)のいずれにも見られる。
Fine Fringe Region ($FFR(2)$)
can be found both in
Figs.\ \ref{Fig09_FourBeamEnlarged}\ [$E(x)$] and
\ref{Fig09_FourBeamEnlarged}\ [$S(x)$]
($x \in \{ a, b, c \}$).
% ナイフエッジのような
% 鋭い線[Knife Edge Line $(KEL)$]はすべての図に見られるが，
Sharp lines [Knife Edge Line $(KEL)$]
are found in all figures.
% Figs.\ \ref{Fig09_FourBeamEnlarged}\ [$E(a)$],
% \ref{Fig09_FourBeamEnlarged}\ [$S(a)$]においてもっとも濃く，
% Figs.\ \ref{Fig09_FourBeamEnlarged}\ [$E(b)$],
% \ref{Fig09_FourBeamEnlarged}\ [$S(b)$]においてもっとも薄く，
% Figs.\ \ref{Fig09_FourBeamEnlarged}\ [$E(c)$],
% \ref{Fig09_FourBeamEnlarged}\ [$S(c)$]においては，その中間である。
These lines found in
Figs.\ \ref{Fig09_FourBeamEnlarged}\ [$E(a)$] and
\ref{Fig09_FourBeamEnlarged}\ [$S(a)$]
are dark
in comparison with
Figs.\ \ref{Fig09_FourBeamEnlarged}\ [$E(b)$] and
\ref{Fig09_FourBeamEnlarged}\ [$S(b)$].
In the cases of
Figs.\ \ref{Fig09_FourBeamEnlarged}\ [$E(c)$] and
\ref{Fig09_FourBeamEnlarged}\ [$S(c)$]
intensities of these lines are intermediate between
the cases of $(a)$ and $(b)$.
% 魚の骨のような模様[Pattern like Fish Born ($PFB$)]，
% アーチ状のライン[Arched Line ($AL$)]，
% 明るい領域[Bright Region ($BR$)]は，
% Figs.\ \ref{Fig09_FourBeamEnlarged}\ [$E(a)$],
% \ref{Fig09_FourBeamEnlarged}\ [$S(a)$]には観察されず，
% その他の図形では観察される。
[Pattern like Fish Born ($PFB$)],
[Arched Line ($AL$)] and
[Bright Region ($BR$)]
are not found
in Figs.\ \ref{Fig09_FourBeamEnlarged}\ [$E(a)$] and
\ref{Fig09_FourBeamEnlarged}\ [$S(a)$].
However, they are found in the cases of
[$E(b)$], [$S(b)$], [$E(c)$] and [$S(c)$]
% 要は，計算機シミュレーションと実験によるトポグラフは，
% 入射X線の偏光状態が同じのとき非常によく一致し，
% 偏光状態に依存して大きく変化することがわかる。
It has been clarified that
the computer-simulated and experimentally obtained
pinhole topograph images coincide with each other
when the polarization state used in the experiment
or assumed in the computer simulation
agreed with each other.

% $(a)$, $(b)$, $(c)$に対応する入射X線は，縦偏光と横偏光の
% 振幅の絶対値の比率は変わらないが，位相差に違いがある。
% この位相差が，$(a)$, $(b)$, $(c)$のトポグラフ図形に
% 大きな差異をもたらしている。
Intensity ratio between the horizontally and vertically polarized
X-rays is the same for
$(a)$, $(b)$ and $(c)$.
However, there are difference in phase
between the amplitudes of horizontally and vertically polarized
X-rays.
This difference in phase caused to
the distinct difference in the topograph images.

\begin{figure}[!t]
% \centering
\begin{center}
\includegraphics[width=0.51\textwidth]{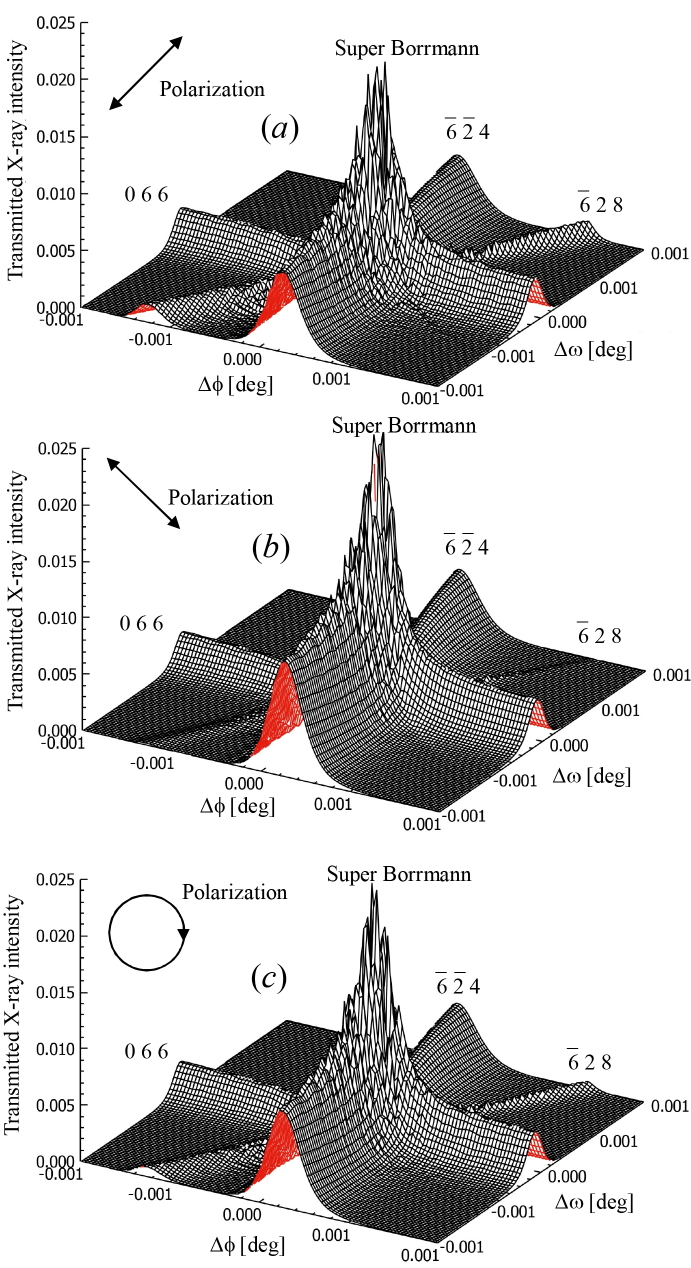}
\end{center}
\captionsetup{width=0.85\textwidth}
\caption{
Transmittance of X-rays
around the condition that
$\overline{6}\ \overline{2}\ 4$-, $\overline{6}\ 2\ 8$- and
$0\ 6\ 6$-reflected X-rays all in Laue geometries
are simultaneously strong.
$\Delta \omega$ and $\Delta \phi$ are angular deviations
around $[\overline{2}\ \overline{1}\ 1]$ and $[0\ 1\ 1]$ axes
from the exact four-beam condition.
}
\label{Fig10_4Beam_RockingCurve_deg_000_Rotate000}
\end{figure}

% Figs.\ \ref{Fig10_4Beam_RockingCurve_deg_000_Rotate000}\ $(a)$,
% \ref{Fig10_4Beam_RockingCurve_deg_000_Rotate000}\ $(b)$,
% \ref{Fig10_4Beam_RockingCurve_deg_000_Rotate000}\ $(c)$は，ビーム下流方向から% 見て，
% それぞれの図の左上に示したように，水平からの傾きが，
% $+45^{\circ}$直線偏光，$-45^{\circ}$直線偏光，右ネジ円偏光の入射X線を仮定して% ，
% E-L理論により計算した前方回折波の強度曲線である。
Figs.\ \ref{Fig10_4Beam_RockingCurve_deg_000_Rotate000}\ $(a)$,
\ref{Fig10_4Beam_RockingCurve_deg_000_Rotate000}\ $(b)$ and
\ref{Fig10_4Beam_RockingCurve_deg_000_Rotate000}\ $(c)$
are rocking curves of the forward-diffracted X-ray intensity
calculated based on the E-L theory
under the assumptions of different polarization states
of the incident X-rays:,
$(a)$ $+45^{\circ}$ and
$(b)$ $-45^{\circ}$-inclined linear polarization and
$(c)$ right-screwed circular polarization.

% $\Delta \omega$は，$[\overline{2}\ \overline{1}\ 1]$
% 軸周りの，
% $\Delta \phi$は，$[0\ 1\ 1]$軸周りの，
% 厳密な4波条件からのズレ角である。
$\Delta \omega$ and $\Delta \phi$
are angular deviations
around axes of $[\overline{2}\ \overline{1}\ 1]$ and
$[0\ 1\ 1]$ directions, respectively,
from the exact four-beam condition.
% 「$\overline{6}\ \overline{2}\ 4$」，「$\overline{6}\ 2\ 8$」，「$0\ 6\ 6$」
% と記入した部分に，X線透過強度の盛り上がりが見られるが，
% これらは，それぞれの反射指数がブラッグ条件を満たしたことによる，
% ボルマン効果(異常透過)
% \cite{borrmann1950}
% によるものである。
Enhancement of the X-ray intensities are found at regions
indicated to be `$\overline{6}\ \overline{2}\ 4$',
`$\overline{6}\ 2\ 8$' and `$0\ 6\ 6$'.
These were caused by the Borrmann effect
(amorous transmission)
\cite{borrmann1950}.
% それぞれの盛り上がりが交差し，4波条件を満たしたところでは，
% 「Super Borrmann」と記入した，スーパーボルマン効果
% \cite{borrmann1965}
% が見られる。
Where these enhanced regions cross with each other to satisfy
the four-beam condition,
further enhancement of the forward-diffracted intensities
owing to super Borrmann effect
\cite{borrmann1965}.
% 「$\overline{6}\ 2\ 8$」の盛り上がりは，
% Fig.\ \ref{Fig10_4Beam_RockingCurve_deg_000_Rotate000}\ $(a)$
% と比較して
% Fig.\ \ref{Fig10_4Beam_RockingCurve_deg_000_Rotate000}\ $(b)$
% では小さくなっており，
The enhancement of `$\overline{6}\ 2\ 8$' found in
Fig.\ \ref{Fig10_4Beam_RockingCurve_deg_000_Rotate000}\ $(a)$
is relatively small in comparison with
that found in
Fig.\ \ref{Fig10_4Beam_RockingCurve_deg_000_Rotate000}\ $(b)$.
% Fig.\ \ref{Fig10_4Beam_RockingCurve_deg_000_Rotate000}\ $(c)$
% は，両者の中間くらいになっている。
In Fig.\ \ref{Fig10_4Beam_RockingCurve_deg_000_Rotate000}\ $(c)$,
the situation is intermediate between them.
% 18.245 keVのX線に対する$\overline{6}\ 2\ 8$反射のブラッグ角は，
% およそ39.64$^{\circ}$であり，
The Bragg angle of $\overline{6}\ 2\ 8$ reflection
for X-rays of 18.245 keV is 39.64$^{\circ}$.
% $\pi$偏光に対する偏光因子を計算すると，
% 0.186程度の小さな値となる。
Then, the polarization factor for $\pi$ polarization
[$\cos(2 \times 39.64^{\circ})$]
is calculated to be 0.186 that is a relatively small value
in comparison with that for $\sigma$ polarization.
% Fig.\ \ref{Fig10_4Beam_RockingCurve_deg_000_Rotate000}\ $(b)$
% の$-45^{\circ}$直線偏光は，
% $\overline{6}\ 2\ 8$反射に対してはほぼ$\pi$偏光となり，
% 「$\overline{6}\ 2\ 8$」の盛り上がりが小さい理由が説明できる。
Then, $-45^{\circ}$-inclined linear polarization for
$\overline{6}\ 2\ 8$ reflection
is almost $\pi$ polarization, from which
the relatively small enhancement of $\overline{6}\ 2\ 8$
can be explained.
% また，Figs.\ \ref{Fig09_FourBeamEnlarged}\ $[S(a)]$,
% \ref{Fig09_FourBeamEnlarged}\ $[S(b)]$に記入した
% 「$KEL$」のX線強度が，入射X線の偏光状態に依存する
% 理由も同様である。
The intensity of $KEL$ depends on the polarization state
of the incident X-rays for the same reason.

\begin{figure}[!t]
% \centering
\begin{center}
\includegraphics[width=0.51\textwidth]{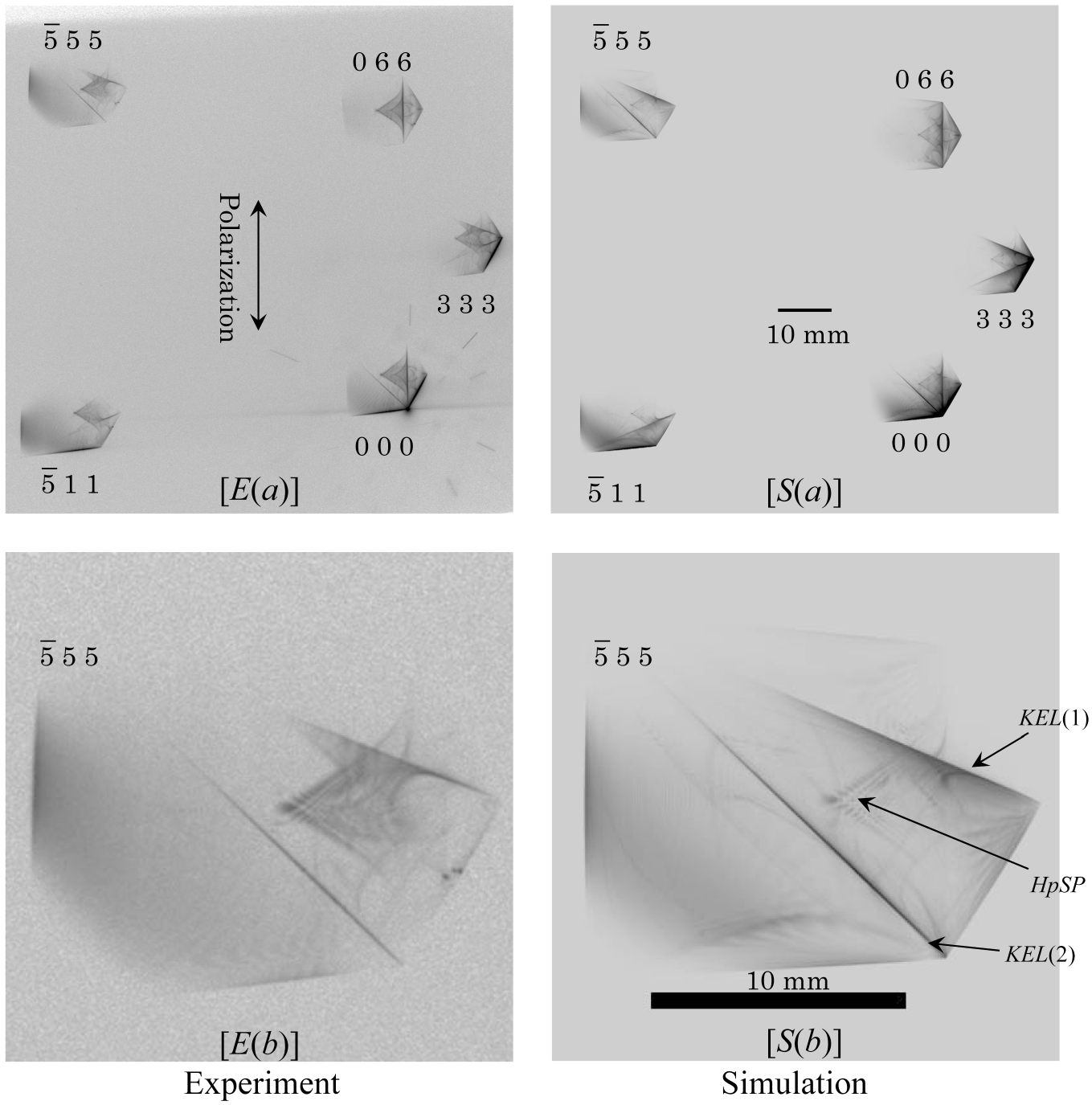}
\end{center}
\captionsetup{width=0.85\textwidth}
\caption[
$E(a)$ and $S(a)$ are experimentally obtained
and computer-simulated five-beam X-ray pinhole topographs
with an incidence of vertical-linearly polarized X-rays whose photon energy
was 18.245 keV.
$E(b)$ and $S(b)$
are $\overline{5} \ 5 \ 5$ reflected X-ray images
enlarged from $E(a)$ and $S(a)$
(reproduction of Fig.\ 8 in Okitsu {\it et al.} (2012)).
        ]
        {
[$E(a)$] and [$S(a)$] are experimentally obtained
and computer-simulated five-beam X-ray pinhole topographs
with an incidence of vertical-linearly polarized X-rays whose photon energy
was 18.245 keV.
[$E(b)$] and [$S(b)$]
are $\overline{5} \ 5 \ 5$ reflected X-ray images
enlarged from [$E(a)$] and [$S(a)$]
(reproduction of Fig.\ 8 in Okitsu {\it et al.} (2012)\cite{okitsu2012}).
}
\label{Fig11_FiveBeam}
\end{figure}

% \subsection{5波ケース}

\subsection{Five-beam case}

% 筆者らの2006年の論文
% \cite{okitsu2006}
% Fig.\ 1に示したように，立方晶の場合，
% 5個の逆格子点がひとつの円周上に存在する場合がある。
There are cases in which
five reciprocal lattice nodes simultaneously exist on a circle
in the reciprocal space
as shown in Fig.\ 1 of the paper published by the
present author and his coauthors
\cite{okitsu2006}.
% Figs.\ \ref{Fig11_FiveBeam}\ [$E(a)$],
% \ref{Fig11_FiveBeam}\ [$S(a)$]
% は，実験と計算機シミュレーションによる，
% 5波ピンホールトポグラフである
% \cite{okitsu2012}。
Fig.\ \ref{Fig11_FiveBeam}\ [$E(a)$] and
\ref{Fig11_FiveBeam}\ [$S(a)$] are
experimentally obtained and computer-simulated
five-beam pinhole topographs.
% 入射X線の偏光状態は，移相子システムで
% 横偏光を縦偏光に変換した。
The vertical-linearly polarized X-rays
that were converted from the horizontal-linearly polarized
synchrotron X-rays.
% Figs.\ \ref{Fig11_FiveBeam}\ [$E(b)$],
% \ref{Fig11_FiveBeam}\ [$S(b)$]は，
% Figs.\ \ref{Fig11_FiveBeam}\ [$E(a)$],
% \ref{Fig11_FiveBeam}\ [$S(a)$]
% の$\overline{5}\ 5\ 5$反射波のイメージを拡大したものである
% \cite{okitsu2012}。
Figs.\ \ref{Fig11_FiveBeam}\ [$E(b)$] and
\ref{Fig11_FiveBeam}\ [$S(b)$] are enlargements of
$\overline{5}\ 5\ 5$-reflected images
\cite{okitsu2012}.
% ナイフエッジのような鋭い線[Knife Edge Line ($KEL(1)$と$KEL(2)$)]，
% 竪琴のようなパターン[Harp-Shaped Patten $(HpSP)$]が
% 実験と計算機シミュレーションの両方に見られる。
[Knife Edge Line ($KEL(1)$, $KEL(2)$)] and
[Harp-Shaped Patten $(HpSP)$] can be found
both in the experimentally obtained and computer-simulated images.

% Figs.\ \ref{Fig11_FiveBeam}\ [$E(b)$],
% \ref{Fig11_FiveBeam}\ [$S(b)$]を見ると，$KEL(1)$と$KEL(2)$の方向は，
% $\overline{5}\ 5\ 5$反射波と$0\ 0\ 0$前方回折波のトポグラフ像を
% 結ぶ方向，および$\overline{5}\ 5\ 5$反射波と$3\ 3\ 3$反射波のトポグラフ像を
% 結ぶ方向に平行である。
The directions of $KEL(1)$ and $KEL(2)$ found in
Figs.\ \ref{Fig11_FiveBeam}\ [$E(b)$] and \ref{Fig11_FiveBeam}\ [$S(b)$]
are parallel to the direction to tie
the topograph images of $\overline{5}\ 5\ 5$- and $3\ 3\ 3$-reflected X-rays.
% このことは，結晶中で
% $\overline{5}\ 5\ 5$反射波と$0\ 0\ 0$前方回折波の間，および
% $\overline{5}\ 5\ 5$反射波と$3\ 3\ 3$反射波の間に，
% エネルギーのやりとりがあることをうかがわせる。
This suggests the energy exchange between
$\overline{5}\ 5\ 5$-reflected and
$0\ 0\ 0$-forward diffracted X-rays and between
$\overline{5}\ 5\ 5$- and
$3\ 3\ 3$-reflected X-rays.
% 同様な鋭い線[Knife Edge Line $(KEL)$]は，
% 3, 4, 6, 8波ケースのピンホールトポグラフにおいても
% 見られる。
Similar [Knife Edge Line $(KEL)$] can be found
also in three-, four-, six- and eight-beam topographs.

% \subsection{6波ケース}

\subsection{Six-beam case}

% 筆者らが，
% 2003年
% \cite{okitsu2003b}，
% 2006年
% \cite{okitsu2006}，
% 2011年
% \cite{okitsu2011a}
% に報告した
% 6波ケースにおいては，
% トポグラフ図形は正6角形であったが，
% この節で記述する6波ケースは，
% トポグラフ図形が正6角形ではない。

In the six-beam case reported by
the present author and his coauthors
\cite{okitsu2003b,okitsu2006,okitsu2011a},
the topograph images were regular hexagons.
However, in the six-beam case
described in the present section,
the topograph images are not regular hexagons.

\begin{figure}[!t]
% \centering
\begin{center}
\includegraphics[width=0.51\textwidth]{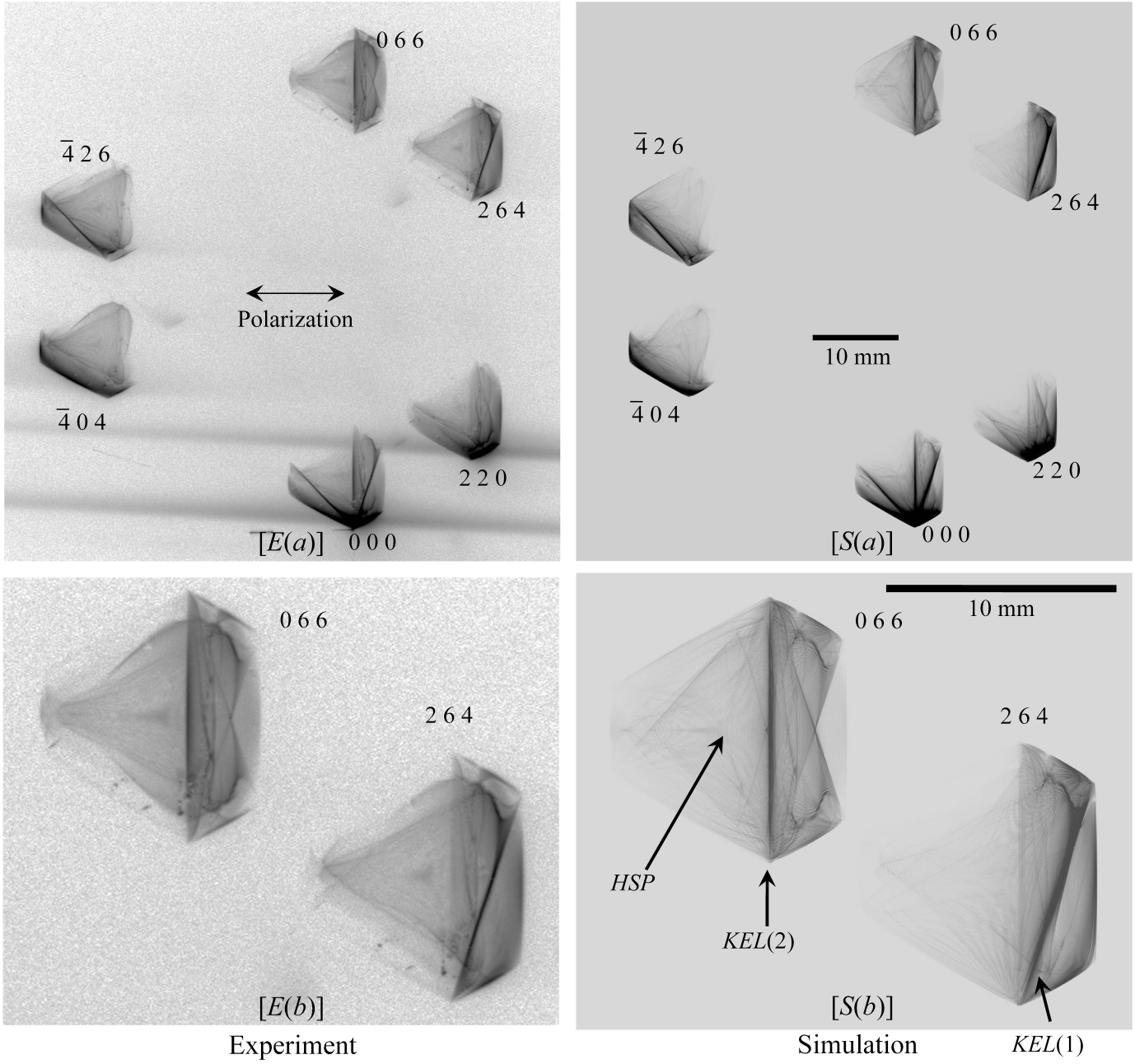}
\end{center}
\captionsetup{width=0.85\textwidth}
\caption[
$E(a)$ and $S(a)$ are experimentally obtained
and computer-simulated six-beam X-ray pinhole topographs
with an incidence of horizontal-linearly polarized X-rays with a photon energy
of 18.245 keV.
$E(b)$ and $S(b)$ are $0 \ 6 \ 6$ and $2 \ 6 \ 4$ reflected X-ray images
enlarged from $E(a)$ and $S(a)$
(reproduction of Fig.\ 9 in Okitsu {\it et al.} (2012))
        ]
        {
[$E(a)$] and [$S(a)$] are experimentally obtained
and computer-simulated six-beam X-ray pinhole topographs
with an incidence of horizontal-linearly polarized X-rays with a photon energy
of 18.245 keV.
[$E(b)$] and [$S(b)$] are $0 \ 6 \ 6$ and $2 \ 6 \ 4$ reflected X-ray images
enlarged from [$E(a)$] and [$S(a)$]
(reproduction of Fig.\ 9 in Okitsu {\it et al.} (2012))
\cite{okitsu2012}].
}
\label{Fig12_Omusubi06_Beam}
\end{figure}

% Fig.\ \ref{Fig12_Omusubi06_Beam}は，水平偏光入射による実験と，
% これを仮定して行った計算機シミュレーションの結果である
% \cite{okitsu2012}。
Fig.\ \ref{Fig12_Omusubi06_Beam} shows
experimentally obtained and computer-simulated topographs
with the incidence of horizontally polarized X-rays
used in the experiment and assumed in the computer simulation
\cite{okitsu2012}.
% Figs.\ \ref{Fig12_Omusubi06_Beam}\ [$E(a)$],
% \ref{Fig12_Omusubi06_Beam}\ [$S(a)$]の
% $0\ 6\ 6$と$2\ 6\ 4$反射波を拡大したのが，
% Figs.\ \ref{Fig12_Omusubi06_Beam}\ [$E(b)$],
% \ref{Fig12_Omusubi06_Beam}\ [$S(b)$]である。
Figs.\ Figs.\ \ref{Fig12_Omusubi06_Beam}\ [$E(b)$] and
\ref{Fig12_Omusubi06_Beam}\ [$S(b)$] are
enlargements of topographs of $0\ 6\ 6$- and $2\ 6\ 4$-reflected X-rays
in Figs.\ \ref{Fig12_Omusubi06_Beam}\ [$E(a)$] and
\ref{Fig12_Omusubi06_Beam}\ [$S(a)$].
% ナイフエッジのような線[Knife Edge Line ($KEL(1)$), ($KEL(2)$)]および
% ハート形のパターン[Hart-Shaped Pattern $(HSP)$]が，
% 実験と計算機シミュレーションの両方に見られる。
[Knife Edge Line ($KEL(1)$) and ($KEL(2)$)] and
[Hart-Shaped Pattern $(HSP)$]
can be found both in images
experimentally obtained and computer-simulated.

% トポグラフ図形が正6角形の6波ケース
% \cite{okitsu2003b,okitsu2006,okitsu2011a}
% の場合には，
% 円錐状のエネルギーフローがあることを示唆する
% リング状のパターンが見られたが，
% Fig.\ \ref{Fig12_Omusubi06_Beam}のケースでは，
% これは観察されなかった。
In the cases of six-beam pinhole topographs
whose shapes are regular hexagons,
circular patterns that suggest the existence of
cone-shaped path of energy flow.
However, such circular patterns cannot be found
in Fig.\ \ref{Fig12_Omusubi06_Beam}.

% \subsection{8波ケース}

\subsection{Eight-beam case}

\begin{figure*}[!t]
% \centering
\begin{center}
\includegraphics[width=0.71\textwidth]{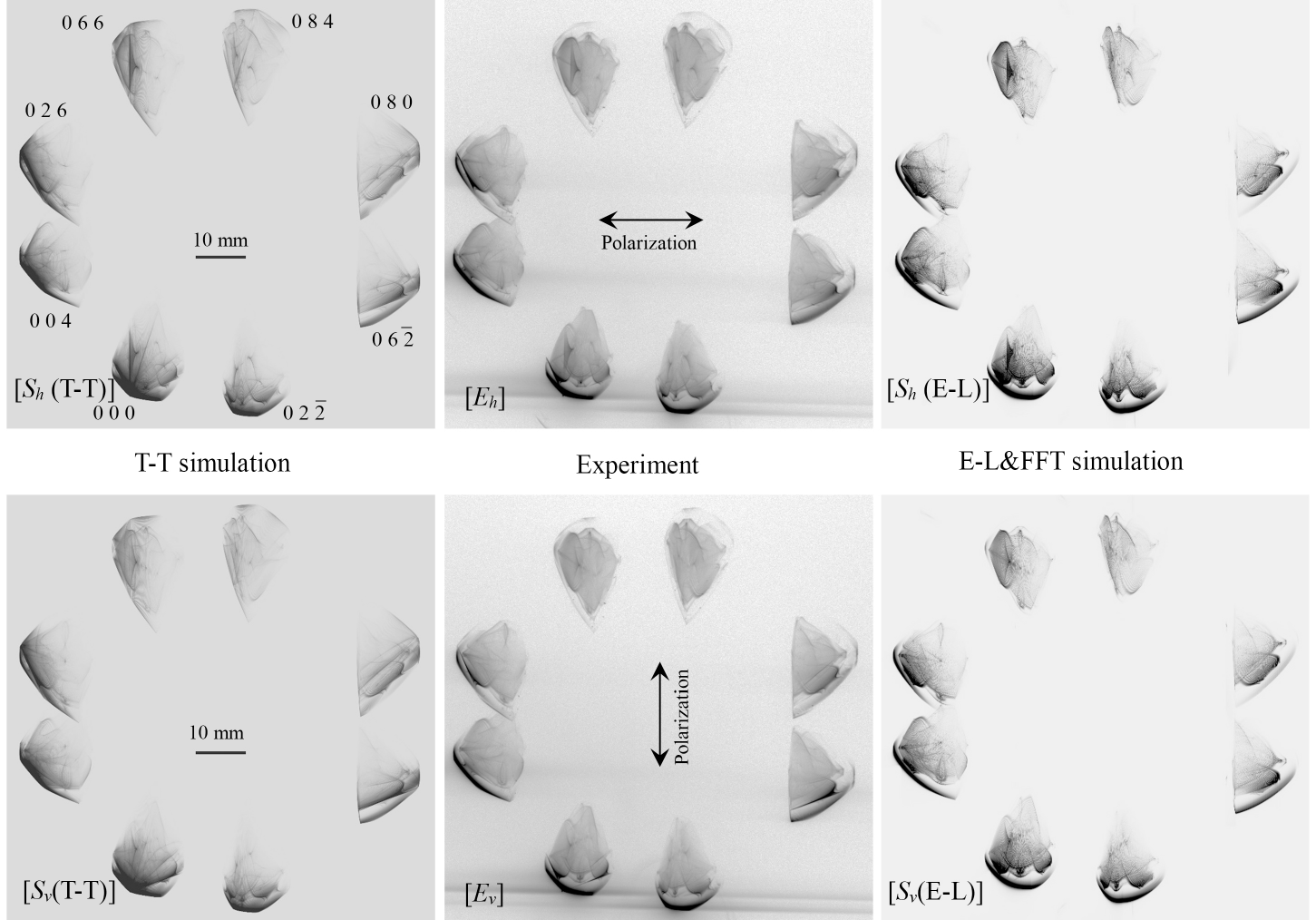}
\end{center}
\caption{
[$S_x$(T-T)], [$E_x$], and [$S_x$(E-L)]
$(x \in \{ h, v \})$ are the T-T simulated, experimentally obtained
and E-L\&FFT simulated
eight-beam pinhole topographs for horizontally ($x = h$) and
vertically ($x = v$) polarized incident X-rays
[reproduction of Fig.\ 5 in Okitsu {\it et al.} (2019)]
}
\label{Fig13_05_8Beam_T-TEq_Expririment_EL-FFT}
\end{figure*}

% Fig.\ \ref{Fig13_05_8Beam_T-TEq_Expririment_EL-FFT}は，
% $8$波ピンホールトポグラフで，
% 各々の図形の反射指数は，
% Fig.\ \ref{Fig13_05_8Beam_T-TEq_Expririment_EL-FFT}\ [$S_h$(T-T)]
% に示したとおりである
% \cite{okitsu2012,okitsu2019b}。
Fig.\ \ref{Fig13_05_8Beam_T-TEq_Expririment_EL-FFT} shows
eight-beam pinhole topographs whose reflection indices
are as shown in
Fig.\ \ref{Fig13_05_8Beam_T-TEq_Expririment_EL-FFT}\ [$S_h$(T-T)]
\cite{okitsu2012,okitsu2019b}.

% Fig.\ \ref{Fig13_05_8Beam_T-TEq_Expririment_EL-FFT}\ [$E_x$]
% $(x \in \{ h, v \})$は，
% 横偏光$(x=h)$と縦偏光$(x=v)$のX線を入射して得られた
% 実験による$8$波ピンホールトポグラフである
% \cite{okitsu2012,okitsu2019b}。
% 横偏光は，移相子システムを外したわけではなく，
% 奇数象限と偶数象限への反射を与える移相子がもたらす
% 位相シフト量の符号を，
% 逆転させることにより得られている。
Fig.\ \ref{Fig13_05_8Beam_T-TEq_Expririment_EL-FFT}\ [$E_x$]
$(x \in \{ h, v \})$
show the experimentally obtained pinhole topographs
obtained with the incidence of horizontally polarized $(x=h)$ and
vertically polarized $(x=v)$ X-rays
\cite{okitsu2012,okitsu2019b}.
The horizontally polarized X-rays were obtained
not by removing the phase-retarder crystals from the X-ray path
but by reversing the sign of phase shift given by
the diamond crystals that reflect the incident X-rays
in the directions of the odd-numbered quadrant and
the even-numbered quadrant directions.

\begin{figure*}[!t]
% \centering
\begin{center}
\includegraphics[width=0.71\textwidth]{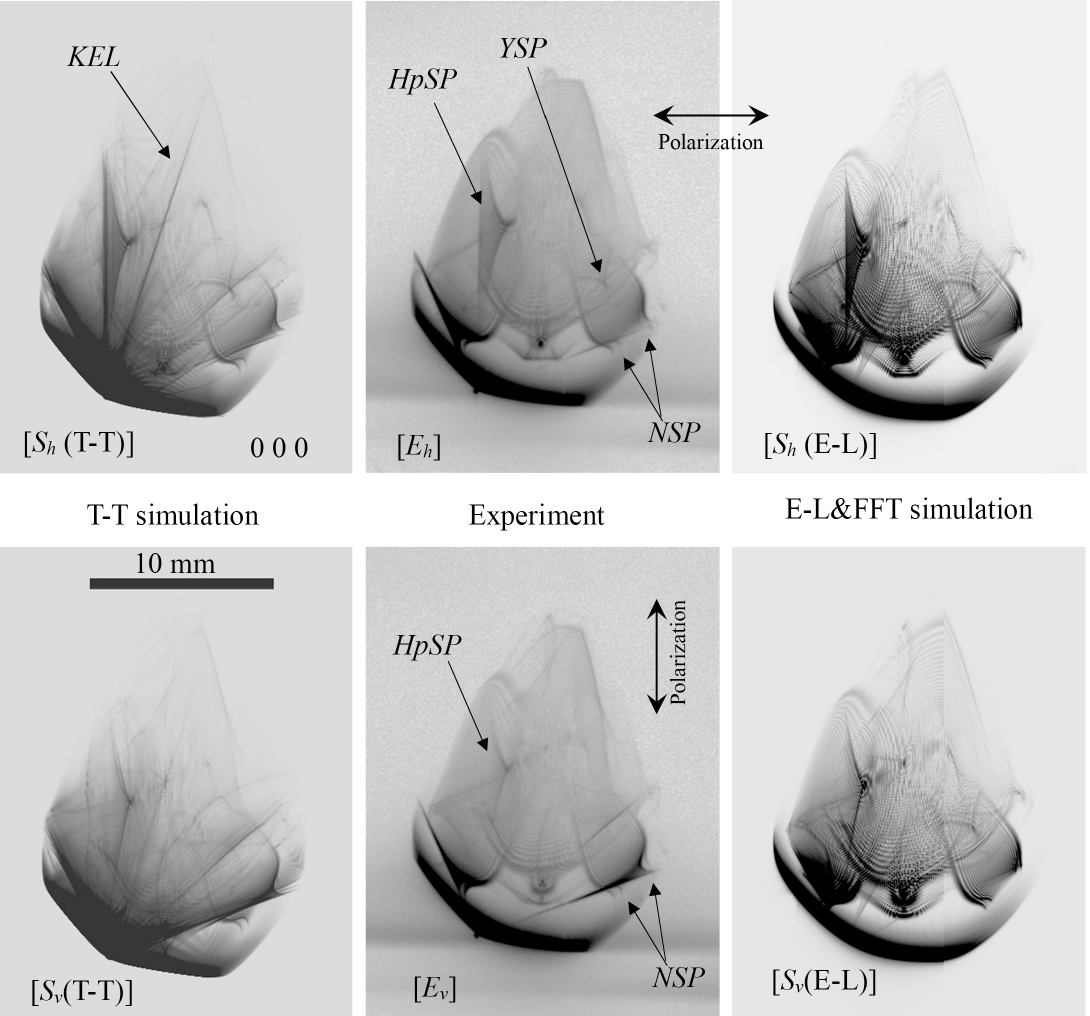}
\end{center}
\caption[
Enlargements of the $0\ 0\ 0$ forward-diffracted images
in Fig.\ \ref{Fig13_05_8Beam_T-TEq_Expririment_EL-FFT}
(reproduction of Fig.\ 6 in Okitsu {\it et al.} (2019))
        ]
        {
Enlargements of the $0\ 0\ 0$ forward-diffracted images
in Fig.\ \ref{Fig13_05_8Beam_T-TEq_Expririment_EL-FFT}
(reproduction of Fig.\ 6 in Okitsu {\it et al.} (2019))
\cite{okitsu2019b}].
}
\label{Fig14_06_8Beam_T-TEq_Expririment_EL-FFT_0+0+0}
\end{figure*}

\begin{figure*}[!t]
% \centering
\begin{center}
\includegraphics[width=0.65\textwidth]{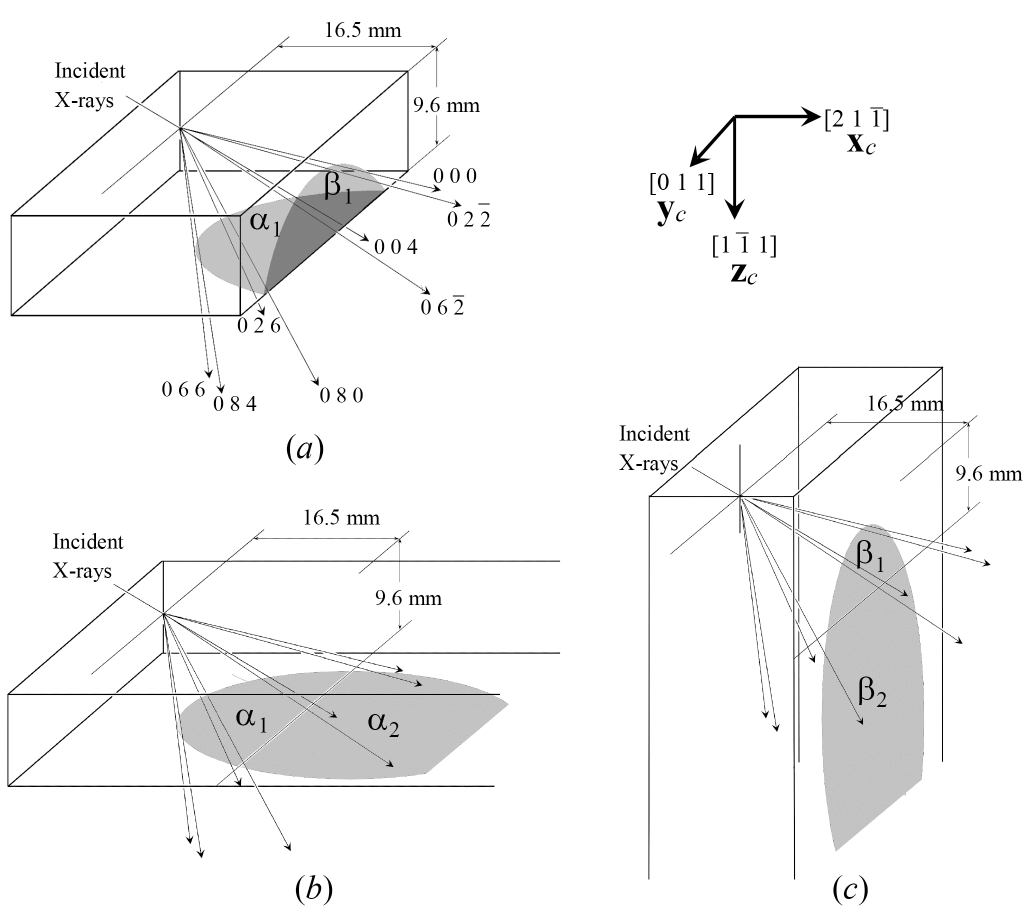}
\end{center}
\caption[
Geometry of the eight-beam pinhole topography.
${\mathbf x}_c$, ${\mathbf y}_c$, and ${\mathbf z}_c$
drawn on the upper right corner are
unit vectors in the directions of
2 1 $\overline{1}$, 0 1 1 and 1 $\overline{1}$ 1, respectively
        ]
        {
Geometry of the eight-beam pinhole topography.
${\mathbf x}_c$, ${\mathbf y}_c$, and ${\mathbf z}_c$
drawn on the upper right corner are
unit vectors in the directions 
$[2\ 1\ \overline{1}]$, $[0\ 1\ 1]$, and $[1\ \overline{1}\ 1]$, respectively
[reproduction of Fig.\ 1 in Okitsu {\it et al.} (2019)]
\cite{okitsu2019b}.
}
\label{Fig15_01_8BeamGeometry}
\end{figure*}

\begin{figure}[!t]
% \centering
\begin{center}
\includegraphics[width=0.27\textwidth]{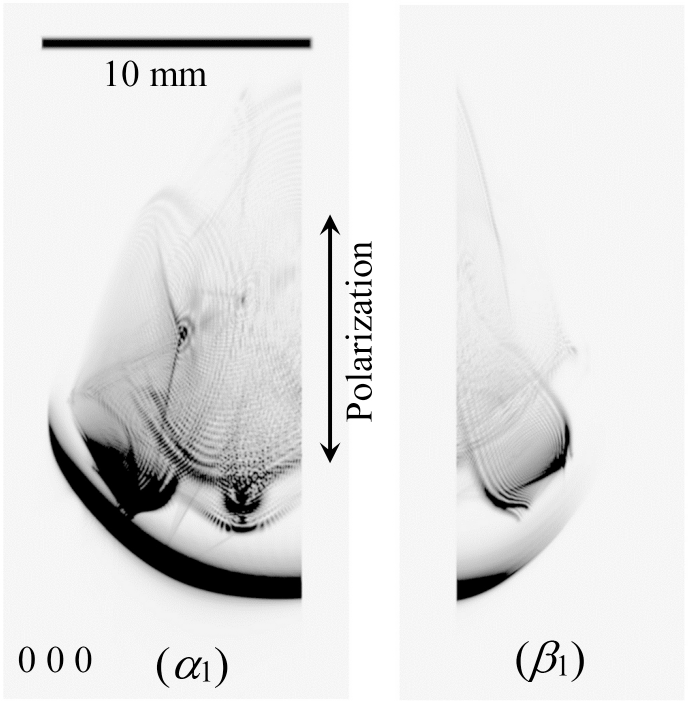}
\end{center}
\captionsetup{width=0.85\textwidth}
\caption[
($\alpha_1$) and ($\beta_1$)
are computed separately
under the assumption of vertically polarized incident X-rays.
These figures have been computed by projecting intensities of
the $0\ 0\ 0$ forward-diffracted X-rays
on the exit planes $\alpha_1$ and $\beta_1$ in Fig.\ \ref{Fig15_01_8BeamGeometry}\ $(a)$
on the imaging plate whose surface was normal to the $1\ 0\ 0$ direction.
X-ray intensities of $\alpha_2$ and $\beta_2$
in Figs.\ \ref{Fig15_01_8BeamGeometry}\ $(b)$ and \ref{Fig15_01_8BeamGeometry}\ $(c)$
have been erased
(reproduction of Fig.\ 8 in Okitsu {\it et al.} (2019)).
        ]
        {
($\alpha_1$) and ($\beta_1$)
are computed separately
under the assumption of vertically polarized incident X-rays.
These figures have been computed by projecting intensities of
the $0\ 0\ 0$ forward-diffracted X-rays
on the exit planes $\alpha_1$ and $\beta_1$ in Fig.\ \ref{Fig15_01_8BeamGeometry}\ $(a)$
on the imaging plate whose surface was normal to the [$1\ 0\ 0$] direction.
X-ray intensities of $\alpha_2$ and $\beta_2$
in Figs.\ \ref{Fig15_01_8BeamGeometry}\ $(b)$ and \ref{Fig15_01_8BeamGeometry}\ $(c)$
have been erased
(reproduction of Fig.\ 8 in Okitsu {\it et al.} (2019)\cite{okitsu2019b}).
}
\label{Fig16_08_8Beam_EL-FFT_0+0+0}
\end{figure}

% Figs.\ \ref{Fig13_05_8Beam_T-TEq_Expririment_EL-FFT}\ [$S_x$(T-T)],
% \ref{Fig13_05_8Beam_T-TEq_Expririment_EL-FFT}\ [$S_x$(E-L)]は，
% 横偏光$(x=h)$\ X線と，縦偏光$(x=v)$\ X線の入射を仮定したときの
% T-T方程式を解くか(T-T simulation)，
Figs.\ \ref{Fig13_05_8Beam_T-TEq_Expririment_EL-FFT}\ [$S_x$(T-T)] are
computer-simulated pinhole topographs for the incidence of
horizontally polarized $(x=h)$ and vertically polarized X-rays
obtained by solving
the $n$-beam T-T equation (T-T simulation).
% E-L理論の解を高速フーリエ変換して得られた(E-L\&FFT simulation)
% 計算機シミュレーションによるトポグラフ図形である。
However, Fig.\ \ref{Fig13_05_8Beam_T-TEq_Expririment_EL-FFT}\ [$S_x$(E-L)]
are obtained by fast Fourier-transforming the calculated X-ray amplitudes
based on the E-L theory (E-L\&FFT simulation).
% 実験を行った際，また計算機シミュレーションで仮定した，結晶の形状と
% X線ビームパスの関係は，Fig.\ \ref{Fig15_01_8BeamGeometry}\ $(a)$
% の通りである。
Fig.\ \ref{Fig15_01_8BeamGeometry}\ $(a)$ shows
the geometrical relation of the crystal shape and
the X-ray path.
% E-L\&FFTシミュレーションは，
% Figs.\ \ref{Fig15_01_8BeamGeometry}\ $(b)$,
% \ref{Fig15_01_8BeamGeometry}\ $(c)$の配置を仮定して行われ，
The E-L FFT simulation was performed under the assumption
of geometry as shown in
Figs.\ \ref{Fig15_01_8BeamGeometry}\ $(b)$ and
\ref{Fig15_01_8BeamGeometry}\ $(c)$ at first.
% それぞれ，$\alpha_2$と$\beta_2$の部分を取り除き，
% Fig.\ \ref{Fig16_08_8Beam_EL-FFT_0+0+0}の
% $(\alpha_1)$と$(\beta_1)$の部分を別々に計算した。
Then, after removing the parts of
$\alpha_2$ and $\beta_2$,
the parts of $(\alpha_1)$ and $(\beta_1)$
were calculated separately.
% Figs.\ \ref{Fig16_08_8Beam_EL-FFT_0+0+0}\ $(\alpha_1)$,
% \ref{Fig16_08_8Beam_EL-FFT_0+0+0}\ $(\beta_1)$を
% つなぎ合わせたのが，
% Fig.\ \ref{Fig14_06_8Beam_T-TEq_Expririment_EL-FFT_0+0+0}
% [$S_v$(E-L)]である。
Figs.\ \ref{Fig16_08_8Beam_EL-FFT_0+0+0}\ $(\alpha_1)$ and
\ref{Fig16_08_8Beam_EL-FFT_0+0+0}\ $(\beta_1)$
were linked to obtain
Fig.\ \ref{Fig14_06_8Beam_T-TEq_Expririment_EL-FFT_0+0+0}
[$S_v$(E-L)].
% Fig.\ \ref{Fig15_01_8BeamGeometry}\ $(b)$と
% Fig.\ Fig.\ \ref{Fig15_01_8BeamGeometry}\ $(b)$を
% 比較すると，入射側表面の法線ベクトルは，互いに直交する。
The downward surface normal of the crystal
in the cases of
Figs.\ \ref{Fig15_01_8BeamGeometry}\ $(b)$ and
\ref{Fig15_01_8BeamGeometry}\ $(c)$,
are mutually perpendicular to each other.
% Fig.\ \ref{Fig15_01_8BeamGeometry}\ $(a)$のような特殊な
% 形状の結晶に対しての，E-L\&FFTシミュレーションの手法を
% 検討する際，気づいたことであるが，
% 正確な$n$波条件からの入射ズレ角によらず，
% X線入射点で入射X線の位相がそろっていること，
% それと，出射側表面の方位とそこまでの距離だけ
% が重要なのである。
The important factor when considering the E-L\&FFT simulation
for such a complex geometry as shown in
Fig.\ \ref{Fig15_01_8BeamGeometry}\ $(a)$,
are that
the plane waves consisting of the incident X-rays should be in phase
at the incidence point on the crystal.
Further, the distances of the incidence point of X-rays
from the edge of the crystal
should be strictly measured i.e.
horizontally 16.5 mm and vertically 9.6 mm
[see Figs.\ \ref{Fig15_01_8BeamGeometry}\ $(a)$, $(b)$ and $(c)$].
% 入射側表面の角度や形状に，
% 高速フーリエ変換(FFT)により得られる解は依存しない。
% The solution obtained by using the FFT
% does not depend on the crystal shape.
% FFTを行うにあたっての詳細は，
% 筆者らの2019年の論文
% \cite{okitsu2019b}
% に記述してある。
Detail of the E-L\&FFT simulation
has been described in a paper published in 2019
\cite{okitsu2019b}.

% Figs.\ \ref{Fig14_06_8Beam_T-TEq_Expririment_EL-FFT_0+0+0}\ [$S_x$(T-T)],
% \ref{Fig14_06_8Beam_T-TEq_Expririment_EL-FFT_0+0+0}\ [$E_x$],
% \ref{Fig14_06_8Beam_T-TEq_Expririment_EL-FFT_0+0+0}\ [$S_x$(E-L)]
% ($x \in \{ h, v \}$)は，
% Figs.\ \ref{Fig13_05_8Beam_T-TEq_Expririment_EL-FFT}\ [$S_x$(T-T)],
% \ref{Fig13_05_8Beam_T-TEq_Expririment_EL-FFT}\ [$E_x$],
% \ref{Fig13_05_8Beam_T-TEq_Expririment_EL-FFT}\ [$S_x$(E-L)]
% $0\ 0\ 0$前方回折波のトポグラフを
% 拡大したものである。
Figs.\ \ref{Fig14_06_8Beam_T-TEq_Expririment_EL-FFT_0+0+0}\ [$S_x$(T-T)],
\ref{Fig14_06_8Beam_T-TEq_Expririment_EL-FFT_0+0+0}\ [$E_x$] and
\ref{Fig14_06_8Beam_T-TEq_Expririment_EL-FFT_0+0+0}\ [$S_x$(E-L)]
($x \in \{ h, v \}$)
are enlargement of topograph images of $0\ 0\ 0$-forward diffracted
X-rays in Figs.\ \ref{Fig13_05_8Beam_T-TEq_Expririment_EL-FFT}\ [$S_x$(T-T)],
\ref{Fig13_05_8Beam_T-TEq_Expririment_EL-FFT}\ [$E_x$] and
\ref{Fig13_05_8Beam_T-TEq_Expririment_EL-FFT}\ [$S_x$(E-L)], respectively.

% 実験による
% Fig.\ \ref{Fig14_06_8Beam_T-TEq_Expririment_EL-FFT_0+0+0}\ [$E_h$]には，
% 竪琴のような模様[Harp-Shaped Pattern ($HpSP$)],
% Y字形の模様[`Y-Shaped' Pattern ($YSP$)],
% 爪のような模様[Nail-Shaped Pattern ($NSP$)]が見られる。
% これらの模様は，
% Figs.\ \ref{Fig14_06_8Beam_T-TEq_Expririment_EL-FFT_0+0+0}\ [$S_h$(E-L)],
% \ref{Fig14_06_8Beam_T-TEq_Expririment_EL-FFT_0+0+0}\ [$S_h$(T-T)]
% にも見られる。
In Fig.\ \ref{Fig14_06_8Beam_T-TEq_Expririment_EL-FFT_0+0+0}\ [$E_h$]
that was experimentally obtained,
[Harp-Shaped Pattern ($HpSP$)],
[Nail-Shaped Pattern ($NSP$)] and
[Nail-Shaped Pattern ($NSP$)] are found.
These patterns are found also in
Figs.\ \ref{Fig14_06_8Beam_T-TEq_Expririment_EL-FFT_0+0+0}\ [$S_h$(E-L)],
and \ref{Fig14_06_8Beam_T-TEq_Expririment_EL-FFT_0+0+0}\ [$S_h$(T-T)].

% Fig.\ \ref{Fig14_06_8Beam_T-TEq_Expririment_EL-FFT_0+0+0}\ [$S_h$(T-T)]
% に見られるナイフエッジのような線
% [Knife Edge Line ($KEL$)]は，
% Figs.\ \ref{Fig14_06_8Beam_T-TEq_Expririment_EL-FFT_0+0+0}\ [$E_h$],
% [$S_h$(E-L)]には，見られない。
[Knife Edge Line ($KEL$)] as found in
Fig.\ \ref{Fig14_06_8Beam_T-TEq_Expririment_EL-FFT_0+0+0}\ [$S_h$(T-T)]
is not found in
Figs.\ \ref{Fig14_06_8Beam_T-TEq_Expririment_EL-FFT_0+0+0}\ [$E_h$] and
[$S_h$(E-L)].
% [$S_h$(T-T)]を計算するにあたっては，結晶のX線入射側表面の
% 1点でだけ，入射波がゼロでない振幅を持つという，
% デルタ関数の境界条件を与えている。
When calculating [$S_h$(T-T)],
Non-zero amplitude of incident X-rays only at
the incidence point of X-rays was given as the boundary condition of
Dirac's delta function.
% これは，無限大の角度広がりを持ったX線が入射している状況を
% 仮定していることを意味する。
This bonday condition means the assumption that
X-rays with infinite angular divergence
are incident on the crystal.
% $KEL$は，非常に細い線であり，これを実空間で合成するには，
% 逆空間において
% 波の進行方向から大きくズレた波の成分が必要である。
$KEL$ is a sharp line that needs plane wave components
whose directions of propagation are far different from the
$n$-beam condition.
% [$S_h$(T-T)]を計算する際には，これを満たす境界条件を与えている。
This bondar condition is given in the computer simulation
to obtain [$S_h$(T-T)].
% [$E_h$]の像を得る実験では，入射X線の角度広がりは有限である。
However, the angular divergence of the incident X-rays
practically used in the experiment
to obtain the image of [$E_h$] is finite.
% [$S_h$(E-L)]の計算では，高速フーリエ変換を
% 有限な角度範囲で打ち切っている。
When calculating [$S_h$(E-L)],
the amplitude of the incident X-rays whose angular deviation
from the exact $n$-beam condition is a finite range.
% このことが，[$E_h$]と[$S_h$(E-L)]に，$KEL$が見られない原因であると
% 考えられる。
This is considered to be the reason for that
$KEL$ is not observed in [$E_h$] and [$S_h$(E-L)].

% 縦偏光で得られた
% Fig.\ \ref{Fig14_06_8Beam_T-TEq_Expririment_EL-FFT_0+0+0}\ [$E_v$]と，
% 縦偏光入射を仮定して得られた
% Figs.\ \ref{Fig14_06_8Beam_T-TEq_Expririment_EL-FFT_0+0+0}\ [$S_v$(T-T)],
% \ref{Fig14_06_8Beam_T-TEq_Expririment_EL-FFT_0+0+0}\ [$S_v$(E-L)]
% においても$HpSP$が見られるが，模様は，
% Figs.\ \ref{Fig14_06_8Beam_T-TEq_Expririment_EL-FFT_0+0+0}\ [$S_h$(T-T)],
% \ref{Fig14_06_8Beam_T-TEq_Expririment_EL-FFT_0+0+0}\ [$E_h$],
% \ref{Fig14_06_8Beam_T-TEq_Expririment_EL-FFT_0+0+0}\ [$S_h$(E-L)]
% と比較して薄くなっている。
Also in Figs.\ \ref{Fig14_06_8Beam_T-TEq_Expririment_EL-FFT_0+0+0}\ [$E_v$],
\ref{Fig14_06_8Beam_T-TEq_Expririment_EL-FFT_0+0+0}\ [$S_v$(T-T)] and
and \ref{Fig14_06_8Beam_T-TEq_Expririment_EL-FFT_0+0+0}\ [$S_v$(E-L)],
vertically polarized incident X-rays were used in the experiment
or assumed in the simulation, $HpSP$ can be found.
However, the intensity is weak
in comparison with
[$S_h$(T-T)], [$E_h$], and [$S_h$(E-L)].

% この8波ケースの場合，
% E-L\&FFTシミュレーションに要した時間は
% 24コアの並列演算で，およそ8分，
% T-Tシミュレーションと比較して，100倍程度
% 高速だった。
% しかし，このことから，
% E-L\&FFTシミュレーションがT-Tシミュレーションと比較して
% 無条件に優れているとは，必ずしも言えない。
% E-L\&FFTシミュレーションの計算時間は，
% 結晶の厚さに依存せず，一定である。
% 一方，T-Tシミュレーションの所要時間は，
% $n$角錐の「ボルマンピラミッド」の中を3次元スキャンするため，
% 結晶の厚さの3乗に比例する。
% 結晶の厚さが$1/10$の1.0 mm程度になれば，
% T-Tシミュレーションは，
% E-L\&FFTシミュレーションより10倍速くなる。

In this eight-beam case,
the time for the E-L\&FFT simulation with
parallel calculation using 24 cores
was $\sim$8 minutes
(about 100 times as fast as the T-T simulation).
However, it cannot be concluded that
the E-L\&FFT is more excellent unconditionally
as compared with the T-T simulation.
The calculation time of the E-L\&FFT simulation
is constant not depending on the thickness of the crystal.
On the other hand,
that of the T-T simulation
is proportional to the third power of the crystal thickness
since the three-dimensional scanning
in the Borrmann pyramid is need for the T-T simulation.
Then, for the crystal whose thickness is 1.0 mm,
the T-T simulation is 10 times as fast as
compared with the E-L\&FFT simulation.

% \subsection{12波ケース}

\subsection{Twelve-beam case}

% 筆者は，立方晶の結晶においてひとつの円周上に
% 存在しうる逆格子点の数は，最大で12だと認識していた。
% しかし，最近になって16個の逆格子点が円周上に存在しうることを見いだした。
% これについては，計算機シミュレーションを行ったが，
% 実験を行っていないので，本稿では記述しない。
The present author recognized that
the largest number of reciprocal lattice nodes
that exist on a circle in the reciprocal space
is twelve for silicon crystal
before publishing the present paper.
On the other hand, he noticed that
sixteen reciprocal lattice nodes exist on a circle
in the reciprocal space.
However, the sixteen-beam case is not described
in the present paper.

\begin{figure}[!t]
% \centering
\begin{center}
\includegraphics[width=0.51\textwidth]{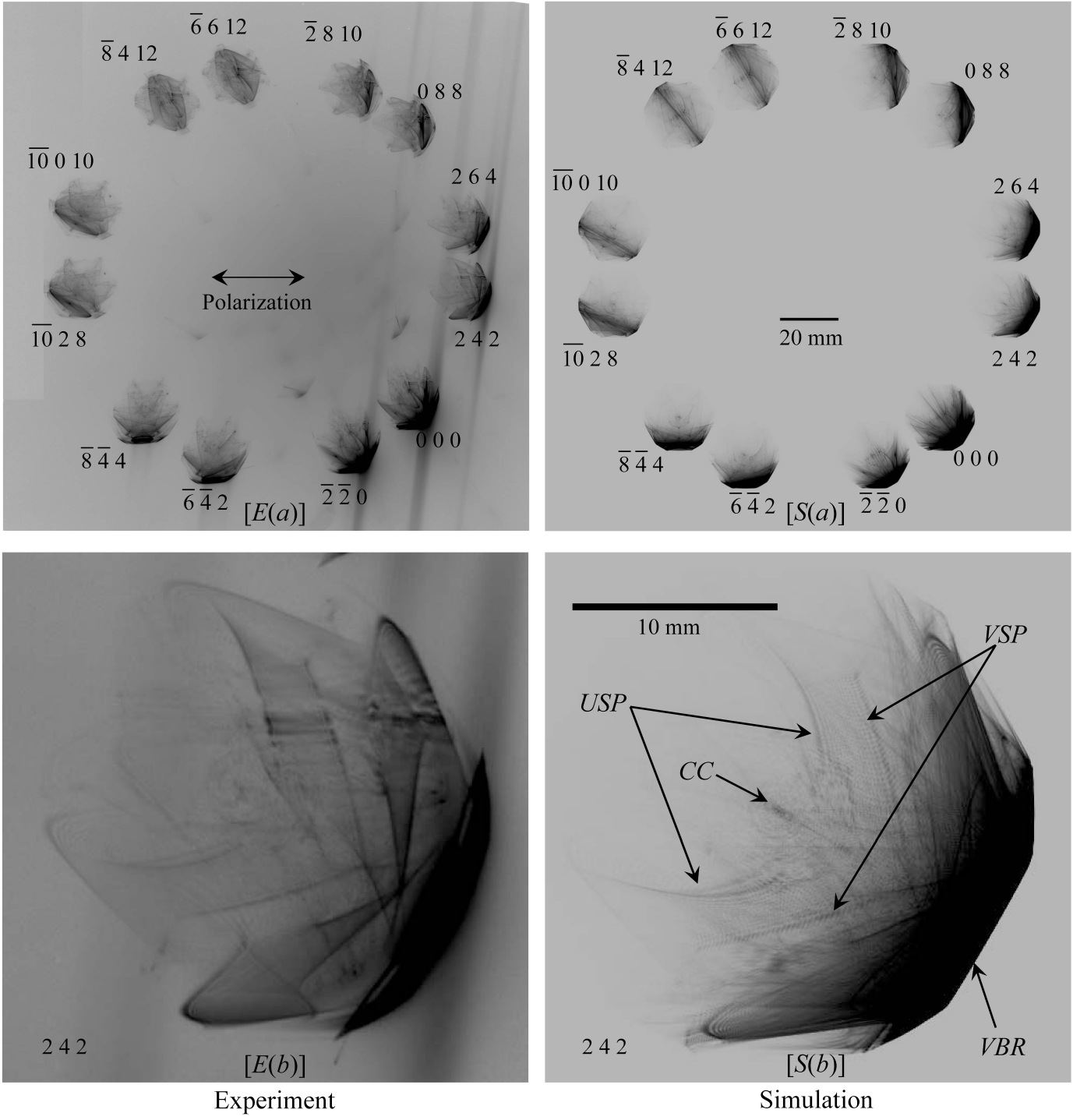}
\end{center}
\captionsetup{width=0.85\textwidth}
\caption[
$E(a)$ and $S(a)$ are experimentally obtained
and computer-simulated twelve-beam X-ray pinhole topographs
with an incidence of horizontal-linearly polarized X-rays whose photon energy
was 22.0 keV.
$E(b)$ and $S(b)$ are $2 \ 4 \ 2$ 
reflected X-ray images enlarged
from $E(a)$ and $S(a)$
(reproduction of Fig.\ 12 in Okitsu {\it et al.} (2012)).
        ]
        {
[$E(a)$] and [$S(a)$] are experimentally obtained
and computer-simulated twelve-beam X-ray pinhole topographs
with an incidence of horizontal-linearly polarized X-rays whose photon energy
was 22.0 keV.
[$E(b)$] and [$S(b)$] are $2 \ 4 \ 2$ 
reflected X-ray images enlarged
from [$E(a)$] and [$S(a)$]
(reproduction of Fig.\ 12 in Okitsu {\it et al.} (2012)\cite{okitsu2012}).
}
\label{Fig17_TwelveBeam}
\end{figure}

% Figs.\ \ref{Fig17_TwelveBeam}\ [$E(a)$],
% \ref{Fig17_TwelveBeam}\ [$S(a)$]は，
% 実験と計算機シミュレーションによる，
% 12波ケースのピンホールトポグラフである
% \cite{okitsu2012}。
% 22.0 keVに単色化された放射光を
% 水平偏光のまま，結晶に入射している。
% 反射指数は，図に示したとおりである。
% Figs.\ \ref{Fig17_TwelveBeam}\ [$E(b)$],
% \ref{Fig17_TwelveBeam}\ [$S(b)$]は，
% Figs.\ \ref{Fig17_TwelveBeam}\ [$E(a)$],
% \ref{Fig17_TwelveBeam}\ [$S(a)$]の
% $2\ 4\ 2$反射波の画像を拡大したものである。
Figs.\ \ref{Fig17_TwelveBeam}\ [$E(a)$] and
\ref{Fig17_TwelveBeam}\ [$S(a)$]
show twelve-beam pinhole topogpraphs
experimentally obtained and computer-simulated
based on the $n$-beam T-T equation
\cite{okitsu2012}.
The horizontally polarized synchrotron X-rays monochromatized
to be 22.0 keV
were directly incident on the silicon crystal.
The indices of reflections
are as shown in figures.
Figs.\ \ref{Fig17_TwelveBeam}\ [$E(b)$] and
\ref{Fig17_TwelveBeam}\ [$S(b)$]
are enlargements of $2\ 4\ 2$-reflected images in
Figs.\ \ref{Fig17_TwelveBeam}\ [$E(a)$] and
\ref{Fig17_TwelveBeam}\ [$S(a)$].

% 計算機シミュレーションによる
% Fig.\ \ref{Fig17_TwelveBeam}\ [$S(b)$]に見られる，
% 明るい領域[Very Bright Region ($VBR$)],
% V字形の模様[`V-Shaped' Pattern ($VSP$)],
% 中心に見える円状の模様[Central Circle ($CC$)],
% U字形の模様[`U-Shaped' Pattern ($USP$)]
% などの模様は，
% 実験によるトポグラフ
% Fig.\ \ref{Fig17_TwelveBeam}\ [$E(b)$]においても認められる。
[Very Bright Region ($VBR$)],
[`V-Shaped' Pattern ($VSP$)],
[Central Circle ($CC$)] and
[`U-Shaped' Pattern ($USP$)]
in Fig.\ \ref{Fig17_TwelveBeam}\ [$S(b)$]
are also found in
the experimentally obtained image in
Fig.\ \ref{Fig17_TwelveBeam}\ [$E(b)$].

% \subsection{18波ケース}

% \label{section_05_004_18BeamCase}

\subsection{18-beam case}

\label{section_05_004_18BeamCase}

\begin{figure}[!t]
% \centering
\begin{center}
\includegraphics[width=0.5\textwidth]{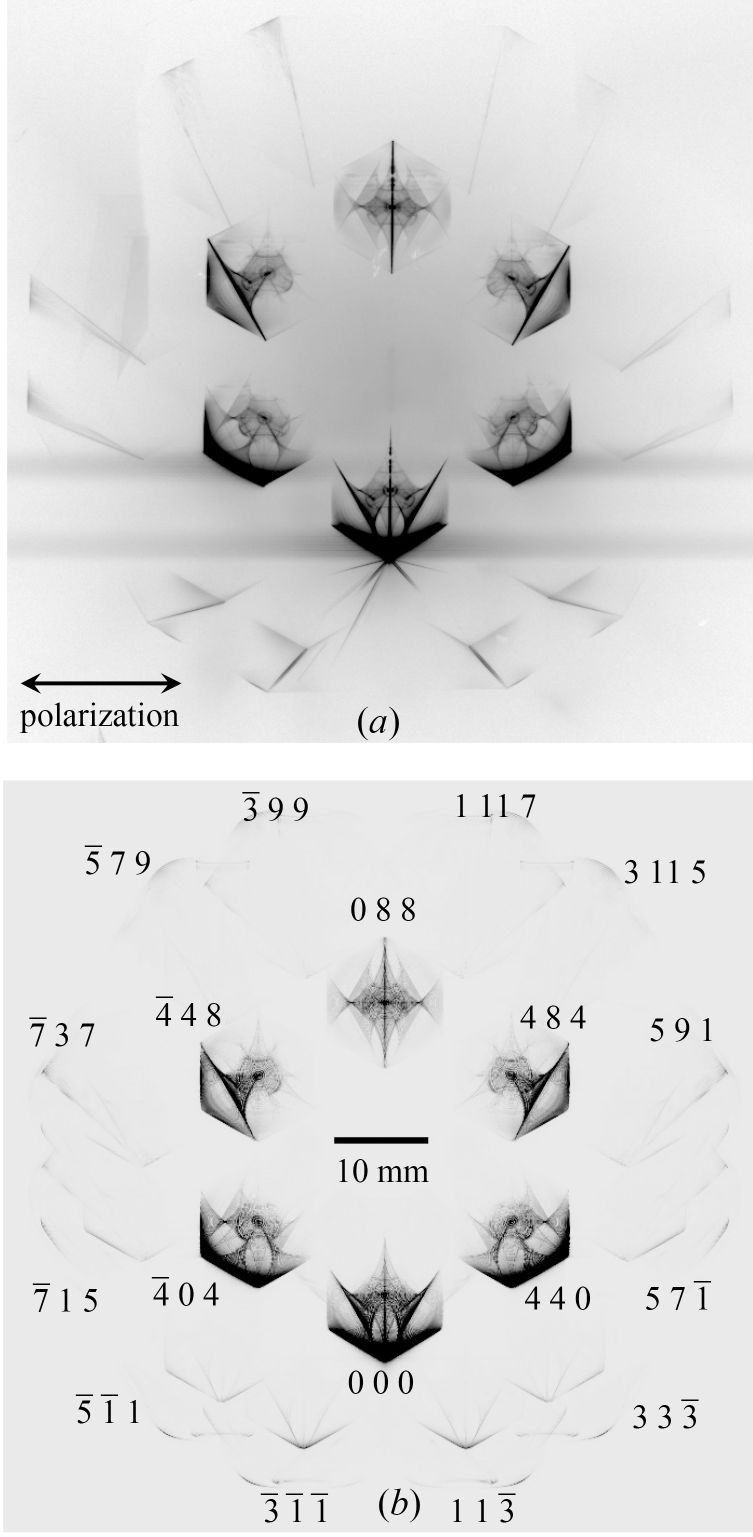}
\end{center}
\captionsetup{width=0.85\textwidth}
\caption[
($a$) Experimentally obtained and ($b$) E-L\&FFT simulated 18-beam pinhole topographs.
($b$) was obtained by the E-L\&FFT simulation under an assumption of
an incidence of X-rays with a photon energy $E = 21.98415$ keV
($\Delta E = E - E_0 = -0.25$ eV, where $E_0 = 21.98440$ keV)
(reproduction of Fig.\ 3 in Okitsu {\it et al.} (2019)).
        ]
        {
($a$) Experimentally obtained and ($b$) E-L\&FFT simulated 18-beam pinhole topographs.
($b$) was obtained by the E-L\&FFT simulation under an assumption of
an incidence of X-rays with a photon energy $E = 21.98415$ keV
($\Delta E = E - E_0 = -0.25$ eV, where $E_0 = 21.98440$ keV)
(reproduction of Fig.\ 3 in Okitsu {\it et al.} (2019)\cite{okitsu2019c}).
}
\label{Fig18_03_18Beam_Exp_Sim_Rev}
\end{figure}

% Figs.\ \ref{Fig18_03_18Beam_Exp_Sim_Rev}\ $(a)$, 
% \ref{Fig18_03_18Beam_Exp_Sim_Rev}\ $(b)$は，
% それぞれ，放射光実験とE-L\&FFTシミュレーションにより得られた，
% 18波ケースのピンホールトポグラフである
% \cite{okitsu2019c}。
Figs.\ \ref{Fig18_03_18Beam_Exp_Sim_Rev}\ $(a)$ and
\ref{Fig18_03_18Beam_Exp_Sim_Rev}\ $(b)$
are 18-beam pinhole topographs
experimentally obtained by using the synchrotron X-rays and
E-L\&FFT-computer-simulated
\cite{okitsu2019c}.
% Fig.\ \ref{Fig18_03_18Beam_Exp_Sim_Rev}\ $(a)$は，
% 22.0 keVの
% 光子エネルギーで6波ケースを狙ったものであったが，
Fig.\ \ref{Fig18_03_18Beam_Exp_Sim_Rev}\ $(a)$ were
experimentally obtained by aiming to take
six-beam pinhole topograph images
with the synchrotron X-rays at 22.0 keV.
% 外周にさらに12個のX線画像が得られた。
However, around the six topograph images
that were aimed to obtained,
additional twelve images were found.
% これを検討したところ，
% Fig.\ \ref{Fig18_03_18Beam_Exp_Sim_Rev}\ $(b)$
% に指数を示したように，エバルト球のごく近傍に
% 12個の逆格子点が存在していたことがわかった。
After careful consideration concerning the geometry
in the reciprocal space,
further twelve reciprocal lattice nodes
were found in the vicinity of the surface of the Ewald sphere.
% 逆格子点の並びを逆空間に作図したのが，
% Fig.\ \ref{Fig19_04_18BeamWithCharacters}である。
The arrangement of 18 ($= 6 + 12$) reciprocal lattice nodes
can be drawn as shown in
Fig.\ \ref{Fig19_04_18BeamWithCharacters}.
% この図からわかるように，
% ${\rm H}_0$, ${\rm H}_1$, ${\rm H}_2$, ${\rm H}_3$, ${\rm H}_4$, ${\rm H}_5$と% ，
% ${\rm H}_j$ $(j \in \{ 6，7,\cdots, 17 \})$から
% ${\rm La}_0$(ラウエ点)までの距離は，同じでない。
In reference to this figure,
The distances of ${\rm H}_i$
$(i \in \{ 0, 1, 2, 3, 4, 5 \})$ and
that of ${\rm H}_j$
$(j \in \{ 6, 7,\cdots, 17 \})$
from the Laue Point ${\rm La}_0$
are not the same.
% このため，光子エネルギーのわずかな変化で
% シミュレーション画像は大きく変化する。
% 21.98415 keVの光子エネルギーを仮定したとき，
% 実験結果をよく再現した。
Therefore,
dynamic change of the 18-beam topograph images
were found by changing slightly the photon energy.
When the photon energy was assumed to be 21.98415 keV,
good agreement between the experimentally obtained
and the E-L\&FFT-simulated pinhole topograph images
was found as shown in
Fig.\ \ref{Fig18_03_18Beam_Exp_Sim_Rev}.
% 式(\ref{eq16:ElementOfMatrixDash})
% の$n$波E-L理論は，
% \S \ref{section_03_002_algorithmEL}に記述した手順で
% 解を求めることができる。
To obtain the E-L\&FFT-simulated pinhole topograph images,
(\ref{eq16:ElementOfMatrixDash}) can be solved
with the procedure described in
\S \ref{section_03_002_algorithmEL}.
% 直接求められるのは，結晶を回転させたときの
% 回折振幅であるが，
Directly calculated based on (\ref{eq16:ElementOfMatrixDash})
were X-ray amplitude profiles when rotating the crystal.
% 8波ケースと同様，
% 結晶の入射側表面のX線がゼロでない振幅を持つという，
% 実空間でのデルタ関数の振幅をシミュレートするには，
% 正確な$n$波条件からのズレ角に依存せず，
% X線入射点で波の位相がそろっている必要がある。
Similarly in the eight-beam case,
it should be considered that
plain wave X-rays that consist of the incident X-rays
with the wave front of the delta function
are in phase at the incident point of the X-rays.

% Fig.\ \ref{Fig19_04_18BeamWithCharacters}を参照すると明らかだが，
% 18波ケースにおいては，
% Fig.\ \ref{Fig02_Omusubi_3D_Caption}\ ($b$)のような
% $n$角錐の「ボルマンピラミッド」を定義できない。
% この故に
% 18波ケースの計算を行うにあたって，
% $n$波T-T理論が，全く無力かというと
% そうではない。
% 式(\ref{eq40_TTPlaneWave2})は平面波入射のケースに使えるので，
% これを差分方程式に置き換えた
% 式(\ref{eq45:sabun})を解くことにより，
% 平面波入射の際の解を求めることができる。
% これを高速フーリエ変換することで，
% ピンホールトポグラフを計算することができる。
% いわばT-T\&FFTシミュレーションが可能である。
% これについては，現在論文を準備中である。
It is clear from the reference to
Fig.\ \ref{Fig19_04_18BeamWithCharacters}
that the Borrmann pyramid as shown in
Fig.\ \ref{Fig02_Omusubi_3D_Caption}\ ($b$)
cannot be defined for this 18-beam case.
However, the T-T simulation
is not completely ineffective.
(\ref{eq40_TTPlaneWave2}) can be applied to
the $n$-beam case for plane-wave incidence.
Difference equation (\ref{eq45:sabun})
derived from (\ref{eq40_TTPlaneWave2}) can be
The amplitude profile calculated by solving (\ref{eq45:sabun})
can be fast Fourier-transformed
to obtain the $n$-beam topograph images
used to simulate the $n$-beam case
even when $n = 18$ like this
(T-T\&FFT simulation).
Regarding this method,
a separate paper is in preparation.

% \section{終わりに}

\section{Concluding remarks}

% Fig.\ \ref{Fig20_Giltch_Si220_Characters}は，
% シリコン$2\ 2\ 0$反射について計算した
% グリッチマップである。
% 横軸は光子エネルギー(eV)，
% 縦軸$(\psi)$は，$[1\ 1\ 0]$軸周りに結晶を回転させた角度で，
% ${\mathbf K}_{000} \times {\mathbf K}_{220}$が
% $[0\ 0\ 1]$方向に平行になるとき，$\psi = 0$である。
% ${\mathbf K}_{000}$と${\mathbf K}_{220}$は，それぞれ，
% 透過波と$2\ 2\ 0$反射波の波数ベクトルである。
% 枠内の曲線はすべてグリッチで，
% これらは，$2\ 2\ 0$以外の逆格子点が，
% エバルト球表面に同時に存在することで，
% 2波近似が破れることにより発生する。
Fig.\ \ref{Fig20_Giltch_Si220_Characters}
is a glitch map (simultaneous reflection map)
calculated for silicon $2\ 2\ 0$ reflection.
The abscissa is the photon energy (eV) of the X-rays.
The ordinate $(\psi)$ is the rotation angle of the crystal
around $[1\ 1\ 0]$ axis.
$\psi = 0$ when
${\mathbf K}_{000} \times {\mathbf K}_{220}$
is parallel to the $[0\ 0\ 1]$ direction.
${\mathbf K}_{000}$ and ${\mathbf K}_{220}$
are the wavevectors of $0\ 0\ 0$-forward diffracted and
$2\ 2\ 0$-reflected X-rays.
All red curves found in the glitches
owing to the simultaneous reflections.
These are caused by the reciprocal lattice nodes
other than $2\ 2\ 0$
existing simultaneously on the surface of the Ewald sphere
to beak the two-beam condition.

% モノクロメーターや，
% 偏光子，検光子，移相子といったX線結晶光学素子を
% 設計するにあたって，グリッチの検討は非常に重要である。
% 例えばシリコン$2\ 2\ 0$反射を用いて
% 波長掃引を行う際，
% $2\ 2\ 0$以外のブラッグ条件を同時に満たすエネルギーでは，
% 2波近似が破れてしまい，
% 光学素子にグリッチ(不具合)が発生する。
% Fig.\ \ref{Fig20_Giltch_Si220_Characters}を見ると，
% グリッチの密度は，低エネルギー領域よりも
% 高エネルギー領域で高いことがわかる。
% 光子エネルギーを固定して，
% $[2\ 2\ 0]$軸周りにエバルト球を$360^{\circ}$回転させると，
% 逆空間にリンゴのような形の立体ができる。
% この立体の中に存在する逆格子点は，
% すべてグリッチの原因となる。
% グリッチ密度は，
% この立体の体積に比例するため，
% 概ね光子エネルギーの3乗に比例することになる。
The consideration on the glitches are important
when designing such X-ray optical device as
the monochromator, polarizer, analyzer and/or
phase retarder.
When scanning the photon energy {e.g.}
by using the silicon $2\ 2\ 0$ reflection,
The two-beam approximation is broken at the photon energy
where reciprocal lattice nodes whose indices are other
than $2\ 2\ 0$ exist on the surface of the Ewald sphere
to cause the glitches (defects).
In reference to Fig.\ \ref{Fig20_Giltch_Si220_Characters},
it can be found that
the density of glitches is higher in the heigh-energy region
compared with the low-energy region.

\begin{figure}[!t]
% \centering
\begin{center}
\includegraphics[width=0.30\textwidth]{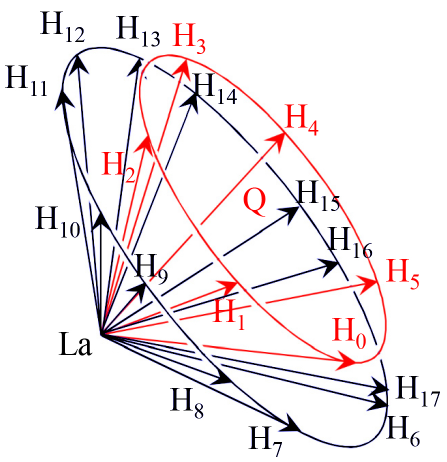}
\end{center}
\captionsetup{width=0.85\textwidth}
\caption[
Six reciprocal lattice nodes are
on a red (smaller) circle in reciprocal space.
Outside of this circle,
a black (larger) circle was observed on which
twelve reciprocal lattice nodes were present.
$Q$ is the center of the red (smaller) circle
(reproduction of Fig.\ 4 in Okitsu {\it et al.} (2019))
        ]
        {
Six reciprocal lattice nodes are
on a red (smaller) circle in reciprocal space.
Outside of this circle,
a black (larger) circle was observed on which
twelve reciprocal lattice nodes were present.
$Q$ is the center of the red (smaller) circle
(reproduction of Fig.\ 4 in Okitsu {\it et al.} (2019)
\cite{okitsu2019c}).
}
\label{Fig19_04_18BeamWithCharacters}
\end{figure}

% X線光学素子のグリッチマップを参照して，
% エネルギースキャン範囲にグリッチが存在しないよう，
% $\psi$の値を予め調整しておくと，
% 2波近似が大きく破れることなく，
% エネルギー掃引を行うことができる。
When an energy-scanning experiment is done
by using the X-ray optical device,
the value of $\psi$ can be adjusted such that
threre is no reciprocal lattice node causing the
two-beam approximation in the vicinity
of the surface of Ewald sphere
% 10 keV程度以下のエネルギー領域では，
% これが行われている。
In the energy ranges bellow 10 keV or so,
such energy scan experiments are usually done
by eliminating the glitches.
% しかし，高エネルギー領域では，
% エネルギーの3乗に比例して高密度となるグリッチを避けることは困難になり，
% 20 keV程度以上では，スペクトロスコピーが
% ほとんど不可能になってしまう。
However, it becomes difficult to eliminate the glitches
whose density is proportional to the third power of the photon energy
in the energy range higher than 20 keV.
% 第3世代の放射光のスペクトルは，
% 高エネルギー側に伸びており，
% 是非この領域を利用したいが，
% 2波の動力学理論のみが頼りでは，
% 結晶光学素子の設計ができなくなってしまう。
X-rays whose intensity is extremely high,
are available at experimental stations of
the third-generation synchrotron radiation sources.
However, the X-ray optical devices cannot be designed
only based on the two-beam dynamical diffraction theories
in such a high energy range
where the two-beam approximation is always broken.
% 2波近似が常に破れているような，
% 高エネルギー領域でのX線結晶光学素子の設計には，
% 式(\ref{eq18:eln-beamDashDash})の多波E-L理論ないしは，
% 式(\ref{eq34_ELtoTTGeneral02})の多波T-T方程式を用いた計算と，
% $\psi$の回転機構を備えたゴニオメーターによる，
% 高度な結晶制御が不可欠になるであろう。
It is considered to be necessary
to design the X-ray optical devices working in the high energy ranges
based on the $n$-beam X-ray dynamical theories
described by (\ref{eq18:eln-beamDashDash})
and/or (\ref{eq34_ELtoTTGeneral02}).
Simultaneously, advanced technique becomes necessary to control
two or three axes {e.g.} $\theta$ and/or $\psi$
of the goniometer on which
the crystal devices are mount.

% 2波近似が破れ，多波回折が無視できなくなるもう一つのケースは，
% 結晶の単位胞が大きく，
% 逆格子点の密度そのものが大きくなる場合である。
On the other hand,
the $n$-beam effect cannot be ignored also when
the two-beam approximation is always broken due to
the large size of crystal lattice {e.g.}
in the protein crystal structure analysis.
% 1990年代後半から，
% 単結晶構造解析において，
% 2次元検出器の利用が一般化し，
% 現在なお，その高度化が進んでいる。
Since the late 1980s,
the use of two-dimensional detector generalized
and are becoming more sophisticated
in the crystal structure analysis.
% 低分子の有機物結晶の場合でも数十個，
% タンパク質の結晶では，数百個から数千個の
% X線回折スポットが，結晶を静止した場合でも，
% 検出器に記録される。
Many diffraction spots;
several dozen even in the cases of small molecular crystals and
several hundred those are found in the cases of protein crystals,
are simultaneously found in general.
% この状況を目の当たりにすると，
% 2波近似が破れていないとは考えにくい。
When seeing such situations,
it is difficult to consider that
the two-beam approximation is not broken.
% 低分子結晶構造解析においてさえ，
% このことは，レニンガー効果
% \cite{renninger1937}
% としてよく知られている
% \cite{ohashi2005}。
such cases where the two-beam approximation is broken
even in the small molecular crystals,
are well known as the Renninger effect
\cite{renninger1937}.
% タンパク質結晶の場合，
% 結晶構造解析完了後に評価される$R$因子が
% $10\%$を下回ることは稀である。
In the case of protein crystals,
it is rare that the reliability factor ($R$-factor) evaluated
after the determination of the molecular structure
is less than $10\%$.
% $R$因子が$10\%$まで下がっても，
% その定義式から考察すると，
% 決定された分子構造から計算されるX線回折強度と，
% 実測される回折強度との間には，
% 加重平均をとって$20\%$にも及ぶ食い違いがあることになる。
Even if the $R$-factor is evaluated to be $10\%$,
it means the weighted discrepancy up to $20\%$
between the X-ray diffraction intensities
experimentally observed and calculated based on the kinematical theory
from the determined molecular structure.
% まだ仮説の段階ではあるが，
% タンパク質結晶に対して$R$因子が下がらないのは，
% 2波近似の破れが原因だ，と筆者は考えている。
% 同時に生じている多くの波を考慮する，
% 式(\ref{eq18:eln-beamDashDash})ないしは
% 式(\ref{eq34_ELtoTTGeneral02})を用いて
% X線回折強度を計算することにより，
% 実測強度との食い違いが大きく軽減されるのではないだろうか。
% だとすれば，2波理論に代わり，
% 多波理論により結晶構造解析を行う時代が，来るかも知れない。

\textbf{
The present author has a hypothesis
concerning the too large values of the $R$-factor for protein crystals
% that the too large values of $R$-factor
% in the protein crystal structure analysis
that this problem
is caused by the bankruptcy of the two-beam approximation
due to the large density of reciprocal lattice nodes
compared with the cases of small-molecule crystals.
If the crystal structure factor $F_c(\mathbf{h})$ for $\mathbf{h}$ reflection
were estimated by using the $n$-beam theory taking into account
the reciprocal lattice nodes in the vicinity of the surface of the Ewald sphere
to be compared with those measured by the experiment $F_o(\mathbf{h})$,
the $R$-factor might be decreased dramatically.
If that is the case for protein crystals,
the crystal structures (and the phase problem) for protein crystals
become to be solved by using the $n$-beam dynamical theory
in place of the two-beam (kinematical) theory.
}

% 加藤範夫の1995年の著書
% \cite{kato1995}
% 第5章冒頭に，次のような記述がある。

In Kato's book published in 1995,
there is a description [in Japanese] as follows:
% 「結晶回折の歴史を概観すると，動力学理論の骨格は，
% % ラウエらの回折現象発見の直後，
% % ダーウィン(C. G. Darwin ; 1914)
% % や
% % エワルト(P. P. Ewald ; 1917)
% によって確立されている。
When overviewing the history pf the X-ray diffraction in crystals,
the backburn of the dynamical theory has been established
by Darwin (1914)
and by Ewald (1917)
just after the discovery of the phenomenon of X-ray diffraction by von.\ Laue.
% ラウエの回折条件(2.30)や
% ブラッグの式(2.29)に代表される
% 運動学理論が安心して用いられたのは，
% 動力学理論による基礎づけがあったからである」。
The kinematical diffraction theory could be felt safe to use
since its foundation
% Laue's and Bragg's diffraction condition
has been given by their dynamical theories.

\begin{figure}[!t]
% \centering
\begin{center}
\includegraphics[width=0.75\textwidth]{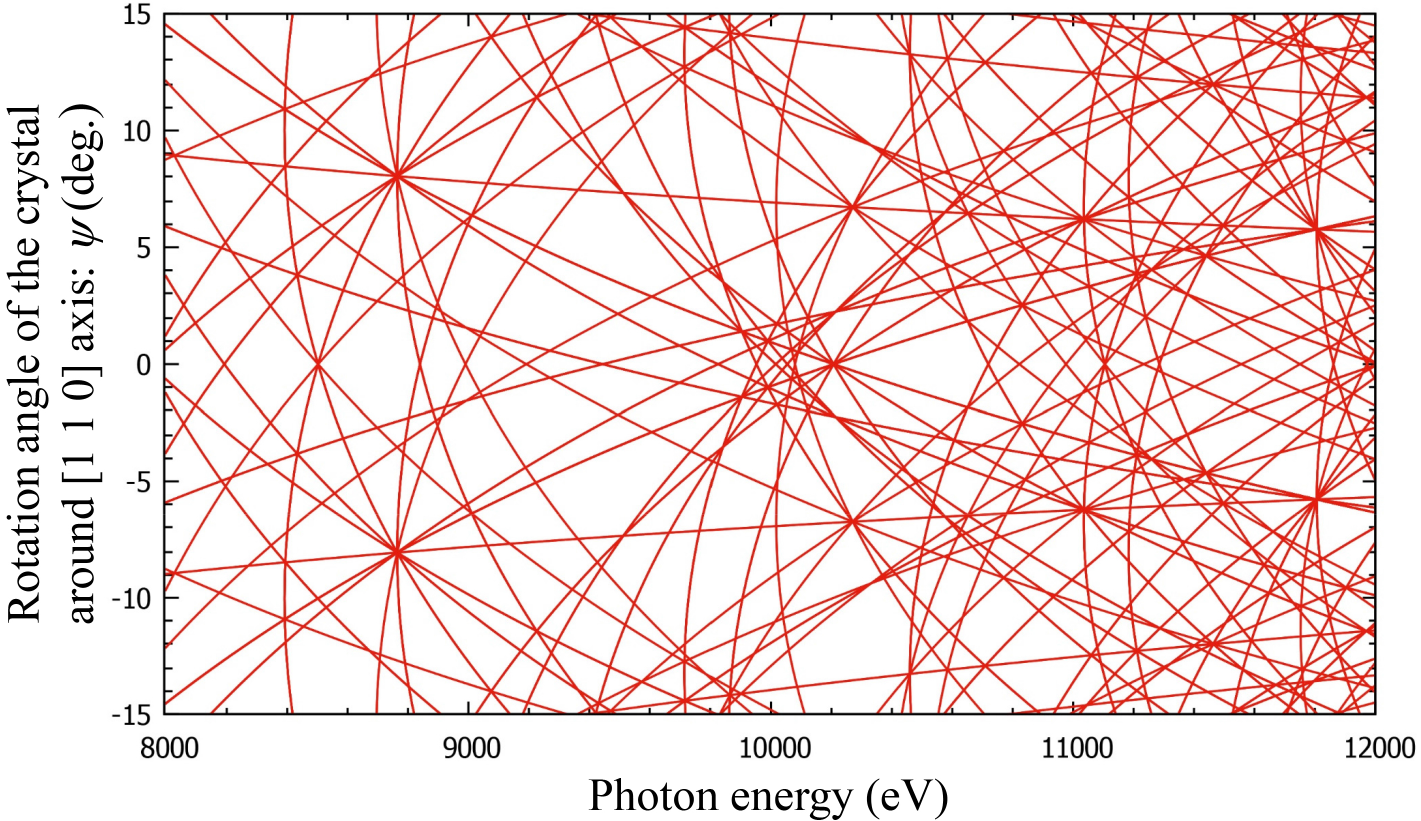}
\end{center}
\captionsetup{width=0.85\textwidth}
\caption{
Glitch map (simultaneous reflection map)
for silicon $2\ 2\ 0$ reflection.
$\psi$ (the ordinate) is rotation angle ($^{\circ}$) around $[1\ 1\ 0]$ axis.
The abscissa is X-ray photon energy (eV).
$\psi$ is $zero$ when ${\mathbf K}_{000} \times {\mathbf K}_{220}$
is parallel to $[0\ 0\ 1]$ direction.
${\mathbf K}_{000}$ and ${\mathbf K}_{220}$
are wave vectors of $0\ 0\ 0$ transmitted and $2\ 2\ 0$ reflected X-rays.
}
\label{Fig20_Giltch_Si220_Characters}
\end{figure}

% 筆者は，2002年に逝去した加藤を深く尊敬していた。
% 彼は，多波ケースの研究には全く手を付けなかった。
The present author had a lot of respect for Kato
who passed away in 2002.
However, he untouched almost at all the $n$-beam diffraction cases.
% 1997年，筆者は，その理由について，
% 彼に訊ねたことがある。
In 1997, the present author asked him about the reason.
% 「ラウエケースとブラッグケースが混在したとき，
% 解が得られなくなってしまうから」
% \cite{kato1997}
% との答えだった。
He answered that he thought the solution of the dynamical theory
cannot be obtained when Laue and Bragg geometries are mixed.
% 当時筆者は，すでに
% ブラッグケースとラウエケースが混在する場合の
% 3波ケースの計算機シミュレーション結果を得ていた。
However, the present author has already obtained the numerical solution
of the T-T dynamical theory
for a three-beam case when Laue and Bragg geometries were mixed.
% このことを告げると，
% 加藤は「え?」という表情を見せた。
When telling him about this,
he seemed to be confused.

% 動力学理論の裏付けがあったから
% 2波近似の運動学理論が安心して使えた，ということは，
% それが破れていることが明らかになれば，
% 安心できないのではないだろうか。
Is it can be said that we cannot feel safe to use
the kinematical theory based on the two-beam approximation when
it is clarified to be broken ?

% 1949年，リプスコムは
% \cite{lipscomb1949}，
% 3波ケースのX線回折強度プロファイルに
% 結晶構造因子の位相情報が含まれていることを指摘した。
In 1949, Lipscomb suggested
\cite{lipscomb1949}
that the phase information of the crystal structure factor
can be extracted from the X-ray diffraction profile
in principle.
% $n$波E-L理論の数値解を初めて報告した，
% 1974年のコレラの論文
% \cite{colella1974}
% では，これを引用し，研究の目的を
% タンパク質結晶の位相決定だとしている。
In the introduction of the famous article published by Collela in 1974
\cite{colella1974},
the purpose of the study was to determine the phases
of crystal structure factor
by referring the Lipscomb's article.
% 以来46年が経過したが，これは実現していない。
However, this has not been realized even today.

% タンパク質結晶の位相決定は，2波近似を前提として，
% 重原子置換法や，
% アミノ酸のひとつであるメチオニンが持つ
% イオウをセレンに置換して行う異常分散法
% \cite{hendrickson1989}
% を用いて，成功を収めてきている。
The phase problem in protein crystallography has been overcome
mainly by using the heavy atom replacement and/or
the method to replace methionine (one of the 20 amino acids included in
molecules of the proteins) with selenomethionine
that has selenium atom in place of sulfur
\cite{hendrickson1989}
based on the two-beam approximation.
The molecular replace methods are usually used
when the similar or partial structures of the molecule
have been determined
by using the above-mentioned phasing methods.
% それでもなお，
% イオウの異常分散を利用する位相決定法が模索されるほど，
%多波ケースを用いた結晶構造解析が一般化すれば，
% ネイティブなタンパク質による位相決定は魅力的である。
However, the phase determination of native protein crystals
using the anomalous dispersion of sulfur
is being surveyed due to its advantage without replacement.
% X線を照射中の結晶方位の検討から，
% エバルト球表面近傍に存在する
% 逆格子点の指数付けのみで，
% 位相が決定できるようになるかも知れない。

\textbf{
When the molecular structures of protein crystals
are obtained based on the n-beam theory in the future,
the phases of the crystal structure factors might be determined only by
indexing the diffraction spots simultaneously recorded
on the two-dimensional detector.
}

% 3波ケースを実現するのは，ほぼ不可能である。
It is impossible to realize the three-beam case
for protein crystals whose density of reciprocal lattice nodes
are extremely high.
% しかし，Fig.\ \ref{Fig19_04_18BeamWithCharacters}
% に示したように，
% 同一円周上にない18個の逆格子点が
% エバルト球表面，
% あるいはエバルト球表面近傍
% に存在する状況が，
% 式(\ref{eq18:eln-beamDashDash})の$n$波E-L理論で記述され，
% これの数値解をフーリエ変換することにより，
% Fig.\ \ref{Fig18_03_18Beam_Exp_Sim_Rev}\ $(b)$が
% 得られている。
However, the $n$-beam Ewald-Laue (E-L) dynamical theory as
described in (\ref{eq18:eln-beamDashDash})
can deal with the cases where
the $n$-reciprocal lattice nodes not on a circle exist
in the vicinity of the surface of the Ewald sphere.
The 18-beam pinhole topography is such a case.
The 18 reciprocal lattice node not on a circle
in the reciprocal space but exist in the vicinity
of the surface of the Ewald sphere as shown in
Fig.\ \ref{Fig19_04_18BeamWithCharacters}
has been computer-simulated and agreed well
with the experimental result
as shown in Fig.\ \ref{Fig18_03_18Beam_Exp_Sim_Rev}\ $(b)$.
% 式(\ref{eq17:eln-beamDash})ないしは
% 式(\ref{eq18:eln-beamDashDash})をフーリエ変換することにより，
% 式(\ref{eq34_ELtoTTGeneral02})や
% 式(\ref{eq38_TTPlaneWave})で記述される，
% $n$波T-T方程式も導出されている。
The $n$-beam T-T equation described as (\ref{eq34_ELtoTTGeneral02})
and/or (\ref{eq38_TTPlaneWave}) has been derived by
Fourier-transforming the E-L theory (\ref{eq17:eln-beamDash})
and/or (\ref{eq18:eln-beamDashDash}).
% これらは，エバルト球表面近傍にある逆格子点を
% すべて考慮して，数値解を求めることができる。
These can numerically be solved by taking into account
the existence of all reciprocal lattice nodes
in the vicinity of the surface of the Ewald sphere.
% 現在，100個程度の波が同時に強い状況で，
% 式(\ref{eq18:eln-beamDashDash})の多波E-L理論，および
% 式(\ref{eq34_ELtoTTGeneral02})，
% 式(\ref{eq35_ELtoTTGeneralWithDisplacement})の多波T-T方程式
% の数値解を求めて，
% X線反射強度を計算するプログラムを開発中である。
The present author is now developing the computer program
to solve the $n$-beam E-L theory described as (\ref{eq18:eln-beamDashDash})
and $n$-beam T-T theory as (\ref{eq34_ELtoTTGeneral02}) and/or
(\ref{eq35_ELtoTTGeneralWithDisplacement}) for $n \sim 100$.
% 重い計算になることが予想される。
It will be time-consuming.

% 計算機の能力は，演算速度，メモリー容量，ハードディスク容量の
% いずれにおいても，
% 5年でおよそ一桁のペースで向上しつつある。
% さらに量子コンピューターが，
% そう遠くない将来，実用化されるだろう。
% こういった現況は，
% 多波($n$波)動力学理論の今後を考えるにあたり，非常に重要である。
The abilities of computers are rapidly being improved
in all cases about calculation speed, memory capacity and hard disk capacity
The quantum computer may be realized in the future.
These situations are important when considering
the perspective of the $n$-beam theory.
% E-L理論が波を逆空間で記述するのに対して，
% T-T理論は，実空間での振る舞いを記述する。
% これらの理論が等価であることを踏まえつつ，
% 引き続き$n$波動力学理論の
% 計算手法を検討してゆきたいと考えている。
The present author will continue to study on the computer simulation
of the $n$-beam X-ray diffraction
with the equivalence in mind between the E-L and T-T dynamical theories.

% \begin{center}
% {\large \bf 謝辞}
% \end{center}
% 本研究は，
% 文部科学省 科学技術振興調整費先導的研究等の推進
% 「アクティブ・ナノ計測基盤技術の確立」プロジェクトの一環として，
% また，東京大学大学院工学系研究科 総合研究機構 ナノ工学研究センターにおいて，
% ナノテクプラットホームプロジェクトの一環として行われた。

\section*{Acknowledgement}

The supercomputer system `sumire', `kashiwa' and `sekirei'
of the Institute for Solid
State Physics, the University of Tokyo
and `TSUBAME 3.0' of Tokyo Institute of Technology
were used for the computer simulations.
The authors are indebted to Dr X.-W. Zhang of KEK Photon Factory,
and Dr T. Oguchi of SPring-8 Service Corporation
for their technical support in the
present experiments
and also to Professor Emeritus S. Kikuta
of The University of Tokyo
and Professor Emeritus H. Hashizume
of Tokyo Institute of Technology
for their encouragement and fruitful discussions regarding
the present work.

The experimental part of the present work
was conducted in cooperation with
Dr. Y. Yoda and Dr. Y. Imai of SPring-8 JASRI
and Dr. Y. Ueji of Rigaku Corporation.

% 計算に用いたスーパーコンピューターは，
% 東京大学物性研究所の
% `sumire', `kashiwa'および`sekirei'，
% 東京工業大学の`TSUBAME 3.0'
% である。

% 実験は，
% SPring-8 BL09XUにおいて，
% 高輝度光科学研究センター（JASRI）の承認
% (Proposal No.
% 2002A 0499-NMD3-np,
% 2003B 0594-NM-np,
% 2004A 0330-ND3c-np,\linebreak
% 2004B 0575-ND3c-np)
% のもと行われた。
% 
% また予備実験は，
% 物質構造科学研究所 Photon Factory ARNE3Aにおいて，
% 放射光共同利用実験審査委員会（PF PAC）の承認
% (Proposal No. 2003G 202, 2003G 203)
% のもと行われた。

% 実験は，高輝度光科学研究センターの今井康彦博士，依田芳卓博士，
% 東京大学大学院新領域創成科学研究科の上ヱ地義徳博士(現 株式会社リガク)
% の協力を得て行われたことを明記し，感謝の意を表します。
% 
% 
% 本研究の意義を理解してくださり，励ましを頂いた，
% 東京大学工学系研究科名誉教授，
% 菊田惺志先生に深く感謝致します。

\section*{Funding information}

The theoretical component and the computer simulations of the
present work were supported by the Nanotechnology Platform
Project (No. 12024046) of the Ministry of Education,
Culture, Sports, Science and Technology (MEXT), Japan.

The preliminary experiments were performed at
BL4A and BL15C of Photon Factory
and AR-NE3A of
Photon Factory AR under the approval of the Photon Factory
Program Advisory Committee
(Proposal Nos. 97G-179, 97G-180, 99S2-003,
2003G202 and
2003G203).
The main experiments were performed at
BL09XU of SPring-8 under the approval of the Japan
Synchrotron Radiation Research Institute (JASRI)
(Proposal No.
2002A 0499-NMD3-np,
2003B 0594-NM-np,
2004A 0330-ND3c-np,
2004B 0575-ND3c-np,
2005B 0714 and
2009B 1384)

% \newpage

% \bibliographystyle{plain}
% \bibliography{000_2024_08_17_001_NBeamTheory}

\clearpage

\medskip

\bibliographystyle{naturemag}%Used BibTeX style is unsrt
\bibliography{000_2024_09_26_001_NBeamTheory}

% \addcontentsline{toc}{chapter}{Bibliography}
% \printbibliography

\end{document}